\documentclass[a4paper]{article}
\usepackage[dvipsnames]{xcolor}
\usepackage[utf8]{inputenc}
\usepackage[english]{babel}
\usepackage{amsmath}
\usepackage{amssymb}
\usepackage{amsthm}

\usepackage{hyperref}
\usepackage{multicol}
\usepackage{tikz}
\usetikzlibrary{shapes,arrows}
\usepackage[nottoc]{tocbibind}
\usepackage{graphicx}
\usepackage{bbold}
\usepackage{csquotes}
\usepackage{comment}

\usepackage[a4paper,width=150mm,top=25mm,bottom=45mm,bindingoffset=6mm]{geometry}
\usepackage{fancyhdr}
\usepackage{mathrsfs}
\usepackage[normalem]{ulem}

\usepackage{import}

\usepackage[capitalize]{cleveref}

\usepackage[doi=false,isbn=false, url=false, date=year, backend=biber,style=numeric,
  ]{biblatex} %Imports biblatex package
\renewbibmacro*{journal+issuetitle}{%
  \usebibmacro{journal}%
  \setunit*{\addspace}%
  \iffieldundef{series}
    {}
    {\newunit
     \printfield{series}%
     \setunit{\addspace}}%
  \usebibmacro{volume+number+eid}%
  \setunit{\bibpagespunct}%
  \printfield{pages}%
  \setunit{\addspace}%
  \usebibmacro{issue+date}%
  \setunit{\addcolon\space}%
  \usebibmacro{issue}%
  \newunit}

\renewbibmacro*{note+pages}{%
  \printfield{note}%
  \newunit}

\newcommand{\E}{\mathbb{E}}
\newcommand{\tr}{\mathrm{tr}}
\newcommand{\e}{\mathrm{e}}
\newcommand{\diam}{\mathrm{diam}}
\renewcommand{\i}{\mathrm{i}}
\renewcommand{\Phi}{\varPhi}
\renewcommand{\Psi}{\varPsi}
\renewcommand{\Sigma}{\varSigma}
\newcommand{\epsi}{\varepsilon}

\newcommand{\g}{\gamma}
\newcommand{\La}{\Lambda}
\newcommand{\tg}{\Tilde{\gamma}}

\newcommand{\w}{\omega}
\newcommand{\N}{\mathbb{N}}
\newcommand{\Z}{\mathbb{Z}}
\newcommand{\C}{\mathbb{C}}

\newcommand{\mA}{\mathcal{A}}
\newcommand{\R}{\mathbb{R}}
\newcommand{\eps}{\varepsilon}
\newcommand{\ODeps}{^{\mathrm{OD} \varepsilon}   }
\newcommand{\OD}[1][]{   ^{\mathrm{OD}#1}   }
\newcommand{\mL}{\mathcal{L}}

\usepackage{mathtools}

\DeclarePairedDelimiter\floor{\lfloor}{\rfloor}

\DeclarePairedDelimiter\norm{\lVert}{\rVert}
\DeclarePairedDelimiter\abs{\lvert}{\rvert}
\DeclarePairedDelimiter\br{\lparen}{\rparen}
\DeclarePairedDelimiter\sq{\lbrack}{\rbrack}

\theoremstyle{plain}
\newtheorem{theorem}{Theorem}[section]
\newtheorem{lemma}[theorem]{Lemma}
\newtheorem{proposition}[theorem]{Proposition}
\newtheorem{corollary}[theorem]{Corollary}

\theoremstyle{definition}
\newtheorem{definition}[theorem]{Definition}

\theoremstyle{remark}
\newtheorem{remark}[theorem]{Remark}

\usepackage{parskip}

\usepackage{colonequals}

\usepackage{fullpage}

\title{Near linearity of the macroscopic Hall current response in infinitely extended gapped fermion systems}
\author{
    Marius Wesle%
    \texorpdfstring{\footnote{\parbox[t]{.7\textwidth}{
                \foreignlanguage{ngerman}{Fachbereich Mathematik, Eberhard Karls Universität Tübingen,\\
                Auf~der~Morgenstelle~10, 72076~Tübingen,} Germany
            }
        }
    }{}%
    \and Giovanna Marcelli%
    \texorpdfstring{%
        \footnote{% chktex 42
            \parbox[t]{.7\textwidth}{
                \foreignlanguage{italian}{Dipartimento di Matematica e Fisica, Università di Roma Tre,\\ L.go S. L. Murialdo 1, 00146 Roma,} Italy
            }
        }
    }{}%
    \and Tadahiro Miyao% 
    \texorpdfstring{%
        \footnote{% chktex 42
            \parbox[t]{.7\textwidth}{
                \foreignlanguage{italian}{Department of Mathematics,  Hokkaido University,
                 Sapporo 060-0810,} Japan
            }
        }
    }{}%
    \and Domenico Monaco%
    \texorpdfstring{%
        \footnote{% chktex 42
            \parbox[t]{.7\textwidth}{
                \foreignlanguage{italian}{Dipartimento di Matematica ``Guido Castelnuovo'', Sapienza Università di Roma,\\ Piazzale Aldo Moro 5, 00185 Roma, Italy}
            }
        }
    }{}%
    \and Stefan Teufel%
    \texorpdfstring{%
        \footnotemark[1]% chktex 42
    }{}%
}

\date{\today}

\usepackage{microtype}
\begin{document}

\maketitle
\begin{center}\large \textit{
Dedicated to the memory of Professor Huzihiro Araki}
\end{center}

\bigskip

\begin{abstract}
We consider an infinitely extended system of fermions on a $d$-dimensional lattice with (magnetic) translation-invariant short-range interactions.  We further assume that the system has a  locally unique gapped ground state. Physically, this is a model for the bulk of a generic topological insulator at zero temperature, and we are interested in the current response of such a system to a constant external electric field. Using the \textit{non-equilibrium almost-stationary states} approach, we prove that the longitudinal current density induced by a constant electric field of strength $\eps$ is of order $\mathcal{O}(\eps^\infty)$, i.e.\ the system is an insulator in the usual sense. For the Hall current density we show instead that it is linear in $\eps$ up to terms of order $\mathcal{O}(\eps^\infty)$. The proportionality factor $\sigma_\mathrm{H}$ is by definition the Hall conductivity, and we show that it is given by a generalization of the well known  double commutator formula to interacting systems. As a by-product of our results, we find that the Hall conductivity is constant within gapped phases, and that for $d=2$ the relevant observable that ``measures'' the Hall conductivity in experiments, the Hall conductance, not only agrees with $
\sigma_{\mathrm{H}}$  in expectation up to $\mathcal{O}(\eps^\infty)$, but also  has vanishing variance.

A notable difference to several existing results on the current response in interacting fermion systems is that we consider a macroscopic system exposed to a small constant electric field, rather than to a small voltage drop.
\end{abstract}

\section{Introduction}

Understanding the experimentally  observed quantization of Hall conductance at low temperatures has  inspired a great deal of work  in mathematics and theoretical physics.  As a theoretical problem, the quantum Hall effect has many different interesting aspects, and our work deals with one of them. In a nutshell, the theorems we prove here show that the conductivity model usually assumed inside the incompressible stripes of a Hall bar can be derived from a semi-realistic microscopic model with very high accuracy. More fundamentally, we also prove that insulators do indeed insulate, even when the applied voltage exceeds the spectral gap by many orders of magnitude.
Conceptually, but not technically,  our results are generalizations of recent results \cite{marcelli2022purely} on fermion systems of non-interacting particles to the case with interactions. 
We postpone
 a more detailed discussion of their physical significance  and their relation to the vast existing literature on this subject to the end of the introduction.

Instead, let us first sketch the mathematical setup and state our main theorems in a  somewhat informal way. 
In our work we consider an infinitely extended electron lattice gas from the outset, instead of the more common approach in related work (e.g.\ \cite{BDF18,de2019persistence,bachmann2021exactness} and many others) of analyzing finite systems uniformly in the system size. To mathematically describe such infinite systems, many pioneers have developed a precise framework based on operator algebras (see, e.g., \cite{ArakiMoriya2002,bratteliI,bratteliII} and references therein), and we adopt this framework in our study. The advantages of this approach in our context will be discussed in detail later in this section.
To model an infinitely extended gas of electrons in a periodic structure with constant density, we take the one-body position  space to be the lattice $\Z^d$, where $d\in\N$ is the spatial dimension, and the one-body Hilbert space to be $\ell^2(\Z^d,\C^n)$, where $\C^n$ models the spin of the electrons and possible other degrees of freedom arising from a non-trivial structure of the unit cell. The observable algebra for the many-body system is the corresponding CAR algebra denoted by $\mA$, which is a subalgebra of the algebra of bounded operators on the fermionic  Fock space $\mathcal{F}(\Z^d,\C^n)$ (see Section~\ref{sec:basic} for details). The Heisenberg dynamics of the unperturbed system is generated by a densely defined derivation $\mathcal{L}_{H}$   of $\mA$ coming from a translation invariant Hamiltonian $H$ with short-range interactions that conserves 
the number of particles. An important novelty in this context is that we require translation invariance only with respect to a very general form of translations, including in particular  magnetic translations. For example, our assumptions also include the many-body version of the Hofstadter model with arbitrary magnetic field strength and weak interactions, see Section~\ref{sec:assumptions} for details.

The essential assumption for our analysis is the so-called incompressibility of the ground state, that is, the assumption that $\mathcal{L}_{H}$ has a  locally unique gapped ground state $\omega_0$ such that the associated GNS Hamiltonian has a spectral gap $g>0$ above its lowest eigenvalue zero (compare Definition~\ref{def:groundstate}). This assumption can be proved so far only for non-interacting or weakly interacting fermion systems, see \cite{GMP2017,Ha19,de2019persistence,K20} and Proposition~\ref{prop:Hofstadter}, but is considered to be the defining property of (topological) insulators.

The perturbation by a constant external electric field of strength $\eps\in\R$   is  modeled by adding to $\mathcal{L}_{H}$ the densely defined derivation $\mathcal{L}_{\eps X_1}$, where $X_1$ denotes the first component of the second quantization of the position operator. Note that neither $H$ nor $X_1$ are bounded operators on Fock space, but $H$ is a bounded sum-of-local-terms (SLT) interaction, while $X_1$ is an unbounded SLT interaction. It has been shown in \cite{Teufel2020,HenheikTeufel2022} that if the system starts in the gapped ground state $\omega_0$ and then the perturbation $\mathcal{L}_{\eps X_1}$ is adiabatically switched on, the system dynamically evolves into a non-equilibrium almost-stationary state (NEASS) $\omega_\eps$. See for example \cite{HT20,marcelli2021new,marcelli2022from} for a discussion of the NEASS approach to linear response theory. Note, however, that our results rely on only two properties of $\omega_\epsi$, namely that it is an almost stationary state for the perturbed dynamics generated by $\mathcal{L}_{H+\epsi X_1}$, and that it is connected to $\omega_0$ by a quasi-local automorphism. For our purposes, it is not necessary to explicitly compute $\w_\epsi$ or its asymptotic expansion in powers of $\epsi$.

Our main result, Theorem~\ref{thm:main}, concerns the current density in the state $\omega_\eps$. Let $\i[H,X_k] = \sum_{x\in\Z^d} j_{x,k}$ be a decomposition of the translation invariant current operator\footnote{Strictly speaking $\i[H,X_k] $ is a bounded SLT-interaction, see Section~\ref{sec:basic} for details.} $\i[H,X_k]$, so
 $j_{x,k}\in \mA$ denotes the $k$-th component of the current operator at location $x\in\Z^d$. Then we show that for all $x\in\Z^d$
\begin{equation}\label{intro:direct}
\w_\eps(j_{x,1}) = \mathcal{O}(\eps^\infty)\,,
\end{equation}
and that for $k\not=1$
\begin{equation}\label{intro:Hall}
\w_\eps(j_{x,k}) =   \eps \,\overline{\w_0}(\i[X_k\OD,X_1\OD])   + \mathcal{O}(\eps^\infty)\,.
\end{equation}
In the above, 
\begin{itemize}
    \item $\overline{\w_0}(\cdot)$ denotes an expectation in the equilibrium ground state taken per unit volume (compare \eqref{eq:omegaPUV} for a precise definition), and
    \item the off-diagonal part $X\OD$ of the position operator (and of any element of the CAR algebra)  with respect to the state $\w_0$ is discussed in Section~\ref{sec:offdiag}, see Definition~\ref{dfn:XjOD}.
\end{itemize}

\noindent Note that while the definition of local current operators for lattice systems in general involves some arbitrariness, in our periodic setting the left-hand sides of \eqref{intro:direct} and \eqref{intro:Hall} are unambiguously defined.

The Hall conductivity, defined as the ratio between the induced current density and the applied electric field in the limit of vanishing field strength, is thus given by 
\begin{equation}\label{intro:DCint}
    \sigma^\mathrm{H}_{k1}= \overline{\w_0}(\i[X_k\OD,X_1\OD])\,.
\end{equation}
Equation \eqref{intro:direct} shows  that the current density in the direction of the applied field  vanishes faster than any power of $\eps$, while \eqref{intro:Hall} shows that the Hall current density is a nearly linear function of the applied field near zero.  Note that the expression on the right-hand side of \eqref{intro:DCint} is the generalization to the interacting case of the so-called double commutator formula for the Hall conductivity:
\begin{equation*}\label{intro:DCF}
\overline \tr \left(\i P_0[ [\hat x_k, P_0], [\hat x_1,P_0]]\right)
\end{equation*}
which holds for non-interacting systems (see \cite{marcelli2022purely} and references therein). Here $\overline\w_0(\cdot)$ replaces the trace per unit volume $\overline\tr(P_0\,\cdot)$, where $P_0$ is the infinite rank Fermi projection, and $X_k\OD$ replaces off-diagonal part of the $k$-th component of  the one-body position operator $\hat x_k$ with respect to $P_0$.

Formula \eqref{intro:DCint} already suggests, as we indeed prove (Corollary \ref{corr:conductivity constant in phase}), that the Hall conductivity depends only on the state~$\omega_0$ but not on the Hamiltonian $H$, although the off-diagonal part $X_k\OD$ is defined using $H$, see Definition~\ref{dfn:XjOD}. Even more, as a simple corollary to a many-body Chern-Simons formula, Lemma~\ref{lem:CS}, which is involved in our proof, we find that the Hall conductivity $\sigma^\mathrm{H}_{k1} $ is constant within gapped phases of ground states (Corollary~\ref{corr:conductivity constant in phase}). 
Our results also allow a complete characterization of the gapped phases of the weakly interacting Hofstadter model and a generalization to irrational magnetic fields of the results of Giuliani, Mastropietro and Porta \cite{GMP2017,G2020}, i.e.\ the stability of the Hall conductivity with respect to the introduction of small interactions, by a completely different approach. Since the proof of the continuity of the Hall conductivity with respect to the magnetic field requires additional technical effort, we provide the details in another paper \cite{MMMTW2025}. However, we want to emphasize that variations of the magnetic field lead to ground states that are not in the same gapped phase with respect to the standard definition based on automorphic equivalence. Therefore, in \cite{MMMTW2025} we also discuss how the standard definition can be extended to cover situations where ground states with different magnetic fields can still be in the same phase.

Furthermore, using the results of Sopenko and Kapustin \cite{kapustin2020hall}, we show  that the Hall conductivity $\sigma^\mathrm{H}_{k1}$ takes quantized
values not only in the non-interacting phase, but in any invertible phase (Corollary~\ref{corr:invertible phase}).

From \eqref{intro:Hall} we also conclude that for $d=2$ the expectation of the Hall conductance observable
\[
G^\mathrm{H}_{21,L}:= \frac{J_L}{\eps L}\,,
\] defined as the ratio of the current operator $J_L$ through a line segment of length $L$ aligned with the applied field and the voltage drop $\eps L$ across this line segment, not only agrees with the Hall conductivity $\sigma^\mathrm{H}_{21}$, but also has a vanishing variance in the limit $L\to \infty$:
\[
\w_\eps\hspace{-2pt}\left(G^\mathrm{H}_{21,L}\right) \,=\, \sigma^\mathrm{H}_{21} \,+\, \mathcal{O}(\epsi^\infty)\,, \quad\quad\mbox{and}\quad\quad
\mathrm{var}\left(G^\mathrm{H}_{21,L}\right) := \omega_\eps\hspace{-2pt}\left(\left(G^\mathrm{H}_{21,L}\right)^2\right) -\left(\omega_\eps\hspace{-2pt}\left(G^\mathrm{H}_{21,L}\right)\right)^2=\mathcal{O}\left(\frac{1}{\epsi^2 L}\right)\,.
\]
Note that such a statement is   relevant in explaining the experimental findings: It is not only the average value $\sigma^\mathrm{H}_{21}$ of $G^\mathrm{H}_{21,L}$ that is found to be quantized, but there are also   no measurable fluctuations around this value for $L$ sufficiently large. As far as we know, this has not been demonstrated before in any model of interacting electrons. Note that the factor $1/\epsi^2$ in the bound on the variance of $G^\mathrm{H}_{21,L}$ is unavoidable, since  even for $\epsi=0$ the fluctuations of $J_L$ in the ground state $\omega_0$ are non-vanishing.

The precise formulation of our results requires some technical preparation  and is therefore postponed to Section~\ref{sec:results}. 
While we explicitly consider  the case of lattice fermions, 
we note that the method presented in this paper can also be applied to quantum spin systems with a conserved charge.

In the remainder of this introduction, we will  comment on the existing literature related to our results, the main new mathematical challenges encountered in our proofs, and their physical significance. 

The mathematical and theoretical physics literature on the quantum Hall effect has grown steadily over the last 40 years and is now vast.
We have therefore chosen to focus on a small selection of results that we consider most relevant to put our results into context.
Our results are concerned with deriving a formula for the Hall conductivity and the Hall conductance at zero temperature from a microscopic model, an enterprise often referred to as proving   linear response or proving Kubo's formula for gapped systems at zero temperature (see e.g.\ \cite{HT20,bachmann2020note} for recent reviews). 
A problem we do not address here is that of proving that the resulting formula gives only quantized values (see \cite{hastings2015quantization,BBDF20,kapustin2020hall} and references therein, or \cite{frohlich1992u,Froehlich2018} for a general approach based on effective actions).
Much of the earlier mathematical work in this area has been concerned with the free electron gas, i.e.\ with non-interacting models, often with a focus on the role of random impurities (see e.g.\ \cite{bellissard1994noncommutative,aizenman1998localization,bouclet2005linear,marcelli2022purely}); some authors have considered weakly interacting electrons~\cite{GMP2017}, but we consider interacting fermions without any smallness assumption on the interaction (let us stress once again that, however, the incompressibility assumption on the existence of a gapped ground state has been proven only in the weakly-interacting regime). Most other results on interacting models consider finite systems with a torus geometry to avoid the problem that, in topological insulators, edge states close the spectral gap.  Due to the lack of a position operator on such a geometry, the external electric field in this setting is introduced by so-called adiabatic  flux threading \cite{koma2004revisiting,bachmann2018quantization,bachmann2020note,bachmann2021exactness,klein1990power}.
Instead, we work directly in the infinite volume and require only a gap in the bulk, a concept introduced in \cite{MO20}. Thus, the main novelty of our setup is that we start from an infinite system of interacting particles, require only a gap in the bulk, and model the driving field by a linear potential~$\eps X_1$. 
It turns out that our approach allows transparent and simple arguments and covers relevant aspects of the real experimental situation. In fact, after establishing the appropriate technical framework and 
proving the
  existence and properties of the NEASS $\w_\eps$ for infinite volume systems without uniqueness assumption on the ground state in Proposition~\ref{prop:NEASS}, the proof of our main theorem, Theorem~\ref{thm:main}, consists only of a short calculation and an application of a new many-body variant of the Chern-Simons lemma, Lemma~\ref{lem:CS}. 

Note that also for interacting lattice fermions it was shown before in \cite{bachmann2021exactness}, based on the approach of \cite{klein1990power}, that power law corrections to Kubo's formula vanish, i.e.\ that the Hall current response is nearly linear in the driving strength $\eps$ up to terms that are asymptotically smaller than any power of $\eps$.
As mentioned above, the authors of \cite{bachmann2021exactness} consider a system of interacting lattice fermions on a finite two-dimensional flat torus, but give error bounds that are uniform in the linear system size $L$.
The main difference from our approach is that, instead of applying a constant electric field of strength $\eps$ throughout the system, the electromotive force is applied in \cite{bachmann2021exactness} by a time-dependent gauge potential that varies at rate~$\eps$. The way this is done in \cite{bachmann2021exactness} translates to applying a potential step of height $\eps$ only over a single line in the system, say the line $\{x_1=0\}$, i.e. to applying an electric field of strength $\eps$ that is supported only on that line. Alternatively, one could choose the time-dependent gauge potential so that an electric field of strength $\eps/L$ is added everywhere.

Mathematically, this is an important difference, since a potential step of height $\eps$ or a constant field of strength $\eps/L$ that decreases with increasing system size are both  regular perturbations that do not close the spectral gap of $H$ for sufficiently small $\eps$. As a consequence, the result in \cite{bachmann2021exactness,klein1990power} follows from an appropriate application of the adiabatic theorem of quantum mechanics. 
Our situation is different: a constant electric field of strength $\eps$ applied to an infinitely extended system is a singular perturbation that closes the spectral gap, no matter how small $\eps\in\R$ is. Thus the standard adiabatic theorem cannot be applied and a local version in space, the so-called space adiabatic theorem, is required, see \cite{Teufel2020,HenheikTeufel2022,PST03}. 

Physically, a linear potential drop over a macroscopic region is also closer to real experimental situations, where in quantum Hall bars the Hall voltage is expected and observed to drop essentially linearly over macro- or at least mesoscopic locally gapped regions, so-called incompressible stripes, see \cite{gerhardts2017effect} for a recent review. An even more basic fact is the following: both in normal and in topological insulators the longitudinal current is observed to vanish even if one applies across the sample a voltage that exceeds the spectral gap by many orders of magnitude. 
Last but not least, to our understanding it is also necessary to  consider the extensive situation of a current flowing in a macroscopic region in order to conclude that the measured Hall conductance has vanishing fluctuations.

The rest of the paper is structured as follows. In Section~\ref{sec:tech} we first introduce the basic technical notions and then collect several technical propositions concerning periodic interactions and, in particular, their commutator with the unbounded position operators. 
In Section~\ref{sec:assumptions} we then formulate the precise assumptions underlying our main theorem and establish in Proposition~\ref{prop:NEASS}  the existence and relevant properties of the NEASS. Moreover, we show in Proposition~\ref{prop:Hofstadter} that the weakly interacting Hofstadter model, which is basically the discrete Landau operator, satisfies all the assumptions of our main theorem. In Section~\ref{sec:offdiag} we need to introduce another object, the off-diagonal map, which selects the off-diagonal part of an interaction with respect to a gapped ground state. In particular, it turns out that the off-diagonal part of the position operator is a bounded SLT-interaction. 
All proofs up to this point are rather technical and should not   surprise   experts. They are therefore 
 postponed to the appendices. In Section~\ref{sec:results} we finally state our main theorem, Theorem~\ref{thm:main}, and several corollaries mentioned above. The proof of Theorem~\ref{thm:main} is given immediately after the statement because, after setting up the technical tools, it is rather short and instructive. Indeed, we consider the simplicity and transparency of the proof to be one of the virtues of our approach.
However, as already mentioned, it uses a new Chern-Simons type lemma, the proof of which is also deferred to an appendix.

\section{Mathematical setup}\label{sec:tech}

We basically use the standard formalism for describing infinitely extended systems of interacting lattice fermions  in the thermodynamic limit, as explained for example in \cite{bratteliI,bratteliII} and, in addition, more specific results for fermionic systems from~\cite{ArakiMoriya2002}. 
However, it turns out that our analysis requires a number of technical results, mostly concerning periodic interactions and position operators, which were not available in the previous literature. 
The proofs of the corresponding lemmas and propositions  are collected in Appendix~\ref{app:B}.

\subsection{Basic notions}\label{sec:basic}

 The anti-symmetric  (or fermionic) Fock space over the lattice $\Z^d$ is given by
    \begin{align*}
       \mathcal{F}(\Z^d) := \bigoplus_{N=0}^{\infty}\ell^2(\Z^d,\C^n)^{\wedge N} ~,
    \end{align*}
where $\ell^2(\mathbb{Z}^d, \mathbb{C}^n)^{\wedge N}$ denotes the $N$-fold anti-symmetric tensor product of $\ell^2(\mathbb{Z}^d, \mathbb{C}^n)$, with the convention $\ell^2(\mathbb{Z}^d, \mathbb{C}^n)^{\wedge 0} = \mathbb{C}$.
    We use $a^*_{x,i}$ and $a_{x,i}$, for $x\in \Z^d$ and $i\in \{1,\dots,n\}$, to denote the fermionic creation and annihilation operators associated to the standard basis of $\ell^2(\Z^d,\C^n)$; recall that they satisfy the canonical anti-commutation relations (CAR).
    The number operator at site $x\in\Z^d$ is defined by
    \begin{align*}
        n_x:=\sum_{i=1}^n a^*_{x,i}a_{x,i}\,.
    \end{align*}
    The algebra of all bounded operators on $\mathcal{F}(\Z^d)$ is denoted by $\mathcal{B}(\mathcal{F}(\Z^d))$. 
    For each $M\subseteq \Z^d$ let $\mA_M$ be the C*-subalgebra of $\mathcal{B}(\mathcal{F}(\Z^d))$ generated by
    \begin{align*}
        \{a^*_{x,i}~|~x\in M,~ i\in \{1,\dots,n\}\}~.
    \end{align*}
    The C*-algebra $\mA := \mA_{\Z^d}$ is the CAR-algebra, which we also call the quasi-local algebra. We define $P_0(\Z^d) := \{M\subseteq \Z^d~|~|M|<\infty \}$ and call
    \begin{align*}
        \mA_0 := \bigcup_{M\in P_0(\Z^d)} \mA_M \subseteq \mA
    \end{align*}
    the local algebra, which is dense in $\mA$ with respect to the norm topology. Consequently, an operator is called quasi-local if it lies in $\mA$ and local if it lies in $\mA_0$.
    For each $\varphi \in \R$ there is a unique automorphism\footnote{In the following the term {\it automorphism} is used in the sense of a $*$-automorphism as defined in the textbook \cite{bratteliI}.} $g_\varphi$ of $\mA$, such that
    \begin{align*}
        g_\varphi(a^*_{x,i}) =  \e^{\i\varphi} \, a^*_{x,i}, \quad \text{for all}~~ x\in \Z^d,~ i\in \{ 1,\dots,n \} ~.
    \end{align*}
One defines the set
    \begin{align*}
        \mA^N := \{A\in \mA~|~  \forall \varphi \in \R :\, g_\varphi (A) = A \}
    \end{align*}
    and calls $\mA^N$ the gauge-invariant sub-algebra of $\mA$. It is the closure of the set of all local observables that commute with the number operator, i.e.\  all local observables $A\in   \mA_0$ that satisfy $A\in \mA_M \Rightarrow [A,N_M]:=[A,\sum_{x\in M}n_x]=0$.  Its part in $M\subseteq\Z^d$  is denoted by $\mA^N_M:= \mA^N\cap \mA_M$.
For disjoint regions $M_1,M_2 \subseteq\Z^d$,  $M_1\cap M_2 = \emptyset$, operators  $A\in \mA_{M_1}^N$ and $B\in \mA_{M_2}$ commute:  $[A,B] = 0$.

In order to define quantitative notions of localization for quasi-local operators, one makes use of the fact that one can localize operators to given regions by means of the fermionic conditional expectation. 
To this end first note that  $\mA$ has a unique state $\w^{\tr}$ (commonly referred to as a tracial state) that satisfies
    \begin{align*}
        \w^{\tr}(AB) = \w^{\tr}(BA)
    \end{align*}
    for all  $A,B \in \mA$ (e.g.\ \cite[Definition 4.1, Remark 2]{ArakiMoriya2002}).

\medskip

\begin{proposition}[{\cite[Theorem 4.7]{ArakiMoriya2002}}]\label{Ex+UniqueExpectation}
    For each $M\subseteq\Z^d$ there exists a unique linear map 
    \begin{align*}
        \E_M:\mA \to \mA_M\,,
    \end{align*}
    called the conditional expectation with respect to $\w^\tr$, such that
    \begin{equation}\label{eq:Conditional expectation defining property}
      \forall A\in \mA \; \;\forall B\in \mA_M \,:\quad \w^{\tr}(AB)=\w^{\tr}(\E_M(A)B) \,.
    \end{equation}
    It is unital, positive and has the properties 
    \begin{eqnarray*}
        \forall M\subseteq \Z^d\;\; \forall A,C\in \mA_M\;\; \forall B\in\mA\, : &&   \E_M \br{A\,B\,C} = A\, \E_M(B)\,C\\[1mm]
      \forall M_1,M_2 \subseteq \Z^d\,:&&   \E_{M_1} \circ \E_{M_2}  = \E_{M_1\cap M_2}\,\\[1mm]
        \forall M \subseteq \Z^d\,:&& \E_M \mA^N \subseteq \mA^N
  \\[1mm]
        \forall M \subseteq \Z^d \; \;  \forall A \subseteq \mA\,:&&  \norm{\E_M(A)}\leq \norm{A}\,.
    \end{eqnarray*}
\end{proposition}
\begin{proof}
    We note that there are two differences between our proposition and \cite[Theorem 4.7]{ArakiMoriya2002}: As can be seen in the proof of the theorem, there is a unique linear map satisfying \eqref{eq:Conditional expectation defining property}, and the property of being a conditional expectation is not required for the uniqueness.  The statement $ \forall M \subseteq \Z^d:$ $\E_M \mA^N \subseteq \mA^N $ is not part of {\cite[Theorem 4.7]{ArakiMoriya2002}}; it however follows by the exact same reasoning as in {\cite[Corollary 4.8]{ArakiMoriya2002}}. One just needs to replace the parity automorphism by $g_\varphi$ for $\varphi \in \R$.
\end{proof}

With the help of $\E$ we can define subspaces of $\mA^N$ that contain operators with well-defined decay properties (cf.\ \cite{MO20} for similar definitions for quantum spin systems).

\medskip

\begin{definition}
    For $\nu \in \N_0$ and $A \in \mA$ let
    \begin{align*}
        \norm{A}_\nu := \norm{A} + \sup_{k\in \N_0}( \norm{A-\E_{\La_k}A} (1+k)^\nu) 
    \end{align*}
    where $\La_k = \{x\in \Z^d \, | \, \norm{x}_\infty \leq k \}$ is the box with side-length $2k$ around $0$ ($\norm{\cdot}_\infty$ is the maximum norm on $\Z^d$).
    We denote the set of all $A\in \mA^N$ with finite $\norm{\cdot}_\nu$ by $D_\nu$ and also define $D_{\infty} := \bigcap_{\nu \in \N_0} D_\nu$. We will sometimes refer to these norms as decay norms.
\end{definition}

\medskip

\begin{lemma}\label{lem:norm properties}
Let $\nu\in\N$ and $A,B\in D_\nu$. Then 
\[ \norm{AB}_\nu \le 2 \, \norm{A}_\nu \, \norm{B}_\nu \quad \text{and therefore} \quad \norm{[A,B]}_{\nu} \le 4 \, \norm{A}_\nu \, \norm{B}_\nu\,,\]
and
\[ \lim_{k\to \infty }\norm{A - \E_{\La_k}A}_{\nu-1} =0 \, . \]
\end{lemma}

The relevant physical dynamics on $\mA$ is generated by densely defined derivations, which in turn are constructed from so-called interactions.
    An interaction is a map $\Phi: P_0(\Z^d) \to \mA^N$, such that $\Phi(\emptyset) = 0$ and for all $M\in P_0(\Z^d)$ it holds that $\Phi(M) \in \mA_M$, $\Phi(M)^* = \Phi(M)$, and the sum 
    \[\sum_{\substack{K\in P_0(\Z^d)\\ M\cap K \neq \emptyset}} \Phi(K)\] 
    converges unconditionally, meaning that the partial sums converge to the same value for all sequences that enumerate the index set. 
    For two interactions   $\Phi$ and $\Psi$ their commutator is given by
    \begin{align*}
        [\Phi,\Psi]: P_0(\Z^d) \to \mA^N\,,\qquad M\mapsto 
        [\Phi,\Psi](M) \;:= \sum_{\substack{M_1,M_2 \subseteq M\\ M_1 \cup M_2= M}} [\Phi(M_1),\Psi(M_2)] \, .
    \end{align*}
    The map $\i[\Phi,\Psi]$ satisfies the definition of an interaction except for the last requirement, which is not always satisfied. All the commutators of interactions appearing in the following will however be interactions (see Proposition \ref{commutator of interactions}).

Interactions  define derivations on the algebra in the following way.    
    For an interaction $\Phi$ let
    \begin{align*}
        \mL_{\Phi}^\circ: \mA_0 \to \mA,~ A\mapsto \sum_{M\in P_0(\Z^d)} [\Phi(M),A] \, .
    \end{align*}
    It follows from \cite[Propositions~3.1.15 and~3.2.22]{bratteliI} that $\mL_{\Phi}^\circ$ is closable. We denote its closure by $\mL_{\Phi}$ and call it the Liouvillian of $\Phi$.
Some basic interactions that will appear frequently in the following are the number operator $N$ and the position operators $X_j$ for $j\in\{1,\dots, d\}$, which are non-vanishing only on one-element sets. They are defined by $N(\{x\}) = n_x$ and $X_j(\{x\})= x_j n_x$ for $x\in\Z^d$.

In the operator-algebraic framework, states are described as positive linear functionals $\w \colon \mA \to \C$ of norm $1$. Given an interaction $\Phi$, a state $\omega$ is called a $\mL_{\Phi}$-ground state, or simply ground state of $\Phi$, if it holds that 
\begin{align*}
 \forall A \in D(\mL_\Phi) : \quad  \w(A^* \mL_\Phi A) \geq 0   \, .
\end{align*}
Here, for a given linear operator $ T $ acting on $\mathcal{A}$, $ D(T) $ denotes its domain.
For a state $\w$ on $\mA$ and an interaction $\Phi$ we define 
   \begin{equation} \label{eq:omegaPUV}
   \overline{\omega}(\Phi) := \lim_{k\to \infty} \frac{1}{|\La_k|}\sum_{M\subseteq \La_k} \w(\Phi(M))\, ,
   \end{equation}
   whenever the limit exists  and call it the per-volume expectation value of $\Phi$ with respect to $\w$.

\medskip

\begin{definition}\label{norm interaction}
    Let $\Phi$ be an interaction and $\nu \in \N_{0}$. Let
    \begin{align*}
        \|\Phi\|_{\nu} := \sup_{x\in \Z^d} \sum_{\substack{M\in
        P_0(\Z^d)\\x\in M}}(1+\mathrm{diam}(M))^\nu  \| \Phi(M)\|  ~.
    \end{align*}
    The set of all interactions with finite $\norm{\cdot}_{\nu}$ is denoted by $B_{\nu}$. We also define $B_{\infty} := \bigcap_{\nu \in \N_0} B_{\nu}$. \\
    For $a>0$ let
    \begin{align*}
        \|\Phi\|_{\exp,a} := \sup_{x\in \Z^d} \sum_{\substack{M\in 
        P_0(\Z^d)\\x\in M}} \exp(a \, \mathrm{diam}(M))  \| \Phi(M)\| 
    \end{align*}
    and denote the set of all interactions with finite $\norm{\cdot}_{\exp,a}$ for some $a>0$ by $B_{\exp}$. By $\mathrm{diam}(M)$ we mean the maximal distance of two elements in $M$ with respect to $\norm{\cdot}_\infty$.
\end{definition}

\medskip

\begin{lemma}\label{lem: sum representation of generator}
    For a $B_\infty$-interaction $\Phi$, $j\in \{ 1,\dots, d \}$ and $p,q\in \R$ it holds that $D_\infty \subseteq D(\mL_{p \Phi + q X_j})$. Let $ A\in D_\infty$, then the sums 
    \[\sum_{M\in P_0(\Z^d)} [\,\Phi(M),A\,] \quad \text{and} \quad \sum_{x\in \Z^d} [\,x_j\,n_x,A\,]\] converge absolutely and  
    \[ \mL_{p \Phi + q X_j}A = p \sum_{M\in P_0(\Z^d)} [\,\Phi(M),A\,] + q \sum_{x\in \Z^d} [\,x_j\,n_x,A\,] \in D_\infty \,. \]
    For each $\nu \in \N_0$ there is a constant $c_\nu$, independent of $\Phi, j, p, q, A$ such that
    \[ \norm{ \mL_{p \Phi + q X_j}A }_\nu \leq c_\nu \, (p \,\norm{\Phi}_{d+1+2\nu} + q \,\norm{n_0}) \, \norm{A}_{d+3+2\nu}\,.\]
\end{lemma}

A cocycle of automorphisms is a family $(\alpha_{u,v})_{(u,v) \in \R^2}$ of automorphisms of $\mA$ such that $\alpha_{u,u} = \mathrm{id}$ for all $u\in \R$ and $\alpha_{u,v}\circ \alpha_{v,w} = \alpha_{u,w} $ for all $u,v,w\in \R$.
We say a cocycle of automorphisms is generated by the family of interactions $(\Phi^{u})_{u\in \R}$ if for all $A \in \mA_0$ it holds that $ \partial_u \alpha_{u,v} A = \alpha_{u,v} \i \mL_{\Phi^{u}} A$. In the case where the generating family is a constant interaction $\Phi$, the family $(\alpha_{u,0})_{u\in \R}$ is a one-parameter group of automorphisms with generator $\i\,\mL_{\Phi}$. One can show that the family of all gauge automorphisms $(g_\varphi)_{ \varphi \in \R}$ is generated by the number operator $N$ in this sense and that, for an observable $A\in D(\mL_N)$, gauge-invariance ($A\in \mA^N$) is equivalent to $\mL_N  A = 0$.

\medskip

\begin{lemma}\label{cocycles}
    If $(\Phi^{u})_{u\in \R}$ is a family of interactions such that $\sup_{u\in \R} \norm{\Phi^{u}}_\nu < \infty $ for all $\nu\in \N_0$ and for each $M \in P_0(\Z^d)$ the map $ u \mapsto \Phi^{u}(M)$ is continuous with respect to the norm topology in $\mA$, then there is a unique cocycle of automorphisms $(\alpha_{u,v})_{(u,v) \in \R^2}$ generated by $(\Phi^{u})_{u\in \R}$.
    This cocycle has the property that for all $\nu\in \N_0$ there is a constant $c_\nu$ such that 
    \begin{align*}
        \forall A\in D_\infty,\, u,v\in \R\colon \quad\norm{\alpha_{u,v} A }_\nu \leq c_\nu \exp(c_\nu\, \sup_{t\in\R}\norm{\Phi^t}_{2d+1+\nu} \, |u-v|) \norm{A}_\nu   \, .
    \end{align*}
    It also holds for all $u,v\in \R$ that $\alpha_{u,v} D_\infty = D_\infty$.
    If additionally $\sup_{u\in \R} \norm{\Phi^{u}}_{\exp,a} < \infty $ for some $a > 0$, then for all $\nu \in \N_0$ there exists a continuous function $b_\nu:\R \to \R$ that grows at most polynomially, such that
    \begin{align*}
        \norm{\alpha_{u,v} A}_\nu \leq b_\nu(|u-v|) \norm{A}_\nu \quad \text{ for all } A\in D_\infty,\, u,v\in \R \,.
    \end{align*}
\end{lemma}

\subsection{Periodic interactions}

From now on we will specialise to periodic interactions and periodic states for which the existence of the per-volume expectation is easy to see. We also formulate additional properties of periodic interactions that are proved in Appendix~\ref{app:B} and will be used in the proof of our main theorems.

\medskip

\begin{definition}\label{def:translation}
We denote by $\mathrm{Aut}(\mA)$ the group of automorphisms on $\mA$.
    A translation is a map $T:\Z^d \to \mathrm{Aut}(\mA)$ that associates to each possible shift vector $\g$ an automorphism $T_\g$ and satisfies the following properties:
    \begin{itemize}
        \item[(i)] For all $\g\in \Z^d$ and $M \subseteq \Z^d$ it holds that $T_{\g}(\mA_M) = \mA_{M+\g}$.
        \item[(ii)] For all $\g\in \Z^d$ and $x\in \Z^d$, it holds that $T_{\g}n_x = n_{x+\g}$.
    \end{itemize}
\end{definition}

Note that this definition of translations includes, in particular, so-called magnetic translations (see \eqref{magtrans} below). The two properties guarantee that translations are compatible with the conditional expectation and gauge automorphisms respectively.

\medskip
\begin{lemma}\label{lem: compatibility of translations}
    Let $T$ be a translation and $\g \in \Z^d$. It holds that 
    \begin{align*}
        \forall M \subseteq \Z^d \colon& \quad T_\g \, \E_M   =  \E_{M + \g} \, T_\g \, ,\\
        \forall \varphi \in \R\colon& \quad g_\varphi \, T_\g = T_\g \, g_\varphi \, .
    \end{align*} 
\end{lemma}

    Let $T$ be a translation. We say a state $\omega$ is $T$-periodic, if
\begin{align*}
\forall \, \gamma \in \Z^d, \, A\in\mathcal{A}:\quad
    \omega(T_\gamma A) = \omega(A)\, .
\end{align*}
    An interaction $\Phi$ is called $T$-periodic, if
    \begin{align*}
\forall \, \gamma \in \Z^d, \, M\in P_0(\Z^d):\quad        T_{\g} \Phi(M) = \Phi(M+\g)\,. 
    \end{align*}

\begin{remark}
    Note that the definition of a translation does not require it be a group homomorphism. Translations do however satisfy a homomorphism property when acting on certain elements like the local number operators or terms of interactions that are periodic with respect to the translation. 
\end{remark}

\medskip

In order to define a standard representation of a periodic interaction in terms of a quasi-local observable, we introduce the following notions. 
    We say $M_1,M_2 \in P_0(\Z^d)$ have the same shape if there is a $\gamma\in \Z^d$ such that $M_1 = M_2 +\g$. For each $M\in P_0(\Z^d)$ we define $s(M) \in \Z^d$ as the vector that shifts the center of mass of $M$ closest to $0$. More precisely it is defined as the unique element of $\Z^d$ such that
    \begin{align}\label{def: shape}
         \sum_{x\in \, M + s(M)} \frac{x}{\abs{M}} \in  {\big ( }-\tfrac{1}{2}\,,\,\tfrac{1}{2}\, {\big]}^d \, .
    \end{align}
    We call $M+s(M)$ the standard representative of $M$ at $0$ and denote the set of all standard representatives at $0$ by $R_0(\Z^d)$.
We call an interaction $\Phi$ absolutely summable at $0$ if 
    \begin{align*}
        \sum\limits_{\substack{M\in R_0(\Z^d)}}  \Phi(M)
    \end{align*}
    converges absolutely. In that case   we denote the limit by $\Phi_0$.

The following construction shows that, under an additional compatibility condition, adding up all the translates of a quasi-local observable yields a $T$-invariant interaction.

\medskip

\begin{definition}
    Let $T$ be a translation. Then $A\in D_\infty$ is called $T$-compatible, iff $T_\g\, T_{\tilde\g} \,A = T_{\g+\tilde\g}\, A $ for all $\g, \tilde\g \in \Z^d $.
\end{definition}

\medskip

\begin{proposition}\label{prop: interaction associated to observable}
    Let $T$ be a translation and $A\in D_\infty$ be self-adjoint and $T$-compatible. Then the map
    \begin{align*}
        \Phi_{A}^T\colon P_0(\Z^d) \to \mA^N  ,\; M\mapsto \Phi_{A}^T(M):=\left\{\begin{array}{cl}
        T_{\g}\E_{\La_{0}}A & \mbox{if  $M=\La_0 +\g$ for some $\g\in\Z^d$}\\
        T_{\g}(\E_{\La_{k}}A - \E_{\La_{k-1}}A)&
        \mbox{if  $M=\La_k +\g$ for some $\g\in\Z^d$ and $k\in\N$}\\
        0 & \mbox{otherwise}
        \end{array}
        \right.
    \end{align*}
   defines a  $T$-periodic $B_\infty$-interaction with $(\Phi_A^T)_0 = A$. Conversely, if $\Phi$ is a $T$-periodic $B_\infty$-interaction, then it is absolutely summable at $0$ and $\Phi_0$ lies in $D_\infty$, is self-adjoint, and $T$-compatible.
\end{proposition}

\medskip

We are now ready to state two technical propositions which will be used in the proof of the main theorem, but which are also relevant to its formulation.

\medskip

\begin{proposition}\label{liouvillian tools}
    Let $T$ be a translation, $\Phi$ a $T$-periodic $B_{\infty}$-interaction, and $\w$ a $T$-periodic state. It holds that:
    \begin{itemize}
        \item[\rm(i)] $\mL_{\Phi}A = \sum_{\g \in \Z^d} [T_\g \Phi_0, A]$ for all $A\in D_\infty$.
        \item[\rm(ii)] $\overline{\omega}(\Phi)$ exists and is equal to $\w (\Phi_0)$.
    \end{itemize}
\end{proposition}

\medskip

\begin{proposition}\label{commutator of interactions}
    Let $\Phi$ and $\Psi$ be $B_\infty$-interactions and $\omega$ a state such that $\Phi$, $\Psi$, and $\w$ are all  $T$-periodic with respect to a translation $T$. Moreover, let $j \in \{1,\dots,d\}$, $p,q \in \R$.
    \begin{itemize}
        \item[\rm(i)] $\i[p\Phi + qX_j, \Psi]$ is a $T$-periodic $B_\infty$-interaction.
        \item[\rm(ii)] $\i[X_j,\Psi]_0  = \i\mathcal{L}_{X_j}(\Psi_0)$.
        \item[\rm(iii)] $\mL_{\i[\Psi,p\Phi+qX_j]}A = \sum_{\mu \in \Z^d} [\i \, \mL_\Psi \, (p\, T_\mu\, \Phi_0 + q \, \mu_j \,  n_\mu) ,\, A\, ]$ for all $A\in D_\infty$.
        \item[\rm(iv)] $\omega(\i[\Phi,\Psi]_0)  =\omega(\i\mathcal{L}_\Phi(\Psi_0)) =-\omega(\i\mathcal{L}_\Psi(\Phi_0))$.
    \end{itemize}
\end{proposition}

\medskip

\section{Model and assumptions}\label{sec:assumptions}

We are now ready to define our model for a (topological) insulator perturbed by a constant electric field directly  for the infinitely extended system using the formalism just described.

The unperturbed system is described by an interaction $H\in B_{\exp}$, which we call the Hamiltonian. 
We assume that $H$ is periodic with respect to a translation $T$ and that $H$ has at least one gapped ground state $\omega_0$.

\medskip

\begin{definition}[{cf.~\cite[Definition~2.3]{tasaki2022lieb}}]\label{def:groundstate}
Let    $H\in B_{\exp}$ be periodic with respect to a translation $T$. We say that a state $\omega_0$ is a $T$-periodic locally unique gapped ground state for $H$ with gap  $g>0$, iff $\omega_0$ is $T$-periodic and it holds for all $A\in \mA_0$  that 
    \begin{equation}\label{eq:gap}
    \w_0( A^* \mL_H A ) \geq g \left( \w_0(A^* A) - \left| \w_0(A) \right|^2 \right)\,.
    \end{equation}
\end{definition}

Systems with a gapped ground state are usually taken to characterize 
systems that are insulators at zero temperature. Our result~\eqref{intro:direct} (compare~\eqref{thm:dc} in Theorem~\ref{thm:main}) supports this view.

 We are interested in how a system that is initially in a gapped ground state responds to the application of a small  constant electric field and possibly also to a small bounded perturbation $V\in B_\infty$.
Assuming that the system is initially in  a ground state $\omega_0$, 
it was shown in \cite{Teufel2020,HenheikTeufel2022}
that when adiabatically switching on  the perturbation $\eps (V+X_1)$, where $V \in B_\infty$ is $T$-periodic, then the system evolves (up to errors of order  $\mathcal{O}(\eps^{\infty})$) into a state $\w_\eps := \w_0 \circ \beta_\eps$ (called NEASS), where $\beta\colon \R \to \mathrm{Aut}(\mA),~ \eps \mapsto \beta_{\eps}$, is a continuous one-parameter family of quasi-locally generated automorphisms. Note that in \cite{HenheikTeufel2022} it is assumed that $\omega_0$ is the unique ground state of $H$, and thus this result does not immediately apply to our more general situation. On the other hand, \cite{Teufel2020} is concerned with finite systems (with bounds that hold uniformly in system size), but allows for non-unique ground states. While we expect the results of \cite{Teufel2020} to hold   for infinitely extended systems as well, the generalization of the adiabatic theorem is not the subject of our present work. However, we emphasize that \cite{Teufel2020} suggests that even in our setting for each gapped ground state $\omega_0$ there is a unique NEASS $\omega_\eps$ into which the system evolves when the perturbation $\eps(V+X_1)$ is adiabatically switched on.

Our main results concern the current response of systems starting in a gapped ground state $\omega_0$, i.e.\ the currents present in  the NEASS $\omega_\eps$ into which the system is driven by applying the small constant electric field. However, since our analysis does not depend on the way such a NEASS is constructed or reached by adiabatic driving, we formulate the properties of NEASS that are relevant for our analysis in a definition, and then prove that a NEASS with these properties always exists for $T$-periodic $H\in B_{\exp}$ with gapped ground state $\omega_0$. 

\medskip

\begin{definition}\label{def:NEASS}
   Let $H\in B_{\exp}$ be $T$-periodic, let $\omega_0$ be a  $T$-periodic gapped ground state for $H$, and let $V\in B_\infty$ be $T$-periodic. We say that a family $[-1,1]\ni\eps \mapsto \w_\eps = \w_0 \circ  \e^{\i \mL_{S_{\eps}}}$ is a NEASS for $H$, $\omega_0$, and $V+X_1$, iff  $[-1,1]\ni \eps \mapsto S_{\eps}$ is a family of $T$-periodic interactions $S_{\eps}\in B_\infty$ such that $\eps \mapsto \frac{1}{\eps}\norm{S_\eps}_\nu$ is   bounded for all $\nu\in\N_0$  and  \begin{equation}\label{NEASSinvariant}
        \sup_{A\in  D_\infty  } \left( \frac{|\w_{\eps}(\mL_{H_{\eps}}A)|}{\norm{A}_{d+3}}\right)  = \mathcal{O}(\eps^{\infty})
    \end{equation}
      for $H_\eps := H + \eps (V+X_1)$.
\end{definition}

Equation~\eqref{NEASSinvariant} states that $\omega_\eps\circ \mathcal{L}_{H_\eps}$ is small in a sufficiently strong sense, and hence  that $\omega_\eps$ is almost invariant for the dynamics generated by~$H_\eps$ for long times.

The next proposition shows that for any gapped ground state $\omega_0$ there exists a  corresponding NEASS  $\omega_\eps$. We expect that under the conditions of the proposition the NEASS $\omega_\eps$ is indeed unique up to terms of order $\mathcal{O}(\eps^\infty)$, but we do not prove this statement. So we note that
 our following results hold for any state $\omega_\eps$ which is a NEASS for $\omega_0$ in the sense of Definition~\ref{def:NEASS}.

\medskip

\begin{proposition}\label{prop:NEASS}
   Let $H\in B_{\exp}$ be $T$-periodic, let $\omega_0$ be a  $T$-periodic gapped ground state for $H$, and let $V\in B_\infty$ be $T$-periodic. Then there exists a family $[-1,1]\ni \eps \mapsto S_{\eps}$ of $T$-periodic interactions $S_{\eps}$ such that $\eps \mapsto \frac{1}{\eps}\norm{S_\eps}_\nu$ is  bounded for all $\nu\in\N_0$  and such that for any $T$-periodic gapped ground state $\omega_0$ of $H$ the state $\w_\eps := \w_0 \circ \beta_\eps$ with $\beta_{\eps} := \e^{\i \mL_{S_{\eps}}}$ is a NEASS for $H$, $\w_0$, and $V+X_1$.
\end{proposition}

Since the proof of Proposition~\ref{prop:NEASS} is merely a technical adaptation of  the original one from  \cite{Teufel2020}, we postpone it to Appendix~\ref{app:NEASS}.
Note that for non-interacting continuum systems the NEASS with similar properties was constructed in \cite{marcelli2022purely} with predecessors going back to \cite{nenciu2002asymptotic,PST03}.

Let us briefly discuss a relevant example of a system (in $d=2$ spatial dimensions and with $n=1$ degrees of freedom per unit cell) for which one can prove the existence of gapped ground states, namely the so-called Hofstadter model  with small interactions. The one-body Hamiltonian of the non-interacting Hofstadter model is just the magnetic discrete Laplacian (discrete Landau operator) $\mathfrak{h}_0^{b}$, i.e.\ the operator on the lattice $\Z^2$ with integral kernel
\[
\mathfrak{h}_0^{b}(x,y) := \left\{\begin{array}{cl}
1 &\mbox{if $x_1=y_1$ and $|x_2-y_2|=1$}\\
\mathrm{e}^{\mathrm{i}bx_2(x_1-y_1)} &\mbox{if $x_2=y_2$ and $|x_1-y_1|=1$}\\
0&\mbox{otherwise.}
\end{array}\right.
\]
Its second quantization reads
\[
H_0^{b,\mu} := \sum_{\substack{x,y\in\Z^2,\\\|x-y\|_{ 1}=1}} \mathrm{e}^{ \mathrm{i}\frac{x_2+y_2}{2} b (x_1-y_1)} a_x^* a_y - \mu\sum_{x\in\Z^2} n_x\,.
\]
Here $b\in\R$ is the strength of a constant external magnetic field and $\mu\in\R$ the chemical potential.  The magnetic translation $T^b$ is the family of automorphisms defined by uniquely extending for each $\gamma\in\Z^2$ the map 
\begin{equation}\label{magtrans}
a_y\mapsto T^b_\gamma(a_y) :=  \mathrm{e}^{-\mathrm{i}b y_1 \gamma_2} a_{y+\gamma} \,,\quad y\in\Z^2\,,
\end{equation}
 to a map $T^b_\gamma\in\mathrm{Aut}(\mA)$. It is straightforward to check that $\gamma\mapsto T^b_\gamma$ defines a translation in the sense of Definition~\ref{def:translation} and that $H_0^{b,\mu}$ is translation invariant with respect to $T^b$.
 To verify this, one can define $h_0^\mu:=a_{(1,0)}^* a_{(0,0)} + a_{(0,1)}^* a_{(0,0)} +a_{(0,0)}^*a_{(1,0)}  + a_{(0,0)}^*a_{(0,1)}- \mu n_{(0,0)}$ and see that $H_0^{b,\mu} = \sum_{\gamma\in\mathbb{Z}^2} T_\gamma^b \,h_0$ holds. Since $T^b_{\tilde \gamma}(T^b_\gamma(a_y))= \mathrm{e}^{-\mathrm{i} b \gamma_1 \tilde\gamma_2} T^b_{\tilde\gamma+\gamma}(a_y)$ we have $T^b_{\tilde \gamma}(T^b_\gamma(h_0^\mu))= T^b_{\tilde\gamma+\gamma}(h_0^\mu)$, thus making the translation invariance explicit.
 The spectrum of the one-body Hofstadter Hamiltonian $\mathfrak{h}_0^{b}$ is very well understood \cite{hofstadter1976energy} and, in particular, it is known that the set of values $(b,\mu)\in\R^2$ for which $\mu $ lies in a gap of $\mathfrak{h}_0^{b}$, i.e.\ $\mu\notin\sigma(\mathfrak{h}_0^{b})$, is open, nonempty and actually of full Lebesgue measure (see \cite{avila2009ten} and, e.g., the discussion in \cite{osadchy2001hofstadter}).

 The following proposition states that for such values of $(b,\mu)$ the many-body Hofstadter interaction $H_0^{b,\mu}$ satisfies the gap condition \eqref{eq:gap} and that, moreover,  this property persists when adding sufficiently weak local many-body interactions to 
 $H_0^{b,\mu}$, e.g.\ a nearest neighbor two-body interaction
 \[
V := \sum_{\substack{x,y\in\Z^2,\\\|x-y\|_{1}=1}} n_xn_y\,.
 \]

\medskip

\begin{proposition}\label{prop:Hofstadter}
Let $b,\mu\in\R$ such that $\mu\notin\sigma(\mathfrak{h}_0^{b})$ and let $V\in B_{\exp}$ be $T^b$-periodic. Then there exists $\lambda_0>0$ such that for all $\lambda\in(-\lambda_0,\lambda_0)$ the \emph{weakly interacting Hofstadter Hamiltonian} 
 \[
H^{b,\mu}_{\lambda} := H_0^{b,\mu} + \lambda V
 \]
 has a $T^b$-periodic locally unique gapped ground state in the sense of Definition~\ref{def:groundstate}.
\end{proposition}

The proof of Proposition~\ref{prop:Hofstadter}, which is postponed to Appendix~\ref{sec:proofHofstadter}, is done in roughly two steps: First we show that $H^{b,\mu}_0$ has indeed a unique gapped ground state in the sense of Definition~\ref{def:groundstate}. Then we use known results \cite{Ha19,de2019persistence,K20} on the stability of the spectral gap for free Fermion systems.

\section{The Off-Diagonal Map}\label{sec:offdiag}

Before we can formulate our main results, we need to introduce a further technical ingredient, namely the map that selects the off-diagonal part of an operator or of an interaction with respect to a ground state~$\omega_0$. This map was  introduced in \cite{hastings2005quasiadiabatic,BMNS12} for finite gapped systems and extended in \cite{MO20} to infinite systems with a gap in the bulk. Here we further generalize it by composing it with automorphisms of the algebra $\mA$.
Its definition involves the choice of a function 
$W_g: \R\to \R$  having the following properties for some $g>0$:
    \begin{itemize}
        \item[(i)] $W_g$ is odd;
        \item[(ii)] the Fourier transform $\widehat W_g$ of $W_g$ satisfies $\widehat{W_g}(k) = \frac{- \i}{\sqrt{2  \pi}\, k}$ for  $k \in \R \setminus [-g,g]$;
        \item[(iii)] $\sup_{s\in \R} \bigl( \lvert s \rvert ^n \lvert W_g(s) \rvert \bigr) < \infty$ for all $n\in\N_0$. 
    \end{itemize}
 It is easy to see that such a function exists for any $g>0$, and an explicit example with additional properties is given in \cite[Lemma 2.6]{BMNS12} (see also \cite[Lemma 2.3 and Equation 2.12]{BMNS12}).

 The following terminology and notation will be useful.

\medskip
 
 \begin{definition}\label{notation:GPS}
We say that a tuple $\br{\omega_0,H,V,W_g,T,\br{S_\eps}_{\eps \in [-1,1]}}$ is a gapped periodic system with perturbation $\eps(V+X_1)$ if $T$ is a translation, $H\in B_{\exp}$ and $V\in B_\infty$ are $T$-periodic, $\omega_0$ is a 
gapped ground state of $H$ with gap $g>0$, the map $W_g$ satisfies the above properties (i)--(iii), and  $\br{S_\eps}_{\eps \in [-1,1]}$ is a family of $T$-periodic interactions having the properties of the one constructed in Proposition~\ref{prop:NEASS}.\\[0.5mm] 
 Whenever a gapped periodic system $\br{\omega_0,H,V,W_g,T,\br{S_\eps}_{\eps \in [-1,1]}}$ is given, then  $\w_\eps := \w_0\circ \beta_\eps$ with $\beta_\eps:=\mathrm{e}^{\mathrm{i}\mL_{S_\eps}}$ denotes the   corresponding NEASS, and $h_0 \coloneq \sum_{M\in R_0(\Z^d)} H(M)$ the part of $H$ located in the origin. Finally, for a general automorphism $\alpha$ of $\mA$ we write $\w_\alpha := \w_0 \circ \alpha$.
 \end{definition} 

 Proposition~\ref{prop:NEASS} shows that, given a $T$-periodic Hamiltonian with a gapped ground state $\omega_0$, then for every $T$-periodic $V\in B_\infty$ there exists a gapped periodic system $\br{\omega_0,H,V,W_g,T,\br{S_\eps}_{\eps \in [-1,1]}}$ for the given data.
 
 The reason for introducing this terminology is that, in general, $H$ does not uniquely determine either $\omega_0$, or $W_g$, or $T$, or $\br{S_\eps}$. Since several of the following definitions and constructions depend not only on $H$, $\omega_0$, and $V$, but also on the choices of $W_g$, $T$ and $\br{S_\eps}$, we make this dependence explicit. However, the final results do not depend on these choices, which we will also make explicit. Actually, it turns out that the Hall conductivity does not even depend on $H$ and $V$, but only on $\omega_0$.

We can now define two more maps that will be relevant for the formulation and the proofs of our results, namely the off-diagonal map and the quasi-local inverse of the Liouvillian.

\medskip

\begin{lemma}\label{lem: almost local obs. for OD and inv. liou.}
    Let $\br{\omega_0,H,V,W_g,T,\br{S_\eps}_{\eps \in [-1,1]}}$ be a gapped periodic system, let $\alpha\in \mathrm{Aut}(\mA)$ be such that $\alpha D_\infty = D_\infty $ and $\alpha T_\g = T_\g \alpha$ for all $\g \in \Z^d$, and $\Psi$ an interaction of the form $\Psi= p\, Y+ q\, X_j$, where $p,q \in \R$, $j\in\{1,\dots,d\}$ and $Y\in B_\infty$ is   $T$-periodic. The operators 
    \begin{align*}
        (\Psi\OD[\alpha])_* :=  \alpha^{-1} \int_{\R}\mathrm{d}s \, W_g(s) \, \e^{\i s\mL_{H}} \,  \alpha \,  \i \,\mL_\Psi \, \alpha^{-1} \, h_0
    \end{align*}
    and
    \begin{align*}
        \mathcal{I}(\Psi)_* \coloneq  - \int_{\R}\mathrm{d}s \,W_g(s) \int_{0}^s \mathrm{d}u  \, \e^{\i u\mL_{H}} \, \i \, \mL_\Psi \, h_0
    \end{align*}
    are in $D_\infty$, self-adjoint, and $T$-compatible. Using Proposition~\ref{prop: interaction associated to observable} we can thus define the $T$-periodic $B_\infty$ interactions
    \begin{equation}\label{dfn:XjOD}
    \Psi\OD[\alpha] \coloneq \Phi_{(\Psi\OD[\alpha])_*}^T\quad\mbox{and}\quad \mathcal{I}(\Psi) \coloneq \Phi_{\mathcal{I}(\Psi)_*}^T\,.
    \end{equation}
For $\alpha=\beta_{\varepsilon}=\e^{\i \mathcal{L}_{S_{\varepsilon}}}$, we denote $\varPsi^{\OD[\beta_{\varepsilon}]}$ simply by $\varPsi^{\OD[\varepsilon]}$, and for $\alpha=\mathrm{id}$, by $\varPsi^{\OD}$.
\end{lemma}

The proofs of the lemmas presented in this section are provided in Appendix \ref{app:OD}.

\medskip
We call the map $\Psi\mapsto  \Psi\OD[\alpha]$ the off-diagonal map, and  refer to $\Psi\OD[\alpha]$ as the off-diagonal part of $\Psi$ (the dependence on $\alpha$ is left implicit if no ambiguity can arise).

\bigskip

\begin{lemma}\label{OD-property}
    Let $\br{\omega_0,H,V,W_g,T,\br{S_\eps}_{\eps \in [-1,1]}}$ be a gapped periodic system, let $\alpha\in \mathrm{Aut}(\mA)$ be such that $\alpha D_\infty = D_\infty $ and $\alpha T_\g = T_\g \alpha$ for all $\g \in \Z^d$, and let $\Psi$ be an interaction that is of the form $\Psi= p\, Y+ q\, X_j$, where $p,q \in \R$, $j\in\{1,\dots,d\}$ and $Y\in B_\infty$ is  $T$-periodic. Then for all $A\in D_\infty$ 
\begin{equation}\label{ODprop}
    \w_{\alpha}(\mL_{\Psi} A) = \w_\alpha(\mL_{\Psi \OD[\alpha] } A)
     \end{equation}
    and 
\begin{equation}\label{Iinvprop}
    \mL_{\i[\mathcal{I}(\Psi), H]} A = \mL_{\Psi \OD } A\,.
    \end{equation}
\end{lemma}

\medskip

\begin{lemma}\label{boundedness of OD}
    Let $\br{\omega_0,H,V,W_g,T,\br{S_\eps}_{\eps \in [-1,1]}}$ be a gapped periodic system. 
    For $j \in \{1,\dots,d\}$ and $\nu \in \N_0$ the map $[-1,1] \to \R, ~ \eps \mapsto \norm{(X_j\ODeps)_0}_\nu$ is bounded.
\end{lemma}

\medskip

\begin{remark}
The terminology ``off-diagonal part'' is motivated by the analogue of Equation~\eqref{ODprop} for pure states in Hilbert spaces: Assume that $P_\alpha$ is a finite-rank orthogonal projection and $A,\Psi$ are bounded operators on a Hilbert space. Then with $\Psi\OD[\alpha]:=P_\alpha\Psi P_\alpha^\perp +P_\alpha^\perp \Psi P_\alpha$ and $\Psi^{\mathrm{D}_\alpha}:=P_\alpha\Psi P_\alpha  +P_\alpha^\perp \Psi P_\alpha^\perp$ 
\[
\w_\alpha(\mL_\Psi A) := \tr (P_\alpha [\Psi,A]) = \tr (P_\alpha [\Psi\OD[\alpha],A]) + \tr (P_\alpha [\Psi^{\mathrm{D}_\alpha},A]) =  \tr (P_\alpha [\Psi\OD[\alpha],A]) =: \w_\alpha(\mL_{\Psi\OD[\alpha]} A)\,,
\]
since
\[
\tr (P_\alpha [\Psi^{\mathrm{D}_\alpha},A]) = \tr (P_\alpha [P_\alpha\Psi P_\alpha  +P_\alpha^\perp \Psi P_\alpha^\perp,A])= \tr(P_\alpha \Psi P_\alpha A P_\alpha) -\tr(P_\alpha A P_\alpha \Psi P_\alpha) =0
\]
by cyclicity of the trace. On the other hand, Equation~\eqref{Iinvprop} implies that $-\i \mathcal{I}$ inverts the action of the Liouvillian $[H,\cdot]$ on $\Psi$ inside $\w_0$ in the sense that for all $A\in D_\infty$
\[
     \w_0(\mL_{-\i \mathcal{I}([H,\Psi])}A)=\w_0(\mL_\Psi A) \,.
\]
This, together with the fact that $\mathcal{I}$ maps $B_\infty$-interactions to $B_\infty$-interactions, 
motivates the name ``quasi-local inverse of the Liouvillian'' for the map $\mathcal{I}$,  see also \cite[Appendix D]{MT19} and references therein.
\end{remark}

\section{Main results}\label{sec:results}

Our main result states that the longitudinal current density in the NEASS $\omega_\eps$ is asymptotically smaller than any power of $\eps$, where physically $\eps$ corresponds to the strength of the applied constant electric field. The transversal or Hall current density is linear in the applied field up to errors that are again asymptotically smaller than any power of $\eps$. The proportionality factor in this linear relation is, by definition, the Hall conductivity. We obtain an explicit formula for the Hall conductivity that depends only on $\w_0$, but neither on $H$ nor on the bounded part $V$ of the perturbation.

The current density in the direction $j\in\{1,\ldots,d\}$ is defined as the expectation per unit volume of the current operator 
\[
J_j := \mathrm{i} [H_\eps,X_j] = \mathrm{i} [H+\eps V,X_j] \in B_\infty
\]
associated with the perturbed Hamiltonian $H_\eps= H+\eps(V+X_1)$ evaluated in the NEASS $\omega_\eps$.

\medskip

\begin{theorem}\label{thm:main} Let $\br{\omega_0,H,V,W_g,T,\br{S_\eps}_{\eps \in [-1,1]}}$ be a gapped periodic system (see Definition~\ref{notation:GPS}). Then
\begin{align}\label{thm:dc}
        \overline{\w_{\eps}}(J_1) =  \mathcal{O}(\eps^{\infty}) \,,
    \end{align}
    and, for $j\in \N$ with $1<j \leq d$,
    \begin{align}\label{thm:hc}
        \overline{\w_{\eps}}(J_j) = -\eps \,\overline{\w_0}(\i[X_1\OD,X_j\OD]) + \mathcal{O}(\eps^{\infty}) ~.
    \end{align}
The Hall conductivity $\sigma^{\mathrm{H}}_{j1} := -\,\overline{\w_0}(\i[X_1\OD,X_j\OD])$ depends only   on $\omega_0$, i.e.\ it agrees for all gapped periodic systems having $\omega_0$ as   ground state. 
\end{theorem}

\begin{proof}
Let $j\in \{ 1,\dots,d \} $.    
We first note that $J_j=\i[H+\eps V,X_j]$ is a $T$-periodic $B_\infty$-interaction with $\i[H+\eps V,X_j]_0 = -\i \mL_{X_j}(h_0+\eps (V)_0)$ (see Proposition~\ref{commutator of interactions}). Thus, with Lemma~\ref{OD-property} and Propositions~\ref{commutator of interactions} and \ref{liouvillian tools} we get
\begin{align*}
    \overline{\w_{\eps}}(\i[H+\eps V,X_j])  = -\i\,\w_\eps(\mL_{X_j}(h_0+\eps (V)_0)) = -\i\,\w_\eps(\mL_{X_j\ODeps}(h_0+\eps (V)_0)) = \i\,\w_\eps(\mL_{H+\eps V} (X_j\ODeps)_0) ~.
\end{align*}
By substituting $H+\eps V = H_\eps - \eps X_1$ we obtain
\begin{align*}
    \overline{\w_{\eps}}(\i[H+\eps V,X_j])&= \i\,\w_\eps(\mL_{H_\eps} (X_j\ODeps)_0) - \i \,\eps\, \w_\eps(\mL_{X_1} (X_j\ODeps)_0) ~.
\end{align*}
The bound
\begin{align*}
    |\w_\eps(\mL_{H_\eps} (X_j\ODeps)_0)| \leq \norm{(X_j\ODeps)_0}_{3d+1} \sup_{A\in D_\infty  } \left\{ \frac{|\w_{\eps}(\mL_{H_{\eps}}A)|}{\norm{A}_{3d+1}}\right\} ~,
\end{align*}
together with the almost stationarity  of the NEASS (Proposition~\ref{prop:NEASS}) and Lemma~\ref{boundedness of OD} yields
\begin{align*}
    \overline{\w_{\eps}}(J_j) &=  - \i\, \eps\, \w_\eps(\mL_{X_1} (X_j\ODeps)_0) + \mathcal{O}(\eps^{\infty}) ~.
\end{align*}
Using Lemma~\ref{OD-property} and Proposition~\ref{commutator of interactions} once more, we obtain
\begin{align*}
    \overline{\w_{\eps}}(J_j)  =  - \i \,\eps\, \w_\eps(\mL_{X_1\ODeps} (X_j\ODeps)_0) + \mathcal{O}(\eps^{\infty}) =  -  \eps \,\overline{\w_\eps}(\i[X_1\ODeps, X_j\ODeps]) + \mathcal{O}(\eps^{\infty}) ~.
\end{align*}
For $j=1$ we thus obtain \eqref{thm:dc}.
To obtain \eqref{thm:hc} for $1<j \leq d$,
it remains to show that   $\overline{\w_\eps}(\i[X_1\ODeps, X_j\ODeps])$ is independent of $\eps$, or, more generally, that expressions of this form  do not change along continuous families of locally generated automorphisms of the algebra. In the non-interacting case this statement is known as the Chern--Simons lemma \cite{klein1990power,marcelli2022purely} and we prove an analogous result in our setting in Lemma~\ref{lem:CS}. With its help we obtain the final result
\begin{align*}
    \overline{\w_{\eps}}(J_j) &=  -  \eps \,\overline{\w_0}( \i[X_1\OD, X_j\OD]) +  \mathcal{O}(\eps^{\infty}) ~.
\end{align*}
For the last claim let  $\br{\omega_0,H_1,V_1,W_1,T_1,\br{S_{1,\eps}}_{\eps \in [-1,1]}}$ and $\br{\omega_0,H_2,V_2,W_2,T_2,\br{S_{2,\eps}}_{\eps \in [-1,1]}}$ both be gapped periodic systems with the same ground state $\w_0$ and let $(X_j)\OD[1]$ and $(X_j)\OD[2]$ for $j\in \{ 2,\dots, d \}$ be the respective off-diagonal position operators. Since both off-diagonal maps satisfy the Property~\eqref{ODprop} from Lemma~\ref{OD-property} we can calculate with Proposition~\ref{commutator of interactions}
    \begin{align*}
        \w_0(\mL_{X_1\OD[1]}(X_j\OD[1])_0)  &= \w_0(\mL_{X_1\OD[2]}(X_j\OD[1])_0) = -\w_0(\mL_{X_j\OD[1]}(X_1\OD[2])_0) = -\w_0(\mL_{X_j\OD[2]}(X_1\OD[2])_0)\\& = \w_0(\mL_{X_1\OD[2]}(X_j\OD[2])_0) 
    \end{align*}
    and therefore, the Hall conductivities are the same.
\end{proof}

The proof of the following lemma, the CAR-algebra version of the Chern--Simons lemma, stating the invariance of the Hall conductivity under locally generated automorphisms is given in Appendix~\ref{app:CSlemma}.

\medskip 
\begin{lemma}\label{lem:CS}
    Let $\br{\omega_0,H,V,W_g,T,\br{S_\eps}_{\eps \in [-1,1]}}$ be a gapped periodic system and let $(\alpha_{u,v})_{(u,v) \in \R^2}$ be a cocycle of automorphisms generated by a family of $T$-periodic interactions $(\Phi^{u})_{u\in \R}$, such that $\sup_{u\in \R} \norm{\Phi^{u}}_\nu < \infty $ for all $\nu\in \N_0$ and for each $M \in P_0(\Z^2)$ the map $ u \mapsto \Phi^{u}(M)$ is continuous. For all $u,v\in \R$ and $j\in \{1,\dots,d\}$  it holds that 
    \begin{align*}
        \overline{\w_{\alpha_{u,v}}}(\i[ X_1\OD[\alpha_{u,v}],X_j\OD[\alpha_{u,v}]]) = \overline{\w_{0}}(\i[X_1\OD,X_j\OD])~.\\
    \end{align*}
\end{lemma}

Given a translation $T$, we say that a state $\w$ is $T$-gapped if there exist $H$, $V$, $W_g$, and $\br{S_\eps}_{\eps \in [-1,1]}$, such that $\br{\omega,H,V,W_g,T,\br{S_\eps}_{\eps \in [-1,1]}}$ is a gapped periodic system. Two $T$-periodic states $\w_1$ and $\w_2$ are said to lie in the same  $T$-periodic phase if there is a cocycle of automorphisms $(\alpha_{u,v})_{(u,v) \in \R^2}$ which satisfies $\w_2 = \w_1\circ \alpha_{1,0}$ and is generated by a family of $T$-periodic interactions $(\Phi^{u})_{u\in \R}$ such that $\sup_{u\in \R} \norm{\Phi^{u}}_\nu < \infty $ for all $\nu \in \N_0$ and for each $M \in P_0(\Z^d)$ the map $ u \mapsto \Phi^{u}(M)$ is continuous.

\medskip

\begin{corollary}\label{corr:conductivity constant in phase}
    $T$-gapped states that lie in the same $T$-periodic phase with respect to some translation $T$ have the same Hall conductivities.
\end{corollary}
\begin{proof}
    Let $\br{\omega_1,H_1,V_1,W_1,T,\br{S_{1,\eps}}_{\eps \in [-1,1]}}$ and $\br{\omega_2,H_2,V_2,W_2,T,\br{S_{2,\eps}}_{\eps \in [-1,1]}}$ be two gapped periodic systems corresponding to two $T$-gapped states $\w_1$ and $\w_2$ in the same $T$-periodic phase. We denote the off-diagonal maps associated to each system by $(\cdot)\OD[1]$ and $(\cdot)\OD[2]$ and the automorphism connecting the states by $\alpha$. The $j$-th Hall conductivity of $\w_2$ is given by $\w_2(\mL_{X_1\OD[2]} (X_j\OD[2])_0)$. The off-diagonal property with respect to $\w_2 = \w_1 \circ \alpha$ is satisfied by both $(\cdot)\OD[2]$ and $(\cdot)\OD[1\alpha]$ (see Lemma~\ref{OD-property}). We can therefore use Proposition~\ref{commutator of interactions} as in the last part of the proof of Theorem~\ref{thm:main} to arrive at
    \begin{align*}
        \w_2(\mL_{X_1\OD[2]} (X_j\OD[2])_0) =\w_2(\mL_{X_1\OD[1\alpha]} (X_j\OD[1\alpha])_0)\,.
    \end{align*}
    Lemma~\ref{lem:CS} tells us that this expression is equal to $\w_1(\mL_{X_1\OD[1]} (X_j\OD[1])_0)$, which is the $j$-th Hall conductivity of $\w_1$.
\end{proof}

\medskip

\medskip

\begin{corollary}\label{corr:invertible phase}
Let $d=2$ and let $\br{\omega_0,H,V,W_g,T,\br{S_\eps}_{\eps \in [-1,1]}}$ be a gapped periodic system. If  $\omega_0$ is invertible (see \cref{app:invertibility}, Definition \ref{def:invertible}), then the Hall conductivity of all states in the same $T$-periodic phase as $\w_0$ lies in $\frac{1}{2\pi} \Z$.
\end{corollary}
\begin{proof}
Proposition~\ref{prop:doublecommutator for half planes} shows that the Hall conductivity of $\w_0$ is equal to a certain ground-state expectation $\w_0(\i[K_{\Sigma_1,\Sigma_1^{\mathsf{C}}}, K_{\Sigma_2,\Sigma_2^{\mathsf{C}}}])$.
In \cite{kapustin2020hall} it is shown that, if $\w_0$ is invertible, this quantity lies in $\frac{1}{2\pi} \Z$  (the relevant statements can be found in \cite[Theorem 1]{kapustin2020hall} and \cite[Equation 55]{kapustin2020hall}). By \cref{corr:conductivity constant in phase} the Hall conductivity is constant within the $T$-periodic phase of $\w_0$.
\end{proof}    

\medskip

Our final corollary shows that the macroscopic Hall conductance has zero variance. Let $d=2$ and let
\[
    J_L :=   \sum_{m=1}^L j_{m,2} := \sum_{m=1}^L T_{(m,0)}\i[H+\eps V,X_2]_0 
\]
denote the Hall current operator through a line of length $L$. Then the Hall conductance observable 
\[
G^{\mathrm{H}}_{21,L} := \frac{J_L}{\epsi L}
\]
 is the ratio of the current flowing through a line of length $L$ to the voltage drop along that line, where the latter is equal to the electric field strength $\epsi$ times the length $L$ of the line in our setup.

\medskip

\begin{corollary}
    Let $d=2$ and let $\br{\omega_0,H,V,W_g,T,\br{S_\eps}_{\eps \in [-1,1]}}$ be a gapped periodic system. It holds for all $L>0$ that
    \begin{equation} \label{eqn:conductance=conductivity}
    \omega_\epsi(G^{\mathrm{H}}_{21,L} ) = \sigma_{21}^\mathrm{H} + \mathcal{O}\left(\eps^\infty\right)
    \end{equation}
    and
    \[
\mathrm{var}\hspace{-2pt}\left(G^{\mathrm{H}}_{21,L}\right) := \omega_\eps\hspace{-2pt}\left(\left(G^{\mathrm{H}}_{21,L}\right)^2\right) -\left(\omega_\eps\hspace{-2pt}\left(G^{\mathrm{H}}_{21,L}\right)\right)^2=\mathcal{O}\left(\frac{1}{\epsi^2 L}\right)\,.
  \]
\end{corollary}
\begin{proof}
The first statement is an immediate consequence of translation invariance and \cref{thm:main},
\[
\omega_\epsi (J_L) =\sum_{m=1}^L \omega_\epsi  (j_{m}) = \sum_{m=1}^L \omega_\epsi \hspace{-2pt} \left( T_{(m,0)} j_{0}\right) = L \, \omega_\epsi ( j_{0} ) = L \,\overline{\w}_\eps (\i[H+\eps V,X_2])\,,
\]
where here and in the rest of the proof we write $j_m:=j_{m,2}$.

The second statement follows from exponential decay of correlations  in the ground state $\omega_0$, which was established in \cite{hastings2006spectral}. 
To see this, first note that 
\begin{eqnarray*}
    \omega_\epsi (J_L^2) -  \omega_\epsi(J_L)^2&=&  
    \sum_{m,\ell=1}^L \big(\omega_\epsi\left(
    j_{m}\,j_{\ell}
    \right)-
\omega_\epsi\left(
    j_{m} 
    \right)\omega_\epsi\left( j_{\ell}
    \right)
    \big)\\
    &=& \sum_{m,\ell=1}^L \big(\omega_0\left( \beta_\epsi(j_{m})\beta_\epsi(
    j_{\ell})
    \right)-
\omega_0\left(
    \beta_\epsi(j_{m} )
    \right)\omega_0\left( \beta_\epsi(j_{\ell})
    \right)
    \big)\,.
\end{eqnarray*}
Since $\beta_\epsi(j_{0})\in D_\infty$, it holds for every $\nu\geq0$ and $k>0$ that   $\|(1- \E_{\Lambda_k})\beta_\epsi(j_{0})\|\leq \|\beta_\epsi(j_{0})\|_\nu \,(1+k)^{-\nu}$, and, by translation invariance, also  $\|(1- \E_{\Lambda_k+(m,0)})\beta_\epsi(j_{m})\|\leq \|\beta_\epsi(j_{0})\|_\nu \,(1+k)^{-\nu}$.  Thus, defining $j_m^{k,\epsi} := \E_{\Lambda_k+(m,0)}\beta_\epsi(j_{m})$ and choosing in each summand $k=\frac{|m-\ell|}{4}$, we find for $\nu>1$
\begin{eqnarray*}
 \omega_\epsi (J_L^2) -  \omega_\epsi(J_L)^2 &=&   \sum_{m,\ell=1}^L \left(\omega_0 ( j^{\frac{|m-\ell|}{4},\epsi}_{m}\,
    j^{\frac{|m-\ell|}{4},\epsi}_{\ell}
    )-
\omega_0 (
    j^{\frac{|m-\ell|}{4},\epsi}_{m} 
     )\omega_0 ( j^{\frac{|m-\ell|}{4},\epsi}_{\ell}) + \mathcal{O}((1+\tfrac{|m-\ell|}{4})^{-\nu}
    )
    \right)\\
    &\leq & 
    \sum_{ m,\ell=1 }^L C\,\e^{-c|m-\ell|}+ \mathcal{O}(  L)  \;=\; \mathcal{O}(L)\,.
\end{eqnarray*}
In the last inequality we used the uniform exponential decay of correlations in gapped ground states of Hamiltonians with short-range interactions proved in \cite[Thm.\ 2.8]{hastings2006spectral}.
\end{proof}

\begin{remark}
Notice how \eqref{eqn:conductance=conductivity} yields in particular the well-known fact that, in the limit $L \to \infty$, Hall conductance and Hall conductivity agree in 2-dimensions (up to terms which are smaller than any arbitrarily large power of the strength of the inducing electric field). In particular, the Hall conductance in this thermodynamic limit is defined in terms of a ``per-length'' (rather than per-volume, or more precisely per-surface in $d=2$) expectation: see \cite[Equation (5.16)]{marcelli2019spin} and references therein for a discussion on this point.
\end{remark}

\medskip\textbf{Acknowledgments}. 
S.T.\ thanks Joscha Henheik for helpful discussions.
This work was funded by the \foreignlanguage{ngerman}{Deutsche Forschungsgemeinschaft} (DFG, German Research Foundation) –
470903074; 
465199066. G.~M. gratefully acknowledges financial support from the Independent Research Fund Denmark--Natural Sciences,
grant DFF–10.46540/2032-00005B and from the European Research Council through the ERC CoG UniCoSM, grant agreement n.724939. 
T.~M.~was supported by JSPS KAKENHI Grant Numbers 20KK0304 and 23H01086. T. M. extends deep gratitude to the Department of Mathematics at the University of T\"ubingen for their generous hospitality. M.W. and T.M. gratefully acknowledge the financial support from the Research Institute for Mathematical Sciences, an International Joint Usage/Research Center at Kyoto University.
D.~M.~gratefully acknowledges financial support from Sapienza Universit\`{a} di Roma within Progetto di Ricerca di Ateneo 2021, 2022 and 2023 and from Ministero dell'Universit\`a e della Ricerca (MUR, Italian Ministry of University and Research) and Next Generation EU within PRIN 2022AKRC5P ``Interacting Quantum Systems: Topological Phenomena and Effective Theories'' and within the activities for PNRR--MUR Project no.~PE0000023-NQSTI.

\appendix
\section{Estimate for the Commutator of Translated Observables}

The following result will be used as an essential ingredient in the proofs of several propositions and lemmas in the other Appendices.

\medskip

\begin{lemma}\label{convergence}
    Let $A \in D_\infty$ and $ B  \in   D_\infty $. Let $T$ be a translation and $\gamma\in \Z^d$. Then for all $\nu,m\in \N_0$
    \begin{align*}
    \norm{[T_{\g}A,B]}_\nu \;\leq\; 4^{\nu+m+3}\frac{\norm{A}_{\nu+m}\norm{B}_{\nu+m}}{(1 + \norm{\g}_{\infty})^{m}}\,.
    \end{align*}
\end{lemma}
\begin{proof}
    The proof is constituted by a somewhat lengthy computation. Our goal is to find the desired upper bound for
    \begin{align*}
        \norm{[T_{\g}A,B]}_\nu &= \norm{[T_{\g}A,B]} + \sup_{k\in\N_0} \norm{(1-\E_{\La_k})[T_{\g}A,B](1+k)^\nu} ~.
    \end{align*}
    We start by bounding $\norm{[T_{\g}A,B]}$. For this we split the commutator into a sum of three terms, where we define $\Tilde{\gamma} = \floor{\frac{\norm{\gamma}_\infty}{2}}$ for better readability:
    \begin{align*}
        [T_{\g}A,B] =~~&[T_\gamma \E_{\Lambda_{\tg}}A,\E_{\Lambda_{\tg}}B] + [T_\gamma (1-\E_{\Lambda_{\tg}})A,\E_{\Lambda_{\tg}}B] + [T_\gamma A,(1-\E_{\Lambda_{\tg}})B]~.
    \end{align*}
    Now we bound each term separately. By definition of $\tg$ it holds that $(\Lambda_{\tg}+\g) \cap \Lambda_{\tg}$ is empty, except when $\g=0$. Since $B$ is gauge-invariant and $\E_{\La_k}$ preserves this property (\cref{Ex+UniqueExpectation}), the first term is $0$ except when $\g=0$:
    \begin{align*}
        \norm{[T_\gamma \E_{\Lambda_{\tg}}A,\E_{\Lambda_{\tg}}B]} &= \delta_{\g,0} \norm{[ \E_{\Lambda_{0}}A,\E_{\Lambda_{0}}B]}\, .
    \end{align*}
    Using that $\norm{\E_{\La_0}}=1$ and $\delta_{\g,0} \leq \frac{1}{\br{1+ \norm{\g}_\infty}^m}$  we get
    \begin{align*}
        \norm{[T_\gamma \E_{\Lambda_{\tg}}A,\E_{\Lambda_{\tg}}B]} \leq 2 \norm{A} \norm{B} \frac{1}{\br{1+ \norm{\g}_\infty}^m} \, .
    \end{align*}
    For the second term we first estimate the commutator and use that $\norm{T_{\g}} = \norm{\E_{\Lambda_{\tg}}} = 1$ to get
    \begin{align*}
        \norm{[T_\gamma (1-\E_{\Lambda_{\tg}})A,\E_{\Lambda_{\tg}}B]} 
        \leq 2 \norm{(1-\E_{\Lambda_{\tg}})A} \norm{B}~.
    \end{align*}
    From the definition of $\norm{\cdot}_m$ and the fact that $(1+\norm{\g}_{\infty})^m \leq 2^m (1+\tg)^m$ we see
    \begin{align*}        
        \norm{[T_\gamma (1-\E_{\Lambda_{\tg}})A,\E_{\Lambda_{\tg}}B]}  \leq 2\norm{A}_m\norm{B} \frac{1}{(1+\tg)^m} \leq 2\norm{A}_m\norm{B} \frac{2^m}{(1+\norm{\g}_{\infty})^m}.
    \end{align*}
    The third term can be handled by the same reasoning:
    \begin{align*}
        \norm{[T_\gamma A,(1-\E_{\Lambda_{\tg}})B]} \leq 2\norm{A} \norm{(1-\E_{\Lambda_{\tg}})B}\leq 2\norm{A}\norm{B}_m \frac{1}{(1+\tg)^m}\leq 2\norm{A}\norm{B}_m \frac{2^m}{(1+\norm{\g}_{\infty})^m}\, .
    \end{align*}
    With all three estimates and using that $\norm{A} \leq \norm{A}_m$ we have shown that
    \begin{align*}
        \norm{[T_{\g}A,B]} \leq &6\norm{A}_m \norm{B}_m\frac{2^m}{(1+\norm{\g}_{\infty})^m}~. 
    \end{align*}
    Next we deal with the supremum part of $\norm{[T_{\g}A,B]}_\nu$. For this we choose a $k\in\N_0$ and consider the expression
    \begin{align*}
        \norm{(1-\E_{\Lambda_k})[T_\gamma A ,B]} (1+k)^\nu ~,
    \end{align*}
    for which we would like to find a suitable $k$-indepentent bound. To do this we again split the commutator as before but this time we further split the second and third therm so that we end up with five terms in total:
    \begin{align*}
        (1-\E_{\Lambda_k}) [T_\gamma A,B]~(1+k)^\nu \;=\;
        &(1-\E_{\Lambda_k}) [T_\gamma \E_{\Lambda_{\tg}}A,\E_{\Lambda_{\tg}}B]~(1+k)^\nu\\
        &+\, (1-\E_{\Lambda_k}) [\E_{\Lambda_k} T_\gamma (1-\E_{\Lambda_{\tg}})A,\E_{\Lambda_{\tg}}B]~(1+k)^\nu\\
        &+\,  (1-\E_{\Lambda_k}) [(1-\E_{\Lambda_k}) T_\gamma (1-\E_{\Lambda_{\tg}})A,\E_{\Lambda_{\tg}}B]~(1+k)^\nu\\
        &+\, (1-\E_{\Lambda_k}) [\E_{\Lambda_k}T_\gamma A,(1-\E_{\Lambda_{\tg}})B]~(1+k)^\nu\\
        &+\, (1-\E_{\Lambda_k}) [(1- \E_{\Lambda_k})T_\gamma A,(1-\E_{\Lambda_{\tg}})B]~(1+k)^\nu \, .
    \end{align*}
    To easily refer to the norms of these five terms we define
    \begin{align*}
        \mathrm{(I)}&:=\norm{(1-\E_{\Lambda_k}) [T_\gamma \E_{\Lambda_{\tg}}A,\E_{\Lambda_{\tg}}B]}~(1+k)^\nu \, ,\\
        \mathrm{(II)}&:=\norm{(1-\E_{\Lambda_k}) [\E_{\Lambda_k} T_\gamma (1-\E_{\Lambda_{\tg}})A,\E_{\Lambda_{\tg}}B]}~(1+k)^\nu \, ,\\
        \mathrm{(III)}&:=\norm{(1-\E_{\Lambda_k}) [(1-\E_{\Lambda_k}) T_\gamma (1-\E_{\Lambda_{\tg}})A,\E_{\Lambda_{\tg}}B]}~(1+k)^\nu \, ,\\
        \mathrm{(IV)}&:=\norm{(1-\E_{\Lambda_k}) [\E_{\Lambda_k}T_\gamma A,(1-\E_{\Lambda_{\tg}})B]}~(1+k)^\nu \, ,\\
        \mathrm{(V)}&:=\norm{(1-\E_{\Lambda_k}) [(1- \E_{\Lambda_k})T_\gamma A,(1-\E_{\Lambda_{\tg}})B]}~(1+k)^\nu~.
    \end{align*}
    We see that the first term can be dealt with in the same way we did before:
    \begin{align*}
         \mathrm{(I)} =\norm{(1-\E_{\Lambda_k}) \delta_{\gamma,0}[\E_{\Lambda_0}A,\E_{\Lambda_0}B]} ~(1+k)^\nu \, .
    \end{align*}
    The expression is equal to zero, because $\La_0$ is a subset of $\La_k$ and $\E_{\Lambda_k}$ acts as the identity on $\mA_{\La_k}$.

    For the second term we can use the first property from \cref{Ex+UniqueExpectation} to move $(1-\E_{\Lambda_k})$ to the second slot of the commutator and then estimate the commutator:
    \begin{align*}
        \mathrm{(II)}&=\norm{(1-\E_{\Lambda_k}) [\E_{\Lambda_k} T_\gamma (1-\E_{\Lambda_{\tg}})A,\E_{\Lambda_{\tg}}B]}~(1+k)^\nu\\
        &=\norm{[\E_{\Lambda_k} T_\gamma (1-\E_{\Lambda_{\tg}})A,(1-\E_{\Lambda_k}) \E_{\Lambda_{\tg}}B]}~(1+k)^\nu\\
        &\leq 2\norm{\E_{\Lambda_k} T_\gamma (1-\E_{\Lambda_{\tg}})A}\norm{(1-\E_{\Lambda_k}) \E_{\Lambda_{\tg}}B}~(1+k)^\nu\,.
    \end{align*}
    Next we use that $\norm{T_\g} = \norm{\E_{\Lambda_k}} = \norm{\E_{\Lambda_{\tg}}} = 1$ and that we can swap $(1-\E_{\Lambda_k})$ and $\E_{\Lambda_{\tg}}$ by \cref{Ex+UniqueExpectation} to get
    \begin{align*}
        \mathrm{(II)}
        &\leq 2\norm{(1-\E_{\Lambda_{\tg}})A}\norm{(1-\E_{\Lambda_k}) B}~(1+k)^\nu~.
    \end{align*}
    We estimate by the decay norm and use $(1+\norm{\g}_{\infty})^m \leq 2^m (1+\tg)^m$:
    \begin{align*}
        \mathrm{(II)}  \leq 2\norm{A}_{m} \norm{B}_{\nu} \frac{(1+k)^\nu}{(1+\tg)^m (1+k)^\nu} \leq 2\norm{A}_m \norm{B}_\nu \frac{2^m}{(1+\norm{\g}_\infty)^{m}} \, .
    \end{align*}
    We deal with the third term by using $\norm{1-\E_{\Lambda_k}} \leq 2 $, and then further estimating via the commutator and $\norm{\E_{\Lambda_{\tg}}}= 1$:
    \begin{align*}
        \mathrm{(III)} &=\norm{(1-\E_{\Lambda_k})[(1-\E_{\Lambda_k}) T_\gamma (1-\E_{\Lambda_{\tg}})A,\E_{\Lambda_{\tg}}B]}(1+k)^\nu\\
        &\leq 2\norm{[(1-\E_{\Lambda_k}) T_\gamma (1-\E_{\Lambda_{\tg}})A,\E_{\Lambda_{\tg}}B]}(1+k)^\nu\\
        &\leq 4\norm{(1-\E_{\Lambda_k}) T_\gamma (1-\E_{\Lambda_{\tg}})A}\norm{B}(1+k)^\nu \, .
    \end{align*}
    By applying \cref{lem: compatibility of translations} we pull $T_{\g}$ past the conditional expectation and use $\norm{T_{\g}}= 1$ to get
    \begin{align*}
        \mathrm{(III)} &\leq 4\norm{(1-\E_{(\Lambda_{k}-\g)}) (1-\E_{\Lambda_{\tg}})A}\norm{B}(1+k)^\nu~.
    \end{align*}
    We observe that using $\norm{1-\E_{\La_k-\g}} \leq 2$ one gets 
    \begin{align*}
        \norm{(1-\E_{(\Lambda_{k}-\g)}) (1-\E_{\Lambda_{\tg}})A}  \leq 2\norm{(1-\E_{\Lambda_{\tg}})A}  \leq \frac{2\norm{A}_{\nu+m}}{(1+\tg)^{\nu+m}}\, .
    \end{align*}
    Now if $k-\norm{\g}_\infty \geq 0$ we have $\La_{k-\norm{\g}_\infty} \subseteq \La_k -\g$, which together with \cref{Ex+UniqueExpectation} and estimating by the decay norm gives us the stronger bound
    \begin{align*}
        \norm{(1-\E_{(\Lambda_{k}-\g)}) (1-\E_{\Lambda_{\tg}})A} &\;\leq\; 2 \norm{(1-\E_{\La_{k-\norm{\g}_\infty}}) (1-\E_{\Lambda_{\tg}})A} \;=\; 2\norm{(1-\E_{\La_{\max(\tg,k-\norm{\g}_\infty)}}) A}\\
        &\;\leq\; \frac{2\norm{A}_{\nu +m}}{(1 + \max(\tg,k-\norm{\g}_\infty))^{\nu + m}}\, .
    \end{align*}
    Combining the two statements, one sees that the stronger bound in fact always holds, which results in
    \begin{align*}
        \mathrm{(III)} &\leq 8\norm{A}_{\nu +m} \norm{B} \frac{ (1+k)^\nu }{(1 + \max(\tg,k-\norm{\g}_\infty))^{\nu + m}}\, .
    \end{align*}
    The function $x \mapsto \frac{ (1+x)^\nu }{(1 + \max(\tg,x-\norm{\g}_\infty))^{\nu + m}}$ is strictly increasing for $x \,<\, \tg + \norm{\g}_\infty$ and strictly decreasing for $x  \,>\, \tg + \norm{\g}_\infty$, thus
    \begin{align*}
        \mathrm{(III)}  \leq 8\norm{A}_{\nu +m} \norm{B} \frac{ (1+\tg + \norm{\g}_\infty)^\nu }{(1 + \max(\tg,\tg + \norm{\g}_\infty -\norm{\g}_\infty))^{\nu + m}} = 8\norm{A}_{\nu +m} \norm{B} \frac{ (1+\tg + \norm{\g}_\infty)^\nu }{(1 + \tg)^{\nu + m}}\, .
    \end{align*}
    We use $(1+\norm{\g}_{\infty})^{\nu+m} \leq 2^{\nu+m} (1+\tg)^{\nu+m}$ and $(1 + \norm{\g}_{\infty}+\tg)^\nu \leq 2^\nu(1+\norm{\g}_{\infty})^\nu$ to obtain
    \begin{align*}
        \mathrm{(III)}
        \leq 8\norm{A}_{\nu+m}\norm{B} \frac{2^\nu 2^{\nu+m}(1+\norm{\g}_{\infty})^\nu}{(1+\norm{\g}_{\infty})^{\nu+m}} \leq 8\norm{A}_{\nu+m}\norm{B} \frac{4^{\nu+m}}{(1+\norm{\g}_{\infty})^{m}}~.
    \end{align*}   
   
    For the fourth term we move $(1-\E_{\La_k})$ inside the commutator, estimate the commutator and use $ \norm{\E_{\La_k}} = \norm{T_{\g}} = 1$, which results in
    \begin{align*}
        \mathrm{(IV)} &\;=\; \norm{(1-\E_{\Lambda_k})[\E_{\Lambda_k}T_\gamma A,(1-\E_{\Lambda_{\tg}})B]}(1+k)^\nu \;=\;\norm{[\E_{\Lambda_k}T_\gamma A,(1-\E_{\Lambda_k})(1-\E_{\Lambda_{\tg}})B]}(1+k)^\nu\\
        &\;\leq\; 2\norm{\E_{\Lambda_k}T_\gamma A}\norm{(1-\E_{\Lambda_k})(1-\E_{\Lambda_{\tg}})B}(1+k)^\nu \;\leq\; 2\norm{A}\norm{(1-\E_{\Lambda_k})(1-\E_{\Lambda_{\tg}})B}(1+k)^\nu~.
    \end{align*}
    Applying \cref{Ex+UniqueExpectation} in the form of $(1-\E_{\Lambda_k})(1-\E_{\Lambda_{\tg}}) = 1-\E_{\Lambda_{\max(\tg,k)}}$ and estimating the result with the decay norm leaves us with
    \begin{align*}
        \mathrm{(IV)}  \;\leq\; 2\norm{A}\norm{(1-\E_{\Lambda_{\max(\tg,k)}})B}(1+k)^\nu \;\leq\; 2\norm{A}\norm{B}_{\nu+m} \frac{(1+k)^\nu}{(1+\max(\tg,k))^{\nu+m}}~.
    \end{align*}
    Now we simply estimate the maximum and then use $(1+\norm{\g}_{\infty})^{m} \leq 2^{m} (1+\tg)^{m}$:
    \begin{align*}
        \mathrm{(IV)}  \;\leq\; 2\norm{A}\norm{B}_{\nu+m} \frac{(1+k)^\nu}{(1+\tg)^{m}(1+k)^{\nu}} \;= \;2\norm{A}\norm{B}_{\nu+m} \frac{1}{(1+\tg)^{m}}\; \leq \;2\norm{A}\norm{B}_{\nu+m} \frac{2^m}{(1+\norm{\g}_\infty)^{m}} \, .
    \end{align*}
    We tackle the fifth term by estimating with $\norm{1-\E_{\La_k}} \leq 2$ and the normal commutator bound. This results in
    \begin{align*}
        \mathrm{(V)} &\;= \;\norm{(1-\E_{\Lambda_k}) [(1- \E_{\Lambda_k})T_\gamma A,(1-\E_{\Lambda_{\tg}})B]}(1+k)^\nu \;\leq \;2\norm{[(1- \E_{\Lambda_k})T_\gamma A,(1-\E_{\Lambda_{\tg}})B]}(1+k)^\nu\\
        &\;\leq\; 4\norm{(1- \E_{\Lambda_k})T_\gamma A}\norm{(1-\E_{\Lambda_{\tg}})B}(1+k)^\nu\,.
    \end{align*}
    Using \cref{lem: compatibility of translations} and $\norm{T_{\g}}=1$ we obtain     \begin{align*}
        \mathrm{(V)} \;\leq\; 4\norm{T_\gamma(1- \E_{(\Lambda_k-\g)})A}\norm{(1-\E_{\Lambda_{\tg}})B}(1+k)^\nu\;\leq \; 4\norm{(1- \E_{(\Lambda_k-\g)})A}\norm{(1-\E_{\Lambda_{\tg}})B}(1+k)^\nu~.
    \end{align*}
    Similarly to before we observe that
    \begin{align*}
        \norm{(1- \E_{(\Lambda_k-\g)})A} \leq 2 \norm{A} \leq 2 \frac{\norm{A}_{\nu+m}}{(1 + 0)^{\nu+m}}
    \end{align*}
    and if $k-\norm{\g}_{\infty} \geq 0$, we have $\Lambda_{(k-\norm{\g}_{\infty})} \subseteq (\Lambda_{k}-\g)$, and therefore get the stronger bound
    \begin{align*}
        \norm{(1- \E_{(\Lambda_k-\g)})A} \;\leq \;2\norm{(1- \E_{\Lambda_{(k-\norm{\g}_{\infty})}})A}\;\leq \;2 \frac{\norm{A}_{\nu+m}}{(1 + k-\norm{\g}_{\infty})^{\nu+m}}~.
    \end{align*}
    Together these two statements show that we always have
    \begin{align*}
        \norm{(1- \E_{(\Lambda_k-\g)})A} \;\leq\;  2 \frac{\norm{A}_{\nu+m}}{(1 + \max(0,k-\norm{\g}_{\infty}))^{\nu+m}}~. 
    \end{align*}
    We use  this observation and the decay norm to further estimate $\mathrm{(V)}$:
    \begin{align*}
        \mathrm{(V)} \;\leq\; 8  \frac{\norm{A}_{\nu+m}}{(1 + \max(0,k-\norm{\g}_{\infty}))^{\nu+m}}\norm{(1-\E_{\Lambda_{\tg}})B}(1+k)^\nu\;\leq  \;\frac{8\norm{A}_{\nu+m}\norm{B}_{\nu+m}(1+k)^\nu}{(1 + \max(0,k-\norm{\g}_{\infty}))^{\nu+m}(1+\tg)^{\nu+m}} \, .
    \end{align*}
    Next we again use $(1+\norm{\g}_{\infty})^{\nu+m} \leq 2^{\nu+m} (1+\tg)^{\nu+m}$ and  
    \begin{align*}
        1 + \max(\norm{\g}_{\infty},k) \leq (1 + \max(0,k-\norm{\g}_{\infty}))(1+\norm{\g}_{\infty})   
    \end{align*}
   to get
    \begin{align*}
        \mathrm{(V)} \;\leq  \;\frac{8\norm{A}_{\nu+m}\norm{B}_{\nu+m}2^{\nu+m}(1+k)^\nu}{(1 + \max(0,k-\norm{\g}_{\infty}))^{\nu+m}(1+\norm{\g}_{\infty})^{\nu+m}}\;\leq  \;\frac{8\norm{A}_{\nu+m}\norm{B}_{\nu+m}2^{\nu+m}(1+k)^\nu}{(1 + \max(\norm{\g}_{\infty},k))^{\nu+m}}~.
    \end{align*}
    By estimating the maximum we conclude
    \begin{align*}
        \mathrm{(V)} \;\leq  \;\frac{8\norm{A}_{\nu+m}\norm{B}_{\nu+m}2^{\nu+m}(1+k)^\nu}{(1 + \norm{\g}_{\infty})^{m}(1 + k)^{\nu}}\;= \; \frac{8\norm{A}_{\nu+m}\norm{B}_{\nu+m}2^{\nu+m}}{(1 + \norm{\g}_{\infty})^{m}}~.
    \end{align*}
    We combine the estimates for all four non-zero terms, this shows that 
    \begin{align*}
       \sup_{k\in \N_0} \norm{(1-\E_{\Lambda_k})[T_\gamma A ,B]} (1+k)^\nu 
       &\leq~ \frac{20\norm{A}_{\nu+m}\norm{B}_{\nu+m}4^{\nu+m}}{(1 + \norm{\g}_{\infty})^{m}}~.
    \end{align*}
    In total we have shown that
    \begin{align*}
        \norm{[T_{\g}A,B]}_\nu
        &\;=\;\norm{[T_{\g}A,B]} + \sup_{k\in \N_0}  \lVert (1-\E_{\Lambda_k})[T_\gamma A ,B] \rVert (1+k)^\nu\\
        &\;\leq\;  6\norm{A}_m \norm{B}_m\frac{2^m}{(1+\norm{\g}_{\infty})^m} + \frac{20\norm{A}_{\nu+m}\norm{B}_{\nu+m}4^{\nu+m}}{(1 + \norm{\g}_{\infty})^{m}} 
         \;\leq \;4^{\nu+m+3}\frac{\norm{A}_{\nu+m}\norm{B}_{\nu+m}}{(1 + \norm{\g}_{\infty})^{m}}~.\;\qedhere
    \end{align*}
\end{proof}

\section{Proofs of Basic Propositions in Section \ref{sec:tech} }\label{app:B}

\begin{proof}[Proof of Lemma \ref{lem:norm properties}]
Let $k,\nu \in \N_0$ and $A,B\in D_\nu$. Then
    \begin{align*}
        \lVert A\,B - \E_{\La_k}  (A\,B) \rVert 
        &\leq \lVert A\,B - \E_{\La_k}  (A\,\E_{\La_k}B) \rVert + \lVert \E_{\La_k} (A\, (B - \E_{\La_k}B)) \rVert\\
        &= \lVert A\,B - \E_{\La_k}  A\,\E_{\La_k}B \rVert + \lVert \E_{\La_k} (A\, (B - \E_{\La_k}B)) \rVert\\
        &\leq \lVert A\, B - A\, \E_{\La_k} B \rVert + \lVert A\, \E_{\La_k} B - \E_{\La_k}  A\,\E_{\La_k}B \rVert + \lVert \E_{\La_k} (A\, (B - \E_{\La_k}B)) \rVert \\
        &\leq \lVert A \rVert \, \lVert B - \E_{\La_k}B \rVert + \lVert A -\E_{\La_k}A \rVert \, \lVert B \rVert + \lVert A \rVert \, \lVert B - \E_{\La_k}B \rVert 
    \end{align*}
    and therefore 
    \begin{align*}
        \hspace{2em}&\hspace{-2em}\lVert A\, B \rVert + \sup_{k\in \N_0} \lVert A\, B - \E_{\La_k}(A\,B) \rVert \, (1+k)^\nu \\
        &\leq 2 \, (\norm{A} + \sup_{k\in \N_0}( \norm{A-\E_{\La_k}A} (1+k)^\nu)) \, (\norm{B} + \sup_{k\in \N_0}( \norm{B-\E_{\La_k}B} (1+k)^\nu) \, .
    \end{align*}
For the second statement note that, by assumption,
    \begin{align*}
        \sup_{l\in \N_0} \lVert A - \E_{\La_l}A \rVert \, (1+l)^{\nu} < \infty \, .
    \end{align*}
    Therefore, $\lim_{l\to \infty} \lVert A - \E_{\La_l}A \rVert \, (1+l)^{\nu-1} = 0$. Hence in
    \begin{align*}
        \lVert A -\E_{\La_k}A \rVert_\nu 
        &= \lVert A -\E_{\La_k}A \rVert+ \sup_{l\in \N_0}\lVert (1-\E_{\La_l})\, (1-\E_{\La_k})\, A \rVert \, (1+l)^{\nu-1}\\
        & = \lVert A -\E_{\La_k}A \rVert+ \sup_{l\in \N_0}\lVert (1-\E_{\La_{\max(l,k)}})\, A \rVert \, (1+l)^{\nu-1}\\
        & = \lVert A -\E_{\La_k}A \rVert+ \sup_{\substack{l\in \N_0 \\ l\geq k}}\lVert (1-\E_{\La_l})\, A \rVert \, (1+l)^{\nu-1}\, 
    \end{align*}
    both terms go to $0$ as $k\to \infty$.
\end{proof}

\begin{proof}[Proof of \cref{lem: sum representation of generator}]
    For $M\in P_0(\Z^d)$ with $0 \notin M$ we define $\La_M $ as the largest box around $0$ that does not intersect $M$ and use this with the fact that $\Phi(M)$ commutes with all operators supported outside $M$ to rewrite the sum
    \begin{align*}
        \sum_{M\in P_0(\Z^d)} \norm{[\Phi(M),A]}_\nu &\;= 
        \sum_{\substack{M\in P_0(\Z^d)\\ 0\in M}} \norm{[\Phi(M),A]}_\nu +\sum_{k=0}^\infty \sum_{\substack{M\in P_0(\Z^d)\\ 0 \notin M \\\La_M = \La_k}} \norm{[\Phi(M),A ]}_\nu  \\
        &\;= \sum_{\substack{M\in P_0(\Z^d)\\ 0\in M}} \norm{[\Phi(M),A]}_\nu + \sum_{k=0}^\infty \sum_{\substack{M\in P_0(\Z^d)\\ 0\notin M \\ \La_M = \La_k}} \norm{[\Phi(M),A - \E_{\La_k}A ]}_\nu\;\eqcolon\; S_1 + S_2 \,.
    \end{align*}
    To estimate $S_1$, we use the first property in \cref{lem:norm properties} and get
    \begin{align*}
        S_1 &\leq \sum_{\substack{M\in P_0(\Z^d)\\ 0\in M}} 4\,\norm{\Phi(M)}_\nu \, \norm{A}_\nu \, .
    \end{align*}
    Now we observe that each $M$ in the sum is contained in a box around $0$ with side length $2\, \mathrm{diam}(M)$, which allows us to estimate $\norm{\Phi(M)}_\nu$ by $3 \,(1+ \diam(M))^\nu \,\norm{\Phi(M)}$. We thus arrive at 
    \begin{equation}\label{eq: decay norm when containing 0}
        S_1  \leq \sum_{\substack{M\in P_0(\Z^d)\\ 0\in M}} 12 \,(1+ \diam(M))^\nu \,\norm{\Phi(M)} \, \norm{A}_\nu \leq 12 \, \norm{\Phi}_\nu \, \norm{A}_\nu \, .
    \end{equation}
    We can estimate $S_2$ by splitting it into parts associated to each lattice site while summing over sets multiple times to get
    \begin{align*}
        S_2 &\leq \sum_{x\in \Z^d} \sum_{k=0}^\infty \sum_{\substack{M\in P_0(\Z^d)\\ 0 \notin M, \, x\in M\\ \La_M = \La_k}} \norm{[\Phi(M),A - \E_{\La_k}A ]}_\nu \, .
    \end{align*}
    The decay norm of the commutator can again be estimated as in \cref{lem:norm properties}, giving us
    \begin{align*}
        S_2 &\leq \sum_{x\in \Z^d} \sum_{k=0}^\infty \sum_{\substack{M\in P_0(\Z^d)\\ 0 \notin M, \, x\in M\\ \La_M = \La_k}} 4 \, \norm{\Phi(M)}_\nu \, \norm{A - \E_{\La_k}A }_\nu\,.
    \end{align*}
    Similarly to before, we observe that $M$ is always contained in a box around $0$ with side length $ 2 \, \left( \norm{x}_\infty + \mathrm{diam}(M) \right) $, which we use to estimate the decay norm of $\Phi(M)$ 
    \begin{align*}
        S_2 &\leq \sum_{x\in \Z^d} \sum_{k=0}^\infty \sum_{\substack{M\in P_0(\Z^d)\\ 0 \notin M, \, x\in M\\ \La_M = \La_k}} 12 \, (1 + \norm{x}_\infty + \mathrm{diam}(M))^\nu \, \norm{\Phi(M)} \, \norm{A - \E_{\La_k}A }_\nu\\
        &\leq \sum_{x\in \Z^d} \sum_{k=0}^\infty \sum_{\substack{M\in P_0(\Z^d)\\ 0 \notin M, \, x\in M\\ \La_M = \La_k}} 12 \, (1 + \norm{x}_\infty)^\nu (1 + \mathrm{diam}(M))^\nu \, \norm{\Phi(M)} \, \norm{A - \E_{\La_k}A }_\nu\,.
    \end{align*}   
    Each $M$ in the sum contains $x$ and some element that has a distance of $k+1$ from the origin, so $\mathrm{diam}(M) \geq \abs{\norm{x}_\infty - k -1}$. This allows us to further estimate the expression as 
    \begin{align*}
        S_2 &\leq \sum_{x\in \Z^d} \sum_{k=0}^\infty \sum_{\substack{M\in P_0(\Z^d)\\ 0 \notin M, \, x\in M\\ \La_M = \La_k}} 12 \, \frac{(1 + \norm{x}_\infty)^\nu (1+\mathrm{diam}(M))^{d+1+2\nu}}{(1 + \abs{\norm{x}_\infty - k - 1} )^{d+1+\nu}}\norm{\Phi(M)} \, \norm{A - \E_{\La_k}A }_\nu\\
        &\leq \sum_{x\in \Z^d} \sum_{k=0}^\infty  24 \, \frac{(1 + \norm{x}_\infty)^\nu}{(1 + \abs{\norm{x}_\infty - k - 1} )^{d+1+\nu}}  \,  \norm{\Phi}_{d+1+2\nu}   \,  \frac{1}{(1+k)^{d+3+\nu}}  \,   \norm{A}_{d+3+2\nu}\\
        &\leq \sum_{x\in \Z^d} \sum_{k=0}^\infty  24 \, \frac{(1 + \norm{x}_\infty)^\nu}{(1 + \abs{\norm{x}_\infty - 1} )^{d+1+\nu}}  \,  \norm{\Phi}_{d+1+2\nu} \, \frac{1}{(1+k)^{2}}  \,   \norm{A}_{d+3+2\nu} \, ,
    \end{align*}
    which is finite. In the first step of the above estimate we used that $ (1+\mathrm{diam}(M))^{d+1+2\nu} \norm{\Phi(M)} \leq \norm{\Phi}_{d+1+2\nu}$ and $\norm{A - \E_{\La_k}A }_\nu \leq \frac{2}{(1+k)^{d+3+\nu}}  \,   \norm{A}_{d+3+2\nu} $. The second estimate is obtained as follows:
    \begin{align*}
        \hspace*{2em}&\hspace*{-2em}\norm{A - \E_{\La_k}A }_\nu \, (1+k)^{d+3+\nu}\\
         &= \norm{A - \E_{\La_k}A  }\, (1+k)^{d+3+\nu} + \sup_{l\in\N_0} \norm{(1-\E_{\La_l})(1-\E_{\La_k})A} (1+l)^\nu(1+k)^{d+3+\nu}\\
         &= \norm{A - \E_{\La_k}A  }\, (1+k)^{d+3+\nu} + \sup_{l\in\N_0} \norm{(1-\E_{\La_{\max(l,k)}})A} (1+l)^\nu(1+k)^{d+3+\nu}\\
         &= \norm{A - \E_{\La_k}A  }\, (1+k)^{d+3+\nu} + \sup_{\substack{l\in\N_0\\ l \geq k}} \norm{(1-\E_{\La_{l}})A} (1+l)^\nu(1+k)^{d+3+\nu}\\
         &\leq \norm{A - \E_{\La_k}A  }\, (1+k)^{d+3+\nu} + \sup_{\substack{l\in\N_0\\ l \geq k}} \norm{(1-\E_{\La_{l}})A} (1+l)^{d+3+2\nu}\\
         &\leq 2 \norm{A}_{d+3+2\nu}\,.
    \end{align*}
    We proceed similarly for the sum corresponding to the position operator
    \begin{align*}
        \sum_{x\in \Z^d} \norm{\sq{x_j\,n_x, A}}_\nu  = \sum_{x\in \Z^d} \norm{\sq{x_j\,n_x, A -\E_{\La_{(\norm{x}_\infty -1)}}A }}_\nu \leq \sum_{x\in \Z^{d}} 4 \abs{x_j} \,\norm{n_x}_\nu \,\norm{A -\E_{\La_{(\norm{x}_\infty -1)}}A }_\nu \,.
    \end{align*}
    Using that $\norm{n_x}_\nu \leq 3 \, \norm{n_0} \left(1+ \left( \norm{x}_\infty - 1 \right) \right)^\nu$ together with the same estimate as above we get
    \begin{align*}
        \sum_{x\in \Z^d} \norm{\sq{x_j\,n_x, A}}_\nu \leq \sum_{x\in \Z^d} 24 \, \abs{x_j} \,\norm{n_0} \, \frac{(1+ \abs{\norm{x}_\infty - 1})^{\nu}}{(1+ \abs{\norm{x}_\infty - 1})^{d+2+\nu}} \norm{A}_{d+2+2\nu}\,,
    \end{align*}
    which is again finite. Thus, we see that both sums converge absolutely and that the limits lie in $D_\infty$.
    Further we observe that for $k\in \N_0$
    \begin{align*}
        \norm{\sum_{M\in P_0(\Z^d)} [(p\Phi + q X_j)(M),A] - \mL_{p\Phi + q X_j}^\circ( \E_{\La_k} A)} \leq \sum_{M\in P_0(\Z^d)} \norm{ [(p\Phi + q X_j)(M),A - \E_{\La_k} A]}_0 \, .
    \end{align*}
    By the above estimates and the second property from \cref{lem:norm properties} the expression converges to $0$ as $k\to \infty$. Since $\mL_{p\Phi + q X_j}$ is the closure of $\mL_{p\Phi + q X_j}^\circ$, we can conclude that $A\in D(\mL_{p\Phi + q X_j})$ and 
    \[
        \mL_{p\Phi + q X_j} A \;= \sum_{M\in P_0(\Z^d)} [(p\Phi + q X_j)(M),A] \,=\, p \sum_{M\in P_0(\Z^d)} [\Phi(M),A] + q \sum_{x\in \Z^d}  [x_j n_x, A] \, .
    \]
    Combining the three estimates results in a constant $c_\nu$ such that
    \[
        \norm{\mL_{p\Phi + q X_j} A}_\nu \leq c_\nu \, (p \,\norm{\Phi}_{d+1+2\nu} + q \,\norm{n_0}) \, \norm{A}_{d+3+2\nu}\, . \qedhere
    \]
\end{proof}

\begin{proof}[Proof of \cref{cocycles}]
    The existence of a unique generated cocycle under the  conditions in \cref{cocycles} is proven in \cite[Corollary 5.2]{BruPedra2017} together with the following Lieb--Robinson Bound. There exists a constant $\tilde{c}_{\nu}$ such that for all $\La_1, \La_2 \in P_0(\Z^d)$ and $A\in \mA_{\La_1}^N$, $B\in \mA_{\La_2}$ it holds that
    \begin{align*}
        \norm{[\alpha_{u,v}A,B]} &\leq 2 \tilde{c}_{\nu}^{-1} \left(\exp\!\Big(2\tilde{c}_{\nu} \, \sup_{t\in \R}\,  \norm{\Phi^t}_{2d+1+\nu} \, |u-v|\Big) -1\right) \, \norm{A} \norm{B} \sum_{x \in  \La_1} \sum_{y\in \La_2} \frac{1}{(1+\norm{x-y}_\infty)^{2d+1+\nu}}\, .
    \end{align*}
    For our purposes it is convenient to work with the weaker bound: 
    \begin{align*}
        \norm{[\alpha_{u,v}A,B]} &\leq 2 \tilde{c}_{\nu}^{-1} \kappa_d \exp\!\Big(2\tilde{c}_{\nu} \, \sup_{t\in \R}\,  \norm{\Phi^t}_{2d+1+\nu} \, |u-v|\Big) \, \norm{A} \norm{B} |\La_1| \,   \frac{1}{(1+\mathrm{dist}(\La_1,\La_2))^{d+\nu}}\, ,
    \end{align*}
    where $\kappa_d = \sum_{y\in\Z^d} \frac{1}{(1+\norm{y}_\infty)^{d+1}}$. To prove the desired bound on 
    \begin{align*}
        \norm{\alpha_{u,v} A }_\nu 
        &= \norm{ A }  + \sup_{k\in \N_0} \norm{(1-\E_{\La_k})\alpha_{u,v} A}(1+k)^\nu\, ,
    \end{align*}
    we estimate the expression under the supremum as 
    \begin{align*}
        \norm{(1-\E_{\La_k})\alpha_{u,v} A}(1+k)^\nu &\leq \norm{(1-\E_{\La_k})\alpha_{u,v} \E_{\La_{\floor{\frac{k}{2}}}}A}(1+k)^\nu + \norm{(1-\E_{\La_k})\alpha_{u,v} (1 - \E_{\La_{\floor{\frac{k}{2}}}})A}(1+k)^\nu\\
        &\leq \norm{(1-\E_{\La_k})\alpha_{u,v} \E_{\La_{\floor{\frac{k}{2}}}}A}(1+k)^\nu + 2\norm{(1 - \E_{\La_{\floor{\frac{k}{2}}}}) A} (1+k)^\nu\\
        &\leq \norm{(1-\E_{\La_k})\alpha_{u,v} \E_{\La_{\floor{\frac{k}{2}}}}A}(1+k)^\nu + 2^{\nu+1} \norm{A}_\nu \, .
    \end{align*}
    Lemma C.2 from \cite{HT20b} tells us that we can use our Lieb--Robinson Bound to get the following estimate:
    \begin{align*}
        \norm{(1-\E_{\La_k})\alpha_{u,v} \E_{\La_{\floor{\frac{k}{2}}}}A} (1+k)^\nu&\leq 2\tilde{c}_{\nu}^{-1} \kappa_d \exp\!\Big(2\tilde{c}_{\nu} \, \sup_{t\in \R}\,  \norm{\Phi^t}_{2d+1+\nu} \, |u-v|\Big) \, \norm{A}  \,|\La_{\floor{\frac{k}{2}}}| \,   \frac{(1+k)^\nu}{(1+\floor{\frac{k}{2}})^{\nu+d}}\\
        &\leq 2\tilde{c}_{\nu}^{-1} \kappa_d \exp\!\Big(2\tilde{c}_{\nu} \, \sup_{t\in \R}\,  \norm{\Phi^t}_{2d+1+\nu} \, |u-v|\Big) \, \norm{A}   \, \frac{2^d(1+k)^\nu}{(1+\floor{\frac{k}{2}})^{\nu}} \\
        &\leq 2\tilde{c}_{\nu}^{-1} \kappa_d \exp\!\Big(2\tilde{c}_{\nu} \, \sup_{t\in \R}\,  \norm{\Phi^t}_{2d+1+\nu} \, |u-v|\Big) \, \norm{A}   \, 2^{d+ \nu } \, .
    \end{align*}
    In total we have
    \begin{align*}
        \norm{\alpha_{u,v} A }_\nu &\leq  \norm{ A } + 2^{d+\nu+1} \,\tilde{c}_{\nu}^{-1} \kappa_d \exp\!\Big(2\tilde{c}_{\nu} \, \sup_{t\in \R}\,  \norm{\Phi^t}_{2d+1+\nu} \, |u-v|\Big) \, \norm{A} + 2^{\nu+1} \norm{A}_\nu \\
        &\leq (1 + 2^{d+\nu + 1} \tilde{c}_{\nu}^{-1} \kappa_d + 2^{\nu+1} ) \exp\!\Big(2\tilde{c}_{\nu} \, \sup_{t\in \R}\,  \norm{\Phi^t}_{2d+1+\nu} \, |u-v|\Big) \, \norm{A}_\nu 
    \end{align*}
    and defining $c_{\nu} = \max((2^{d+\nu} \tilde{c}_{\nu}^{-1} \kappa_d + 2^{\nu+1} + 1), 2\tilde{c}_{\nu})$ gives us the desired bound. The bound for  the case $\sup_{u\in \R} \norm{\Phi^{u}}_{\exp,a} < \infty $ can be found in Lemma B.3 of \cite{HenheikTeufel2022}.

    It remains to prove $\alpha_{u,v} D_\infty = D_\infty$. With the bound we have already shown and the fact that $\alpha_{u,v}^{-1} = \alpha_{v,u}$ this reduces to showing that $\alpha_{u,v} \mA^N \subseteq \mA^N$. To this end, let $\varphi \in \R$ and $A\in \mA_0$. It holds that 
    \begin{align*}
        -\i \, \partial_u \, g_\varphi\alpha_{u,v}g_\varphi^{-1} A &\;=\; g_\varphi\alpha_{u,v} \mL_{\Phi^u} g_\varphi^{-1} A \;=\; g_\varphi\alpha_{u,v} \sum_{M\in P_0(\Z^{d})} [\Phi^u(M),g_\varphi^{-1} A]\\
        &\;=\; g_\varphi\alpha_{u,v} g_\varphi^{-1} \sum_{M\in P_0(\Z^{d}}) [\Phi^u(M), A] \;=\; g_\varphi\alpha_{u,v} g_\varphi^{-1} \mL_{\Phi^u}  A\, .
    \end{align*}
    By the uniqueness of the cocycle generated by $(\Phi^{u})_{u\in \R}$ it follows that $g_\varphi\alpha_{u,v} g_\varphi^{-1} = \alpha_{u,v}$ and therefore  $\alpha_{u,v} \mA^N \subseteq \mA^N$.
\end{proof}

\begin{proof}[Proof of \cref{lem: compatibility of translations}]
    We show that for all $M\subset \Z^d$ the map $T_\g^{-1} \, \E_{M+\g} \, T_\g$ is equal to $\E_M$. Let $M\subseteq \Z^d$, $A \in \mA$ and $B\in \mA_M$ and consider
    \begin{align*}
        \w^\tr\! \left( \left( T_\g^{-1} \,  \E_{M+\g} \, T_\g \, A\right) B\right)\, .
    \end{align*}
    The uniqueness of the tracial state implies that it must be invariant under all automorphisms, which we use to rewrite
    \begin{align*}
        \w^\tr\! \left( \left( T_\g^{-1} \,  \E_{M+\g} \, T_\g \, A\right) B\right) &= \w^\tr\! \left( \left( \E_{M+\g} \, T_\g \, A\right) \left(T_\g  \, B\right)\right)\,.
    \end{align*}
    Property $(\i)$ of the translation implies that $T_\g \, B \in \mA_{M+\g}$, which allows us to use the defining property of the conditional expectation from \cref{Ex+UniqueExpectation} to get
    \begin{align*}
        \w^\tr\! \left( \left( T_\g^{-1} \,  \E_{M+\g} \, T_\g \, A\right) B\right)  = \w^\tr\! \left( \left( T_\g \, A\right) \left(T_\g  \, B\right)\right) 
         = \w^\tr\! \left(  T_\g \left( A  \, B \right) \right) = \w^\tr\! \left(   A  \, B  \right)\, .
    \end{align*}
    By the uniqueness from \cref{Ex+UniqueExpectation}, we have shown $T_\g^{-1} \, \E_{M+\g} \, T_\g = \E_M$.\\

    For the second statement we show that the one-parameter group of automorphisms $\left( T_\g^{-1} \, g_\varphi \, T_\g \right)_{\varphi\in \R}$ is generated by the number operator $N$. To this end let $A \in \mA_M$ for some $M\in P_0(\Z^{d})$ and consider
    \begin{align*}
       -\i\, \partial_\varphi \left( T_\g^{-1} \, g_\varphi \, T_\g A \right)  =  T_\g^{-1}  g_\varphi \mL_N T_\g A  = \sum_{x\in M+\g} T_\g^{-1} g_\varphi T_\g \, [ T_\g^{-1} n_x,  A]\, . 
    \end{align*}
    With property ($\i \i$) of the translation, this gives us
    \begin{align*}
        -\i\,\partial_\varphi \left( T_\g^{-1} \, g_\varphi \, T_\g A \right)  =  \sum_{x\in M+\g} T_\g^{-1} g_\varphi T_\g\,[ n_{x-\g}, A] = T_\g^{-1} g_\varphi T_\g\,\mL_N A \, .
    \end{align*}
    Therefore, $\left( T_\g^{-1} \, g_\varphi \, T_\g \right)_{\varphi\in \R}$ is generated by $N$ and by the uniqueness from \cref{cocycles}, we have \\
    $T_\g^{-1} \, g_\varphi \, T_\g = g_\varphi$ for all $\varphi \in \R$.
\end{proof}

\begin{proof}[Proof of Proposition \ref{prop: interaction associated to observable}]
    Since translations and the conditional expectation preserve gauge-invariance and self-adjointness (see Proposition \ref{Ex+UniqueExpectation} and Lemma \ref{lem: compatibility of translations}), we see that $\Phi_A^T(M)$ is self-adjoint and gauge-invariant for all $M\in P_0(\Z^d)$. Next we check the $T$-periodicity: Let $k\in \N_0$, $\g,\mu\in \Z^d$. Due to the $T$-compatibility of $A$ and Lemma \ref{lem: compatibility of translations}, it holds that
    \begin{align*}
        T_\g \, T_\mu \, \E_{\La_k} \, A & = \E_{\La_k + \mu + \g}\, T_\g \, T_\mu\, A\\
        & = \E_{\La_k + \mu + \g}\, T_{\g + \mu} \, A\\
        & = T_{\g + \mu} \, \E_{\La_k }\, A\, ,
    \end{align*}
    which implies the $T$-periodicity of $\Phi_A^T$. We show that for each $\nu \in \N_0$ the interaction norm $\lVert \Phi_A^T \rVert_\nu$ is finite: Let $\nu \in \N_0$. By the definition of the norm and $\Phi_A$, we have that
    \begin{align*}
        \lVert \Phi_A^T \rVert_\nu &= \sup_{x\in \Z^d} \sum_{\substack{M\in P_0(\Z^d)\\ x\in M}} (1+\mathrm{diam}(M))^\nu \, \lVert \Phi_A^T(M) \rVert\\
        &= \sum_{k=1}^\infty \sum_{\g \in \La_k} (1+2\, k)^\nu \, \lVert T_{-\g}\,(\E_{\La_k}\, A - \E_{\La_{k-1}}\, A )\rVert  + \lVert \E_{\La_0}\, A \rVert \, .
    \end{align*}
    Using properties of the conditional expectation from Proposition \ref{Ex+UniqueExpectation} and the definition of the decay norm, we arrive at
    \begin{align*}
        \lVert \Phi_A^T \rVert_\nu &\leq \sum_{k=1}^\infty \frac{(1 + 2\, k )^d\,(1+2\, k)^\nu}{(1 + (k-1))^{d+\nu+2}} \, \lVert A \rVert_{\nu +d+2} + \lVert \E_{\La_0}\, A \rVert < \infty \, . 
    \end{align*}
    We show that $(\Phi_A^T)_0 = A$. Since $\Phi_A^T$ is supported only on the boxes $\La_k + \g$ for $k\in \N_0$ and $\g \in \Z^d$ and the representative at $0$ of such a box is $\La_k$, we get
    \begin{align*}
        (\Phi_A^T)_0 &= \sum_{M\in R_0(\Z^d)}\Phi_A^T(M) \\
        &= \sum_{k=1}^\infty (\E_{\La_k} \, A - \E_{\La_{k-1}} \, A) + \E_{\La_0}\, A\\
        &= \lim_{k\to \infty} \E_{\La_k} \, A = A \,.
    \end{align*}
    The limit in the last line converges to $A$ as shown in Lemma \ref{lem:norm properties}.
    
    Now let $\Phi$ be a $T$-periodic $B_\infty$-interaction. For $\nu \in \N_0$ it holds that
    \begin{align*}
        \sum_{M \in R_0(\Z^d)} \norm{\Phi(M)}_\nu &\leq \sum_{M \in R_0(\Z^d)} 3 \, (1 + \mathrm{diam}(M))^\nu \norm{\Phi(M)}\, ,
    \end{align*}
    since each $M$ is a standard representative and thus contained in a box around $0$ of side length $2\, \mathrm{diam}(M) $. By periodicity, we can shift each $M$ such that it contains $0$,     giving us the estimate
    \begin{equation}\label{eq:decay norm of represenative estimate}
        \sum_{M \in R_0(\Z^d)} \norm{\Phi(M)}_\nu  \leq \sum_{\substack{M \in P_0(\Z^d)\\ 0\in M}} 3\, \left(1+ \mathrm{diam}(M)\right)^\nu  \norm{\Phi(M)} \leq  3\norm{\Phi}_\nu < \infty \, .
    \end{equation}
    Therefore, $\Phi$ is absolutely summable at $0$ and $\norm{\Phi_0}_\nu < \infty$. Since all local terms of $\Phi$ are gauge-invariant and self-adjoint we get that $\Phi_0$ lies in $D_\infty$ and is self-adjoint.
    It remains to show the $T$-compatibility of $\Phi_0$. Let $\g, \mu \in \Z^d$ and consider
    \begin{equation*}
        T_{\g + \mu} \, \Phi_0 = \sum_{M\in R_0(\Z^{d})} T_{\g + \mu} \, \Phi(M)= \sum_{M\in R_0(\Z^{d})}  \Phi(M + \g + \mu)= \sum_{M\in R_0(\Z^{d})}  T_{\g} \, T_\mu \, \Phi(M)= T_{\g} \, T_\mu \, \Phi_0 \, . \qedhere
    \end{equation*}
\end{proof}

\begin{proof}[Proof of Proposition~\ref{liouvillian tools}]

    Let $A \in D_\infty$. \cref{lem: sum representation of generator} tells us that
    \begin{align*}
        \mL_\Phi A = \sum_{M \in P_0(\Z^d)} [\Phi(M),A]\,,
    \end{align*}
    where the sum converges absolutely. We can thus rearrange the sum and use the fact that $\Phi$ is absolutely summable at 0 from Proposition \ref{prop: interaction associated to observable}, to get
    \begin{align*}
        \mL_\Phi A = \sum_{\g \in \Z^d} \sum_{M \in R_0(\Z^d)}  [T_\g \Phi(M),A] &=  \sum_{\g \in \Z^d}   [T_\g \sum_{M \in R_0(\Z^d)} \Phi(M),A] = \sum_{\g \in \Z^d}   [T_\g \Phi_0,A] \, .
    \end{align*}

    For the proof of (ii), for each $M \in  P_0(\Z^d)$ and $k\in \N_0$ we define 
    \begin{align*}
        n(M,k) = |\{\g\in \Z^d~|~  M+\g\subseteq\Lambda_k\}| \, .
    \end{align*}
    Let $M$, $m\in \N$ be such that $M\subseteq \Lambda_m$. We find that for all $k\in \N$ with $k\geq m$
    \begin{align*}
        (2(k-m)+1)^d = n(\Lambda_m,k) \leq n(M,k) \leq |\Lambda_k| = (2k+1)^d~,
    \end{align*}
    thus
    \begin{align*}
        0 \leq \frac{n(M,k)}{|\Lambda_k|} \leq 1
    \qquad\mbox{and}\qquad 
        \lim_{k\to \infty} \frac{n(M,k)}{|\Lambda_k|} = 1 \, .
    \end{align*}
    With this we can calculate as follows
    \begin{align*}
        \frac{1}{|\Lambda_k|}\sum_{M\subseteq \Lambda_k}& \omega(\Phi(M)) = \frac{1}{|\Lambda_k|}\sum_{M\in R_0(\Z^d)} \sum_{\g\in \Z^d}  \chi_{(M+\g)\subseteq \La_k} ~ \omega(\Phi(M + \g))\\
        &= \frac{1}{|\Lambda_k|}\sum_{M\in R_0(\Z^d)} \sum_{\g\in \Z^d}  \chi_{(M+\g)\subseteq \La_k} ~ \omega(T_{\g}\Phi(M)) = \frac{1}{|\Lambda_k|}\sum_{M\in R_0(\Z^d)} \sum_{\g\in \Z^d}  \chi_{(M+\g)\subseteq \La_k} ~ \omega(\Phi(M))\\
        &= \sum_{M\in R_0(\Z^d)} \frac{n(M,k)}{|\Lambda_k|} ~  \omega(\Phi(M)) ~.
    \end{align*}
    Using dominated convergence and continuity of $\w$ we arrive at our desired result.
\end{proof}

\begin{proof}[Proof of Proposition~\ref{commutator of interactions}]
    To show that $\i[\Phi, \Psi]$ is $T$-periodic, let $\g \in \Z^d$ and $M\in P_0(\Z^d)$. We have
    \begin{align*}
        \i[\Phi, \Psi](M+\g)  = \sum_{\substack{M_1,M_2 \subset (M + \g)\\ M_1 \cup M_2= (M+\g)}} \i[\Phi(M_1),\Psi(M_2)] = \sum_{\substack{M_1,M_2 \subset M \\ M_1 \cup M_2 = M}} \i[\Phi(M_1+\g),\Psi(M_2+\g)]~.
    \end{align*}
    By the $T$-periodicity of $\Phi$ and $\Psi$ we get
    \begin{align*}
        \i[\Phi, \Psi](M+\g)  = \sum_{\substack{M_1,M_2 \subset M \\ M_1 \cup M_2 = M}} \i[T_\g \Phi(M_1),T_\g\Psi(M_2)] = T_\g \sum_{\substack{M_1,M_2 \subset M \\ M_1 \cup M_2 = M}} \i[\Phi(M_1),\Psi(M_2)] = T_\g \i[\Phi, \Psi](M)~,
    \end{align*}
    showing that $\i[\Phi, \Psi]$ is $T$-periodic. To show that $\i[\Phi, \Psi] \in B_\infty$  let $\nu\in \N_0$ and consider
    \begin{align*}
        \norm{\i[\Phi, \Psi]}_\nu &= \sup_{x\in \Z^d} \sum_{\substack{M\in P_0(\Z^d)\\ x \in M}} (1+\mathrm{diam}(M))^\nu \norm{\i[\Phi, \Psi](M)} \\
        &\leq \sup_{x\in \Z^d} \sum_{\substack{M\in P_0(\Z^d)\\ x \in M}} \sum_{\substack{M_1,M_2 \subset M \\ M_1 \cup M_2 = M}} (1+\mathrm{diam}(M))^\nu  \norm{[\Phi(M_1),\Psi(M_2)]} \\
        &\leq \sup_{x\in \Z^2} \sum_{\substack{M\in P_0(\Z^d)\\ x \in M}} \sum_{\substack{M_1,M_2 \subset M \\ M_1 \cup M_2 = M \\ M_1 \cap M_2 \neq \emptyset}} (1+\mathrm{diam}(M))^\nu  2\norm{\Phi(M_1)} \norm{\Psi(M_2)}
    \end{align*}
    From here one can change the summation index and then further estimate to get
    \begin{eqnarray*}
        \norm{\i[\Phi, \Psi]}_{\nu} &\leq   &\sup_{x\in \Z^d} \sum_{\substack{M_1,M_2 \in P_0(\Z^d)\\ x \in M_1 \cup M_2 \\ M_1 \cap M_2 \neq \emptyset}} (1+\mathrm{diam}(M_1 \cup M_2))^\nu  2\norm{\Phi(M_1)} \norm{\Psi(M_2)} \\
        &\leq & \sup_{x\in \Z^{d}} \sum_{\substack{M_1 \in P_0(\Z^{d})\\ x \in M_1  }} \sum_{\substack{M_2 \in P_0(\Z^d) \\M_1 \cap M_2 \neq \emptyset}} (1+\mathrm{diam}(M_1 \cup M_2))^\nu  2\norm{\Phi(M_1)} \norm{\Psi(M_2)}\\
        &&+\, \sup_{x\in \Z^d} \sum_{\substack{M_2 \in P_0(\Z^{d})\\ x \in M_2  }} \sum_{\substack{M_1 \in P_0(\Z^{d})\\ M_1 \cap M_2 \neq \emptyset}} (1+\mathrm{diam}(M_1 \cup M_2))^\nu  2\norm{\Phi(M_1)} \norm{\Psi(M_2)} \\
        &\eqcolon&  S_1 + S_2 ~.
    \end{eqnarray*}
    We proceed by treating each summand separately. For $S_1$ we find
    \begin{align*}
        S_1 &\leq \sup_{x\in \Z^d} \sum_{\substack{M_1 \in P_0(\Z^d)\\ x \in M_1  }} \sum_{y \in M_1 } \sum_{\substack{M_2 \in P_0(\Z^d) \\y\in M_2}} (1+\mathrm{diam}(M_1 \cup M_2))^\nu  2\norm{\Phi(M_1)} \norm{\Psi(M_2)}\\
        &\leq \sup_{x\in \Z^d} \sum_{\substack{M_1 \in P_0(\Z^d)\\ x \in M_1  }} \sum_{y \in M_1 } \sum_{\substack{M_2 \in P_0(\Z^d) \\y\in M_2}} (1+\mathrm{diam}(M_1))^\nu (1+\mathrm{diam}(M_2))^\nu  2\norm{\Phi(M_1)} \norm{\Psi(M_2)}\\
        &\leq 2\sup_{x\in \Z^d} \sum_{\substack{M_1 \in P_0(\Z^d)\\ x \in M_1  }} (1+\mathrm{diam}(M_1))^\nu \norm{\Phi(M_1)} \sum_{y \in M_1 } \sum_{\substack{M_2 \in P_0(\Z^d) \\y\in M_2}}  (1+\mathrm{diam}(M_2))^\nu  \norm{\Psi(M_2)}\\
        &\leq 2\sup_{x\in \Z^d} \sum_{\substack{M_1 \in P_0(\Z^d)\\ x \in M_1  }} (1+\mathrm{diam}(M_1))^\nu \norm{\Phi(M_1)}  |M_1| \sup_{y \in \Z^d } \sum_{\substack{M_2 \in P_0(\Z^d) \\y\in M_2}}  (1+\mathrm{diam}(M_2))^\nu  \norm{\Psi(M_2)} ~.
    \end{align*}
    We use that on the $d$-dimensional square lattice $|M_1| \leq (2 \, \mathrm{diam}(M_1) + 1)^d $ and periodicity of $\Phi$ and $\Psi$ to see
    \begin{align*}
        S_1 &\leq 2^{d+1} \sup_{x\in \Z^d} \sum_{\substack{M_1 \in P_0(\Z^d)\\ x \in M_1  }} (1+\mathrm{diam}(M_1))^{\nu+d}  \norm{\Phi(M_1)}   \sup_{y \in \Z^d } \sum_{\substack{M_2 \in P_0(\Z^d) \\y\in M_2}}  (1+\mathrm{diam}(M_2))^\nu  \norm{\Psi(M_2)} \\
        &= 2^{d+1} \sum_{\substack{M_1 \in P_0(\Z^d)\\ 0 \in M_1  }} (1+\mathrm{diam}(M_1))^{\nu+d}  \norm{\Phi(M_1)}   \sum_{\substack{M_2 \in P_0(\Z^d) \\0\in M_2}}  (1+\mathrm{diam}(M_2))^\nu  \norm{\Psi(M_2)} ~.
    \end{align*}
    From which we can see that
    \begin{align*}
        S_1 
        &\leq  2 ^{d+1}  \,\norm{\Phi}_{\nu +d} \,\norm{\Psi}_{\nu} \, < \, \infty\, .
    \end{align*}
    An analogous calculation shows that $S_2 < \infty$, hence $\i[\Phi,\Psi] \in B_\infty$.
    
    We prove the same statements for $\i[X_j,\Psi]$. Let $\g\in \Z^d$ and $M\in P_0(\Z^d)$. To prove $T$-periodicity we consider
    \begin{align*}
        \i[X_j, \Psi](M + \g) = \sum_{\substack{M_1,M_2 \subset M + \g\\ M_1 \cup M_2= M+\g}} \i[X_j(M_1),\Psi(M_2)]
    \end{align*}
    and see that since $X_j$ is only supported on one-element sets, the expression reduces to
    \begin{align*}
        \i[X_j, \Psi](M + \g) = \sum_{\substack{x \in M + \g}} \i[x_j n_x, \Psi(M + \g)] =\sum_{\substack{x \in M}} \i[(x+\g)_j n_{x+\g}, \Psi(M + \g)]~.
    \end{align*}
    From there we calculate
    \begin{align*}
        \i[X_j, \Psi](M + \g) &=\sum_{\substack{x \in M}} \i[x_j n_{x+\g}, \Psi(M + \g)] + \sum_{\substack{x \in M}} \i[\g_j n_{x+\g}, \Psi(M + \g)]\\
        &=T_\g \sum_{\substack{x \in M}} \i[x_j n_{x}, \Psi(M)] + T_\g \sum_{\substack{x \in M}} \i[\g_j n_{x}, \Psi(M)]\\
        &=T_\g \i[X_j, \Psi](M) + T_\g  \g_j  \i\mL_N(\Psi(M))
    \end{align*}
    and see, that since the terms of an interaction are always gauge-invariant, $\i[X_j, \Psi]$ is $T$-periodic. Now let $\nu \in \N_0$. We look at
    \begin{align*}
        \norm{\i[X_j, \Psi]}_\nu
        &= \sup_{x\in \Z^d} \sum_{\substack{M\in P_0(\Z^d)\\ x \in M}} (1+\mathrm{diam}(M))^\nu \norm{\mL_{X_j}\Psi(M)}{}\\
        &= \sum_{\substack{M\in P_0(\Z^d)\\ 0 \in M}} (1+\mathrm{diam}(M))^\nu \norm{\mL_{X_j}\Psi(M)}
    \end{align*}
    and use \cref{lem: sum representation of generator} (with $\nu=0$) to get the existence of a constant $c$, such that
    \begin{align*}
        \norm{\i[X_j, \Psi]}_\nu &\leq  \sum_{\substack{M\in P_0(\Z^d)\\  0 \in M}} (1+\mathrm{diam}(M))^\nu  c \norm{n_0} \norm{\Psi(M)}_{d+3}\\
        &\leq  \sum_{\substack{M\in P_0(\Z^d)\\  0 \in M}} 3\,(1+\mathrm{diam}(M))^{\nu+d+3} c \norm{\Psi(M)} ~,
    \end{align*}
    where we used the same estimate as in \ref{eq: decay norm when containing 0} in the last step. This expression is finite since $\Psi\in B_\infty$. We observe that the property of being a $T$-periodic $B_\infty$-interaction is preserved when taking real linear combinations, hence we have shown statement  (i).
        
    For statement  (ii) we proceed by using \cref{convergence} (with $\nu=0$ and $m= d+2$) on 
    \begin{align*}
        \sum_{M\in R_0(\Z^d)} \sum_{x\in \Z^d} \norm{[x_j n_x, \Psi(M)]} &\leq \sum_{M\in R_0(\Z^d)} \sum_{x\in \Z^d}  4^{d+5}\frac{|x_j| \norm{n_0}_{d+2}}{(1+\norm{x}_\infty)^{d+2}} \norm{ \Psi(M)}_{d+2}\\
        &\leq \sum_{M\in R_0(\Z^d)} \sum_{x\in \Z^d}  4^{d+5}\frac{|x_j| \norm{n_0}_{d+2}}{(1+\norm{x}_\infty)^{d+2}} \,3\,(1+\mathrm{diam}(M))^{d+2} \norm{ \Psi(M)}
    \end{align*}
    and see that the expression is finite due to $\Psi$ being a $B_\infty$-interaction. Hence, we can reorder the sum to see
    \begin{align*}
        \i[X_j,\Psi]_0  = \sum_{M\in R_0(\Z^d)} \sum_{x\in \Z^d} \i[x_j n_x, \Psi(M)] = \sum_{x\in \Z^d} \sum_{M\in R_0(\Z^d)}  \i[x_j n_x, \Psi(M)] = \i\mathcal{L}_{X_j}\Psi_0~.
    \end{align*}

    To prove statement (iii) we use \cref{convergence} twice to show absolute convergence of the sum
    \begin{align*}
        \sum_{M\in P_0(\Z^d)} \sum_{x\in \Z^{d}} &\norm{[[x_j n_x,\Psi(M)],A]} = \sum_{M\in R_0(\Z^d)} \sum_{\g\in \Z^d} \sum_{x\in \Z^d} \norm{[[x_j n_x,T_\g \Psi(M)],A]}\\
        &= \sum_{M\in R_0(\Z^d)} \sum_{\g\in \Z^d} \sum_{x\in \Z^d} \norm{[T_\g [x_j n_{x-\g}, \Psi(M)],A]}\\
        &\leq \sum_{M\in R_0(\Z^d)} \sum_{\g\in \Z^d} \sum_{x\in \Z^d} 4^{2d+6}\frac{\norm{A}_{2d+3}}{(1+\norm{\g}_\infty)^{2d+3}}\norm{[x_j n_{x-\g}, \Psi(M)]}_{2d+3}\\
        &\leq \sum_{M\in R_0(\Z^d)} \sum_{\g\in \Z^d} \sum_{x\in \Z^d} 4^{5d+14}\frac{\norm{A}_{2d+3}}{(1+\norm{\g}_\infty)^{2d+3}} \frac{|x_j| \norm{n_0}_{3d+5}}{(1+\norm{x-\g}_\infty)^{d+2}} \norm{\Psi(M)}_{3d+5} \\
        &\leq \sum_{M\in R_0(\Z^d)} \sum_{\g\in \Z^d} \sum_{x\in \Z^d} 4^{5d+14}\frac{\norm{A}_{2d+3}}{(1+\norm{\g}_\infty)^{d+1}} \frac{|x_j| \norm{n_0}_{3d+5}}{(1+\norm{x}_\infty)^{d+2}} \norm{\Psi(M)}_{3d+5} \;<\;\infty ~.
    \end{align*}
    Therefore, we can again reorder the sum and get
    \begin{align*}
        \mL_{\i[\Psi,X_j]}A  = \sum_{M\in P_0(\Z^d)} \sum_{x\in \Z^d} \Big[\i[\Psi(M),x_j n_x],A\Big] = \sum_{x\in \Z^d} \sum_{M\in P_0(\Z^d)}  \Big[\i[\Psi(M),x_j n_x],A\Big] = \sum_{x \in \Z^d} [\i\mL_\Psi x_j n_x ,A]~.
    \end{align*}
    Similarly we look at
    \begin{align*}
        \mL_{\i[\Psi,\Phi]} A = \sum_{\g \in \Z^d} \sum_{\substack{M\in R_0(\Z^d)}} \sum_{\substack{M_1,M_2 \subset M \\ M_1 \cup M_2 = M}}  [\,T_\g\, \i\, [\, \Psi(M_1),\, \Phi(M_2)\, ], \, A ]\, ,
    \end{align*}  
    which we will now show absolute convergence of. We first use Lemma \ref{convergence} to get
    \begin{align*}
        \hspace{2em}& \hspace{-2em}\sum_{\g \in \Z^d} \sum_{\substack{M\in R_0(\Z^d)}} \sum_{\substack{M_1,M_2 \subset M \\ M_1 \cup M_2 = M}} \lVert[\,T_\g\, \i\,[\, \Psi(M_1),\, \Phi(M_2)\, ], \, A ] \rVert \\
        &\leq \sum_{\g \in \Z^d} \sum_{\substack{M\in R_0(\Z^d)}} \sum_{\substack{M_1,M_2 \subset M \\ M_1 \cup M_2 = M}} 4^{d+4}\, \frac{\lVert [\, \Psi(M_1),\, \Phi(M_2)\, ] \rVert_{d+1} \, \lVert A \rVert_{d+1} }{(1+\lVert \g \rVert_\infty)^{d+1}} 
    \end{align*}
    Observe that, by the same reasoning as in  \ref{eq:decay norm of represenative estimate} we have
    \begin{align*}
        \sum_{\substack{M\in R_0(\Z^d)}} \sum_{\substack{M_1,M_2 \subset M \\ M_1 \cup M_2 = M}}  \norm{[\Phi(M_1),\Psi(M_2)]}_{d+1} 
        \leq \sum_{\substack{M\in P_0(\Z^d)\\ 0 \in M}} \sum_{\substack{M_1,M_2 \subset M \\ M_1 \cup M_2 = M}}  3\,(1+ \mathrm{diam}(M)^\nu) \, \norm{[\Phi(M_1),\Psi(M_2)]} 
    \end{align*}
    and that therefore we have already shown when we proved statement (i), that the sum converges absolutely. Hence, we can reorder the sum arbitrarily. We exploit this by performing a change in summation index via the bijection
    \begin{align*}
        &\{(M_1,M_2) | M_1\in P_0(\Z^d), \,  M_2\in R_0(\Z^d) \} \to \{(M,M_1,M_2)| M\in R_0(\Z^d),\, M_1,M_2 \subset M,\,  M_1 \cup M_2 = M\}\\
        &(M_1,M_2) \mapsto (M_1 \cup M_2 + s(M_1 \cup M_2),M_1 + s(M_1 \cup M_2), M_2 + s(M_1 \cup M_2)) ~,
    \end{align*}
    where $s(M_1 \cup M_2)$ is the shift vector defined in \cref{def: shape}
    and moving the left sum to the right. This gives us
    \begin{align*}
        \mL_{\i[\Psi,\Phi]} A 
        & =  \sum_{\substack{M\in R_0(\Z^d)}} \sum_{\substack{M_1,M_2 \subset M \\ M_1 \cup M_2 = M}} \sum_{\g \in \Z^d} [\,T_\g\, \i\,[\, \Psi(M_1),\, \Phi(M_2)\, ], \, A ]\\
        & =  \sum_{\substack{M_1\in P_0(\Z^d)}} \sum_{\substack{M_2 \in R_0(\Z^d)}} \sum_{\g \in \Z^d} [\, T_\g\,\i[\Psi(M_1 + s(M_1 \cup M_2)),\Phi(M_2 +s(M_1 \cup M_2))], \, A\,]\, .
    \end{align*}
    Now we can exploit the periodicity of $\Phi$ and $\Psi$ together with an index shift to get 
    \begin{align*}
        \mL_{\i[\Psi,\Phi]} A 
        &=  \sum_{\substack{M_1\in P_0(\Z^d)}} \sum_{\substack{M_2 \in R_0(\Z^d)}} \sum_{\g \in \Z^d} [\, T_{\g+s(M_1 \cup M_2)}\,\i[\Psi(M_1),\Phi(M_2)], \, A\,]\\
        &=  \sum_{\substack{M_1\in P_0(\Z^d)}} \sum_{\substack{M_2 \in R_0(\Z^d)}} \sum_{\g \in \Z^d} [\, T_{\g}\,\i[\Psi(M_1),\Phi(M_2)], \, A\,]\, .
    \end{align*}
    Rearranging the sum again, we finally find
    \begin{align*}
        \mL_{\i[\Psi,\Phi]} A \
        & = \sum_{\g \in \Z^d} \sum_{\substack{M_1\in P_0(\Z^d)}} \sum_{\substack{M_2 \in R_0(\Z^d)}}  [\, T_{\g}\,\i[\Psi(M_1),\Phi(M_2)], \, A\,]\\
        & = \sum_{\g \in \Z^d} [\, T_\g \,  \i\, \mL_{\Psi} \, \Phi_0, \, A\, ]\,.
    \end{align*}

    To prove statement (iv), we employ the same bijection to change the summation index:  
    \begin{align*}
        \omega(\i[\Phi,\Psi]_0) &= \omega(\sum_{\substack{M\in R_0(\Z^d)}} \sum_{\substack{M_1,M_2 \subset M \\ M_1 \cup M_2 = M}}  \i[\Phi(M_1),\Psi(M_2)]\,)\\
        &= \omega(\sum_{\substack{M_1\in P_0(\Z^d)}} \sum_{\substack{M_2 \in R_0(\Z^d)}}  \i[\Phi(M_1 + s(M_1 \cup M_2)),\Psi(M_2 +s(M_1 \cup M_2))]\,)~.
    \end{align*}
    At this point we use the $T$-periodicity of $\w$, $\Phi$ and $\Psi$ to see
    \begin{align*}
        \omega(\i[\Phi,\Psi]_0) &= \omega(\sum_{\substack{M_1\in P_0(\Z^d)}} \sum_{\substack{M_2 \in R_0(\Z^d)}} T_{s(M_1 \cup M_2)} \i[\Phi(M_1),\Psi(M_2 )]\,)\\
        &= \omega(\sum_{\substack{M_1\in P_0(\Z^d)}} \sum_{\substack{M_2 \in R_0(\Z^d)} }\i[\Phi(M_1),\Psi(M_2 )]\,)\\
        &= \omega(\sum_{\substack{M_1\in P_0(\Z^d)}}   \i\Big[\Phi(M_1),\sum_{\substack{M_2 \in R_0(\Z^d)}}\Psi(M_2)\,\Big]\,)\\
        &= \omega(\i\mathcal{L}_\Phi(\Psi_0)) ~.
    \end{align*}
    By instead using the bijection
    \begin{align*}
        &\{(M_1,M_2) | M_1\in R_0(\Z^d), \,  M_2\in P_0(\Z^d) \} \to  \{(M,M_1,M_2)| M\in R_0(\Z^d),\, M_1,M_2 \subset M,\,  M_1 \cup M_2 = M\} \\
        &(M_1,M_2) \mapsto (M_1 \cup M_2 + s(M_1 \cup M_2),M_1 + s(M_1 \cup M_2), M_2 + s(M_1 \cup M_2))  ~,
    \end{align*}
    one gets 
    \[
        \omega(\i[\Phi,\Psi]_0) = -\omega(\i\mathcal{L}_\Psi(\Phi_0)) ~. \qedhere
    \]
\end{proof}

\section{Proofs related to the Off-diagonal map}\label{app:OD}

\begin{proof}[Proof of Lemma \ref{lem: almost local obs. for OD and inv. liou.}]
    It can be easily seen that $(\Psi\OD[\alpha])_*$ and $\mathcal{I}(\Psi)_*$ are gauge-invariant and self-adjoint, since $h_0$ lies in $D_\infty$ and is self-adjoint (Proposition \ref{prop: interaction associated to observable}) and the maps $\alpha$, $\alpha^{-1}$, $\e^{\i s\mL_{H}}$ and $\i \, \mL_{\Psi}$ all map $D_\infty$ to itself (\cref{lem: sum representation of generator} and \cref{cocycles}) and preserve self-adjointness. Let $\nu\in \N_0$ and consider $\norm{\alpha \mL_{\Psi} \alpha^{-1} h_0}_\nu$, which is finite since $\alpha$, $\alpha^{-1}$ and $\i\, \mL_{\Psi}$ all map $D_\infty$ to itself. With \cref{cocycles} we can see that 
    \begin{align*}
        \left\|\int_{\R}\mathrm{d}s \,W_{g}(s)\, \e^{\i s\mL_{H}} \alpha \i \mL_\Psi \alpha^{-1} h_0\right\|_\nu & \leq \int_{\R}\mathrm{d}s \norm{ \,W_g(s)\, \e^{\i s\mL_{H}} \alpha \i\mL_\Psi \alpha^{-1} h_0}_\nu  \\&\leq \int_{\R}\mathrm{d}s \,|W_g(s)| \, b_{\nu}(|s|) \norm{ \alpha \i\mL_\Psi \alpha^{-1} h_0}_\nu  < \infty
    \end{align*}
    and since $\alpha^{-1}$ maps $D_\infty$ to itself we arrive at $(\Psi\OD[\alpha])_* \in D_\infty$. Analogously one finds that $\mathcal{I}(\Psi)_* \in  D_{\infty}$. To show that $(\Psi\OD[\alpha])_*$ is $T$-compatible i.e., that for $\g , \mu \in \Z^d$ we have $T_{\g + \mu} \, (\Psi\OD[\alpha])_* =  T_\g \, T_\mu \,  (\Psi\OD[\alpha])_*$, we observe that,
    since $\e^{\i s \mL_H}$, $\alpha$ and $\alpha^{-1}$ commute with the translation, we can pull $T_{\g + \mu}$ all the way through the expression to get
    \begin{align*}
        T_{\g + \mu} \, (\Psi\OD[\alpha])_* & = \alpha^{-1} \int_{\R}\mathrm{d}s \, W_g(s) \, \e^{\i s\mL_{H}} \,  \alpha \,  \i \, T_{\g + \mu} \, \mL_\Psi \, \alpha^{-1} \, h_0\, .
    \end{align*}   
    The position operator is periodic up to a shift proportional to the number operator, so for the last part of the expression we have 
    \begin{align*}
        T_{\g + \mu} \mL_{\Psi} \alpha^{-1} h_0  &= (\mL_{p\Phi}  + \mL_{qX_j} - q \,(\g + \mu)_j \, \mL_{N}) T_{\g + \mu}  \alpha^{-1} h_0 \\
        &= (\mL_{p\Phi}  + \mL_{qX_j} - q \,(\g + \mu)_j \, \mL_{N}) \alpha^{-1} T_{\g + \mu} h_0 \, .
    \end{align*}
    The periodicity of $H$ implies that $h_0$ is $T$-compatible, as shown in Proposition \ref{prop: interaction associated to observable}. Moving $T_{\g} \,  T_\mu$ all the way to the left of the expression with the same steps in reversed order produces
    \begin{align*}
        T_{\g + \mu} \, (\Psi\OD[\alpha])_* = T_{\g} \, T_\mu \, (\Psi\OD[\alpha])_* \, .
    \end{align*}
    One can proceed in the same way to show $T_{\g + \mu} \, \mathcal{I}(\Psi)_* =  T_\g \, T_\mu \,  \mathcal{I}(\Psi)_*$.
\end{proof}

\begin{proof}[Proof of \cref{OD-property}]

    To prove the first claim, let $A\in D_\infty$ and consider $\mL_{\Psi\OD[\alpha]} A$. Proposition \ref{liouvillian tools} tells us that
    \begin{align*}
        \w_\alpha(\mL_{\Psi\OD[\alpha]} A) &= \sum_{\g\in \Z^d} \w_\alpha ([T_\g (\Psi\OD[\alpha])_0 , A])= \sum_{\g\in \Z^d} \w_\alpha\left(\Big[\i \alpha^{-1} \int_{\R}\mathrm{d}s \,W_g(s) \,e^{\i s\mL_{H}} \alpha \mL_\Psi \alpha^{-1} T_\g h_0 , \,A\,\Big]\right)\\
        &= \sum_{\g\in \Z^d} \int_{\R}\mathrm{d}s \sum_{\mu \in \Z^d} \w_0\left(\Big[\i   W_g(s) \,e^{\i s\mL_{H}} [ \alpha \, (p \, T_\mu \, \Phi_0 + q \, \mu_j \, n_\mu), T_\g h_0] ,\, \alpha\, A\, \Big]\right) \, .
    \end{align*}
    Similar to previous arguments we can show that this expression converges absolutely by using \cref{convergence} and \cref{cocycles}. Because of this, we can change the order of summation/integration to get
    \begin{align*}
\w_\alpha(\mL_{\Psi\OD[\alpha]} A) &= \sum_{\mu \in \Z^d} \int_{\R}\mathrm{d}s \sum_{\g\in \Z^d} \w_0\left(\Big[\i   W_g(s)\, \e^{\i s\mL_{H}} [  \alpha \, (p \, T_\mu \, \Phi_0 + q \, \mu_j \, n_\mu), T_\g h_0] , \,\alpha \, A\,\Big]\right)\\
        &=\sum_{\mu\in \Z^d} \int_{\R}\mathrm{d}s \, \w_0\left(\Big[- \i  W_g(s) \,\e^{\i s\mL_{H}} \mL_H\,\alpha \, (p \, T_\mu \, \Phi_0 + q \, \mu_j \, n_\mu)  , \, \alpha \, A\,\Big]\right)\\
        &=\sum_{\mu\in \Z^d} \w_0([-\i\, \mL_H \, \mathcal{I}_H \, \alpha \, (p \, T_\mu \, \Phi_0 + q \, \mu_j \, n_\mu)  , A]) \, ,
    \end{align*}
    where we define $\mathcal{I}_H$ as in \cite{HenheikTeufel2022} as
    \begin{align*}
        \mathcal{I}_H \, A  \coloneq \int_\R \mathrm{d}s\, W_g(s) \, \e^{\i s \mL_H}\, A
    \end{align*}
    for $ A \in D_\infty$. Now it follows from Proposition 3.3 in \cite*{HenheikTeufel2022} that 
    \begin{align*}
        \w_0([-\i\, \mL_H \, \mathcal{I}_H \, \alpha \, (p \, T_\mu \, \Phi_0 + q \, \mu_j \, n_\mu)  , \alpha \, A]) = \w_0([\alpha (p \, T_\mu \, \Phi_0 + q \, \mu_j \, n_\mu)  ,\, \alpha \, A])\, .
    \end{align*}
    Therefore by Proposition \ref{liouvillian tools} it holds that 
    \begin{align*}
        \w_\alpha(\mL_{\Psi\OD[\alpha]} A) = \sum_{\mu\in \Z^d} \w_0([\alpha\, (p \, T_\mu \, \Phi_0 + q \, \mu_j \, n_\mu)  ,  \, \alpha\, A])  = \w_\alpha(\mL_\Psi\, A)\, .
    \end{align*}
    For the proof of  the second claim, let $A\in D_\infty$ and consider 
    \begin{align*}
        \mL_{-\i[H,\mathcal{I}(\Psi)]}\, A = - \sum_{\g \in \Z^d} [\, T_\g \, \i\,\mL_H \, \mathcal{I}(\Psi)_0\, ,\, A \, ]\, .
    \end{align*} 
    This identity holds by Proposition \ref{commutator of interactions}. By Propositions \ref{liouvillian tools}  and \ref{prop: interaction associated to observable}  it holds that
    \begin{align*}
        \i \, \mL_H \, \mathcal{I}(\Psi)_0 = - \sum_{\g \in \Z^d} \int_\R \mathrm{d}s \int_0^s \mathrm{d}u \, W_g(s) \, \i \, [ \, T_\g\, h_0, \, \e^{\i u \mL_H} \, \i \, \mL_\Psi \, h_0 \, ]\, ,
    \end{align*}
    which is absolutely summable/integrable by the bounds from Lemmas \ref{convergence} and \ref{cocycles}. We can therefore change the order of summation and integration to get
    \begin{align*}
        \i \, \mL_H \, \mathcal{I}(\Psi)_0 & = - \int_\R \mathrm{d}s \int_0^s \mathrm{d}u \, W_g(s) \,  \e^{\i u \mL_H} \, \i\,\mL_H\, \i\,\mL_\Psi \, h_0\\
        & = -\int_\R \mathrm{d}s  \, W_g(s) \,  (\e^{\i s \mL_H} -1 ) \, \i\,\mL_\Psi \, h_0 \\
        & = \int_\R \mathrm{d}s  \, W_g(s) \,  \e^{\i s \mL_H}  \, \i\,\mL_\Psi \, h_0 \\
        & = (\Psi\OD)_0\, ,
    \end{align*}
    where we used the fact that $W_g$ is an odd function in the third equality.
\end{proof}

\begin{proof}[Proof of \cref{boundedness of OD}]
    Let $\nu\in\N_0$. We estimate
    \begin{align*}
        \sup_{\eps \in [-1,1]} \norm{(X_j\ODeps)_0}_\nu &\leq \sup_{\eps \in [-1,1]} \int_{\R}\mathrm{d}s \, |W_g(s)|\, \norm{\beta_\eps^{-1} \, \e^{\i s\mL_{H}} \beta_\eps \mL_{X_j} \beta_\eps^{-1} h_0}_\nu
    \end{align*}
    and use the bounds from \cref{cocycles} repeatedly for each of the automorphisms together with the bound from \cref{lem: sum representation of generator} to get constants $c_1$, $c_2$, $c_3$ and an at most polynomially growing function $b_\nu$, such that
    \begin{align*}
        \sup_{\eps \in [-1,1]} &\norm{(X_j\ODeps)_0}_\nu\\
         &\leq \sup_{\eps \in [-1,1]} \int_{\R}\mathrm{d}s \, \abs{W_g(s)} \, c_1^2 \, \exp\!\Big(2 \, c_1 \, \norm{S_\eps}_{2d+1+\nu}\Big) \, b_\nu\br{\abs{s}} \, c_2\,\norm{n_0} \, c_3 \exp \!\Big( c_3 \, \norm{ S_\eps}_{3d+4+2\nu} \Big) \norm{h_0}_{d+3+2\nu}  \,.
    \end{align*}
    Due to  definition of a gapped periodic system $\norm{S_\eps}_{2d+1+2\nu}$ and $\norm{S_\eps}_{3d+4+2\nu}$ are bounded in $\eps$ and with the fact that $|W_g(s)|$ decays faster than any polynomial power of $|s|$ as $|s| \to \infty$, this expression is finite.
\end{proof}

\section{Proof of Lemma \ref{lem:CS} }\label{app:CSlemma}

    If $u$ is equal to $v$ the identity holds. The proof strategy is to show that the left-hand side is differentiable with respect to $u$ and that the derivative is zero everywhere.  
    
    By Propositions \ref{liouvillian tools}, \ref{commutator of interactions}, and \cref{OD-property} we can rewrite the left-hand side as
    \begin{align*}
        \overline{\w_{\alpha_{u,v}}}(\i[ X_1\OD[\alpha_{u,v}],X_j\OD[\alpha_{u,v}]]) &= \i\w_{\alpha_{u,v}}(\mL_{X_1\OD[\alpha_{u,v}]}(X_j\OD[\alpha_{u,v}])_0)= \i\w_{\alpha_{u,v}}(\mL_{X_1}(X_j\OD[\alpha_{u,v}])_0)\\
        &= \i\w_0(\alpha_{u,v} \mL_{X_1}  (X_j\OD[\alpha_{u,v}])_0) ~.
    \end{align*}
    Inserting the definitions of $\mL_{X_1}$ and $(X_j\OD[\alpha_{u,v}])_0$ we get
    \begin{align*}
        \overline{\w_{\alpha_{u,v}}}(\i[ X_1\OD[\alpha_{u,v}],X_j\OD[\alpha_{u,v}]]) &= \i\w_0(\alpha_{u,v} \sum_{x\in \Z^d} \Big[x_1 n_x , \i \alpha_{u,v}^{-1} \int_{\R}\mathrm{d}s \,W_g(s)\, \e^{\i s\mL_{H}} \alpha_{u,v} \mL_{X_j} \alpha_{u,v}^{-1} h_0\Big])\\
        &=  -\sum_{x\in \Z^d} \int_{\R}\mathrm{d}s\, \w_0( \Big[\alpha_{u,v} x_1 n_x ,  W_g(s)\, \e^{\i s\mL_{H}} \alpha_{u,v} \mL_{X_j} \alpha_{u,v}^{-1} h_0\Big]) \\
        &=  -\sum_{x\in \Z^d} \int_{\R}\mathrm{d}s\, \w_0( \Big[\alpha_{u,v} x_1 n_x ,  W_g(s)\, \e^{\i s\mL_{H}} \alpha_{u,v} \sum_{y \in \Z^d} [y_j n_y, \alpha_{u,v}^{-1} h_0]\,\Big]) \\
        &=  -\sum_{x\in \Z^d} \int_{\R}\mathrm{d}s\, \sum_{y \in \Z^d} \w_0( \Big[\alpha_{u,v} x_1 n_x ,  W_g(s)\, \e^{\i s\mL_{H}}   [\alpha_{u,v} y_j n_y,  h_0]\,\Big]) \, ,
    \end{align*}
    where we used that the sum and the integral converge in norm and that therefore the  continuous linear maps $\alpha_{u,v}:\mA\to \mA$ and $\w_0:\mA\to \C$ can be exchanged with the limits involved in the definition of the sum and the integral.
    Next we want to differentiate the expression with respect to $u$. To prove that we can commute the derivative past the sum and integral, we show that the derivative is uniformly summable/integrable on bounded intervals. Meaning that for bounded intervals $I \subseteq \R$, it holds that
    \begin{align*}
        \sum_{x\in \Z^d} \int_{\R}\mathrm{d}s \sum_{y \in \Z^d} \sup_{u \in I} \norm{\partial_u \, \w_0( \big[\alpha_{u,v} x_1 n_x ,  W_g(s)\, \e^{\i s\mL_{H}}   [\alpha_{u,v} y_j n_y,  h_0]\,\big])} < \infty ~.
    \end{align*}    
    Using the chain rule and that $(\alpha_{u,v})_{(u,v) \in \R^2}$ is generated by $(\Phi^{u})_{u\in \R}$ we see that
    \begin{align*}
        \partial_u \, \w_0( & \left[\alpha_{u,v} x_1 n_x ,  W_g(s)\, \e^{\i s\mL_{H}}   [\alpha_{u,v} y_j n_y,  h_0]\,\right]) \\
         \quad \quad &=\w_0( \left[\alpha_{u,v} \i \mL_{\Phi^u} x_1 n_x ,  W_g(s)\, \e^{\i s\mL_{H}}   [\alpha_{u,v} y_j n_y,  h_0]\,\right]) + \w_0( \left[\alpha_{u,v} x_1 n_x ,  W_g(s)\, \e^{\i s\mL_{H}}   [\alpha_{u,v} \i \mL_{\Phi^u} y_j n_y,  h_0]\,\right]) \\
         &\eqcolon S_1(u,x,s,y) + S_2(u,x,s,y) ~.
    \end{align*}
    We can treat the two summands separately. The fact that states have unit norm and \cref{convergence} (with $\nu=0$ and $m=d+2$) leads us to estimate
    \begin{align*}
        \sup_{u \in I} \abs{S_1(u,x,s,y)}  =& \;\sup_{u \in I} \abs{\w_0( \left[\alpha_{u,v} \mL_{\Phi^u} x_1 n_x ,  W_g(s)\, \e^{\i s\mL_{H}}   [\alpha_{u,v} y_j n_y,  h_0]\,\right])} \\
        \leq&\; \sup_{u \in I} \norm{ \left[\alpha_{u,v} \mL_{\Phi^u} x_1 n_x ,  W_g(s)\, \e^{\i s\mL_{H}}   [\alpha_{u,v} y_j n_y,  h_0]\,\right]}\\
        \leq& \;\sup_{u \in I} 4^{d+5}  \frac{\norm{ \alpha_{u,v} \mL_{\Phi^u} x_1 n_0}_{d+2}}{(1+ \norm{x}_{\infty})^{d+2}} |W_g(s)| \, \norm{  \e^{\i s\mL_{H}}   \left[\alpha_{u,v} y_j n_y,  h_0\right]}_{d+2} ~.
    \end{align*}
    Applying \cref{cocycles} to both $\alpha_{u,v}$ and $\e^{\i s\mL_{H}}$ tells us that there is a constant $c$ and an at most polynomially growing function $b$, such that
    \begin{align*}
        \sup_{u \in I} \abs{S_1(u,x,s,y)} \leq& \;\sup_{u \in I}  c \, \exp(c \,  |u-v|) \, \frac{\norm{ \mL_{\Phi^u} x_1 n_0}_{d+2}}{(1+ \norm{x}_{\infty})^{d+2}} |W_g(s)| \, b(\abs{s}) \norm{     [\alpha_{u,v} y_j n_y,  h_0]}_{d+2} ~,
    \end{align*}
    from where we use \cref{convergence} (with $\nu=d+2$ and $m=d+2$) again to arrive at
    \begin{align*}
        \sup_{u \in I} \abs{S_1(u,x,s,y)} \leq& \;\sup_{u \in I}  c \, \exp(c \,  |u-v|) \, \frac{\norm{ \mL_{\Phi^u} x_1 n_0}_{d+2}}{(1+ \norm{x}_{\infty})^{d+2}} |W_g(s)| \, b(\abs{s})\, 4^{ 2d+7}\frac{\norm{\alpha_{u,v} y_j n_0}_{2d+4}}{(1+\norm{y}_\infty)^{d+2}} \norm{h_0}_{2d+4}\, .
    \end{align*}
    Another use of \cref{cocycles} gives us the existence of a constant $\tilde{c}$ such that
    \begin{align*}
        \sup_{u \in I} \abs{S_1(u,x,s,y)} &\leq  \sup_{u \in I}  \tilde{c} \, \exp(\tilde{c} \,  |u-v|) \, \frac{\norm{ \mL_{\Phi^u} x_1 n_0}_{d+2}}{(1+ \norm{x}_{\infty})^{d+2}} |W_g(s)| \, b(\abs{s})\frac{\norm{ y_j n_0}_{2d+4}}{(1+\norm{y}_\infty)^{d+2}} \norm{h_0}_{2d+4}\,.
    \end{align*}
    Finally, we exploit the assumption that $\sup_{u\in \R} \norm{\Phi^u}_{3d+9} <\infty $ with the bound from \cref{lem: sum representation of generator} to get that $ \norm{ \mL_{\Phi^u} n_0}_{d+4}$ is bounded in $u$. Since the exponential in the expression is bounded on $I$ we get the existence of a constant $C$, such that
    \begin{align*}
        \sup_{u \in I} \abs{S_1(u,x,s,y)} \leq  C \frac{|x_1|}{(1+ \norm{x}_{\infty})^{d+2}} |W_g(s)| \, b(\abs{s})  \frac{\norm{ y_j n_0}_{2d+4}}{(1+\norm{y}_\infty)^{d+2}} \norm{h_0}_{2d+4} ~,
    \end{align*}
    which is summable/integrable in $x$, $s$ and $y$. An analogous calculation shows that this also holds for $S_2(u,x,s,y)$.
    Thus, the map \[\R\ni u \mapsto   \i\w_0(\alpha_{u,v} \mL_{X_1}  (X_j\OD[\alpha_{u,v}])_0)\in\R\]
    is differentiable and 
    \begin{align*}
        \partial_u \i \w_0(\alpha_{u,v} \mL_{X_1}  (X_j\OD[\alpha_{u,v}])_0) = - \sum_{x\in \Z^d} \int_{\R}\mathrm{d}s \sum_{y \in \Z^d} \left(S_1(u,x,s,y) + S_2(u,x,s,y)\right)~.
    \end{align*}
    We find that by definition of the off-diagonal map (see Lemma \ref{lem: almost local obs. for OD and inv. liou.})  and Proposition \ref{prop: interaction associated to observable}
    \begin{align*}
        - \sum_{x\in \Z^d} \int_{\R}\mathrm{d}s \sum_{y \in \Z^d} S_1(u,x,s,y) &= -\w_0( \alpha_{u,v} \sum_{x\in \Z^d} \Big[ \i \mL_{\Phi^u} x_1 n_x ,\alpha_{u,v}^{-1} \int_{\R}\mathrm{d}s W_g(s)\, \e^{\i s\mL_{H}}  \alpha_{u,v} \sum_{y \in \Z^d} [ y_j n_y, \alpha_{u,v}^{-1} h_0]\Big])\\
        &=\i \w_0( \alpha_{u,v} \sum_{x\in \Z^d}\left[ \i\mL_{\Phi^u} x_1 n_x ,(X_j\OD[\alpha_{u,v}])_0\right])~,
    \end{align*}
    which by \cref{commutator of interactions} and Lemma \ref{OD-property} is equal to
    \begin{align*}
        \i \, \w_{\alpha_{u,v}}( \mL_{\i[\Phi^u,X_1]} \, (X_j\OD[\alpha_{u,v}])_0) 
        &= -\i \, \w_{\alpha_{u,v}}( \mL_{X_j} \, (\i\, [\, \Phi^u,\, X_1\, ] )_0) \\
        &= -\w_{\alpha_{u,v}}( \mL_{X_j} \, \mL_{X_{ 1}}\,\Phi^u_0) 
    \end{align*}
    We also have
    \begin{align*}
        - \sum_{x\in \Z^d} \int_{\R} \mathrm{d}s \sum_{y \in \Z^d} S_2 (u,x,s,y) 
         = -  \w_0( \alpha_{u,v} \sum_{x\in \Z^d} \Big[ x_1 n_x ,\alpha_{u,v}^{-1} \int_{\R}\mathrm{d}s  W_g(s)\, \e^{\i s\mL_{H}} \alpha_{u,v} \sum_{y \in \Z^d} [ \i \mL_{\Phi^u} y_j n_y,  \alpha_{u,v} ^{-1} h_0]\,\Big])~.
    \end{align*}
    Using  Propositions \ref{commutator of interactions} and \ref{liouvillian tools} we can see that this is equal to
    \begin{align*}
        \i \, \w_0( \alpha_{u,v} \sum_{x\in \Z^d} \Big[ x_1 n_x ,\alpha_{u,v}^{-1} \int_{\R}\mathrm{d}s  W_g(s)\, \e^{\i s\mL_{H}} \alpha_{u,v} \i\mL_{\i[\Phi^u,X_j]}\,\alpha_{u,v} ^{-1} h_0\,\Big])
        &= \i\, \w_{\alpha_{u,v}}(\mL_{X_1} \, (\i\,[\, \Phi^u,\,X_j\,]\OD[\alpha_{u,v}])_0)\, .
    \end{align*}
    Further using Proposition \ref{commutator of interactions} and Lemma \ref{OD-property} lets us rewrite the expression as follows:
    \begin{align*}
        \i\, \w_{\alpha_{u,v}}(\mL_{X_1} \, (\i\,[\, \Phi^u,\,X_j\,]\OD[\alpha_{u,v}])_0) 
        &= -\i\, \w_{\alpha_{u,v}}(\mL_{\i[ \Phi^u,X_j]\OD[\alpha_{u,v}]} \, (X_1\OD[\alpha_{u,v}])_0)\\
        &= -\i\, \w_{\alpha_{u,v}}(\mL_{\i[ \Phi^u,X_j]} \, (X_1\OD[\alpha_{u,v}])_0)\\
        &= \i\, \w_{\alpha_{u,v}}(\mL_{X_1} \, (\i\,[\, \Phi^u,\,X_j\,])_0)\\
        &= \w_{\alpha_{u,v}}(\mL_{X_1} \, \mL_{X_j}\, \Phi^u_0)\, .
    \end{align*}
    It therefore holds that 
    \begin{equation*}
        \partial_u \i\w_0(\alpha_{u,v} \mL_{X_1}  (X_j\OD[\alpha_{u,v}])_0) = -\w_{\alpha_{u,v}}( \mL_{X_j} \, \mL_{X_{ 1}}\,\Phi^u_0) + \w_{\alpha_{u,v}}(\mL_{X_1} \, \mL_{X_j}\, \Phi^u_0) = 0\,. \qedhere
    \end{equation*}

\section{Proof of Corollary \ref{corr:invertible phase}} \label{app:invertibility}

For $m \in \N_0$ we denote the CAR-algebra associated to the Hilbert space $\ell^2(\Z^d,\C^m)$ by $\mathrm{CAR}(\ell^2(\Z^d,\C^m))$. For $m,k \in \N_0$ we can identify $\mathrm{CAR}(\ell^2(\Z^d,\C^{m+k}))$ with the $\Z_2$-graded tensor product \[\mathrm{CAR}(\ell^2(\Z^d,\C^m)) \otimes_{\Z_2} \mathrm{CAR}(\ell^2(\Z^d,\C^k)).\]

\begin{remark}
    Since the CAR-algebra of a separable Hilbert space is nuclear, there is a unique tensor product C$^*$-algebra $\mathrm{CAR}(\ell^2(\Z^d,\C^m)) \otimes \mathrm{CAR}(\ell^2(\Z^d,\C^k))$. The $\Z_2$-graded tensor product is obtained from the tensor product by modifying multiplication and star operation as explained in the following. 
    An element $A \in \mathrm{CAR}(\ell^2(\mathbb{Z}^d, \mathbb{C}^m))$ is said to be even (resp. odd) if it satisfies $g_{ \pi}(A) = A$ (resp. $g_{ \pi}(A) = -A$). Here, the automorphism $g_{\pi}$ of $\mathrm{CAR}(\ell^2(\mathbb{Z}^d, \mathbb{C}^m))$ is  given  by $g_{ \pi}(a_{x, i}^*) =\e^{\i \pi}a_{x, i}^*= -a_{x, i}^*$.
    For an odd or even element $A$ of one of the CAR algebras  we define 
    \[ \abs{A} = \begin{cases*}
        0 & \text{if $A$ is even}\\
        1 & \text{if $A$ is odd.}
    \end{cases*}\]
    Now let $A_1,B_1 \in \mathrm{CAR}(\ell^2(\Z^d,\C^m))$ and $A_2,B_2 \in \mathrm{CAR}(\ell^2(\Z^d,\C^k))$ each be odd or even. We define the modified star operation and multiplication by
    \begin{align*}
        (A_1 \otimes A_2)^* &= (-1)^{\abs{A_1} \, \abs{A_2}} \,A_1^* \otimes A_2^*, \\
        (A_1\otimes A_2)(B_1 \otimes B_2) &= (-1)^{\abs{A_2} \, \abs{B_1}} \,A_1\, B_1 \otimes A_2 \, B_2
    \end{align*}
    and linear extension. The resulting C$^*$-algebra $\mathrm{CAR}(\ell^2(\Z^d,\C^m)) \otimes_{\Z_2} \mathrm{CAR}(\ell^2(\Z^d,\C^k))$ is isomorphic to $\mathrm{CAR}(\ell^2(\Z^d,\C^{m+k}))$.
\end{remark}

\medskip

\begin{definition} \label{def:invertible}
Let $\omega$ on $\mA = \mathrm{CAR}(\ell^2(\Z^d,\C^n))$ be the unique gapped ground state of a Hamiltonian $H$ in $B_{\exp}$. We call $\w$ invertible if there are $m\in \N_0$ and a state $\tilde{\omega}$ on $\mathrm{CAR}(\ell^2(\Z^d,\C^m))$ that is the unique gapped ground state of a Hamiltonian $\tilde{H}$ in $B_{\exp}$, such that there exists a family of interactions $(\Phi^u)_{u\in \R}$ on $\mathrm{CAR}(\ell^2(\Z^d,\C^{n+m}))$ in $B_{\exp}$ with unique gapped ground states $(\omega_u)_{u\in\R}$ that satisfies:
\begin{itemize}
    \item[(i)] $\Phi^0 = H \otimes \mathbb{1} + \mathbb{1} \otimes \tilde{H}$, 
     i.e. $\Phi^0(M) = H(M) \otimes \mathbb{1} + \mathbb{1} \otimes \tilde{H}(M)$ for all $M\in P_0(\Z^d)$. 
    \item[(ii)] $\forall M \in P_0(\Z^d)$ the map $\R \to \mathrm{CAR}(\ell^2(\Z^d,\C^{n+m})),\, u\mapsto \Phi^u(M)$ is differentiable.
    \item[(iii)] $\Phi^1$ is supported only on singletons.
\end{itemize} 
\end{definition}

\medskip

\begin{proposition}\label{prop:doublecommutator for half planes}
    Let $\br{H,W,T,\br{S_{\eps}}_{\eps \in \R}}$ be a gapped periodic system  in dimension $d=2$. For $j\in \{1,2\}$  we denote the right/upper half plane respectively by
    \begin{align*}
        \Sigma_j \coloneq \{ (x_1,x_2)\in \Z^2 \, | \, x_j \geq 0\}\,.
    \end{align*}
    We further define  
    \begin{align*}
        K_{x,x'} = \int_{\R} \mathrm{d}s \, W_g(s)\, \e^{\i s\mL_H} \left( \i[T_x h_0,n_{x'}] - \i [T_{x'} h_0, n_x]\right) 
    \end{align*}
    for $x,x' \in \Z^2$ and
    \begin{align*}
        [K_{\Sigma_1,\Sigma_1^{\mathsf{C}}}, K_{\Sigma_2,\Sigma_2^{\mathsf{C}}}] = \sum_{x\in \Sigma_1} \sum_{x'\in \Sigma_1^{\mathsf{C}}}\sum_{y\in \Sigma_2} \sum_{y'\in \Sigma_2^{\mathsf{C}}} [K_{x,x'},K_{y,y'}]\,.
    \end{align*}
    It holds that
    \begin{equation} \label{eq:DCF=Kapustin}
            \overline{\w_0}(\i[X_1\OD,X_2\OD]) = \w_0(\i[K_{\Sigma_1,\Sigma_1^{\mathsf{C}}}, K_{\Sigma_2,\Sigma_2^{\mathsf{C}}}])\, .
        \end{equation}
\end{proposition}

In \cite{kapustin2020hall} it is shown that, if $\w_0$ is invertible in the sense of Definition \ref{def:invertible}, the right-hand side of \eqref{eq:DCF=Kapustin} lies in $\frac{1}{2\pi} \Z$ (the relevant statements can be found in \cite[Theorem 1]{kapustin2020hall} and \cite[Equation 55]{kapustin2020hall}). The previous result then allows to conclude the proof of Corollary \ref{corr:invertible phase} on quantization of the Hall conductivity in phases connected to an invertible one.

\begin{proof}[Proof of \cref{prop:doublecommutator for half planes}]
    We observe that
    \begin{align*}
        \sum_{y\in \Sigma_2}& \sum_{y'\in \Sigma_2^{\mathsf{C}}} [K_{x,x'},K_{y,y'}]\\
        &=\sum_{y\in \Sigma_2} \sum_{y'\in \Sigma_2^{\mathsf{C}}} \Big[K_{x,x'},\int_{\R} \mathrm{d}s \, W_g(s)\, \e^{\i s\mL_H} \i[T_y h_0,n_{y'}]\Big] - \sum_{y\in \Sigma_2} \sum_{y'\in \Sigma_2^{\mathsf{C}}} \Big[K_{x,x'},\int_{\R} \mathrm{d}s \, W_g(s)\, \e^{\i s\mL_H} \i [T_{y'} h_0, n_{y}]\Big]\, .
    \end{align*}
    Changing the order of summation/integration by using \cref{convergence} and \cref{cocycles} to show absolute convergence results in 
    \begin{align*}
        \sum_{y\in \Sigma_2} &\sum_{y'\in \Sigma_2^{\mathsf{C}}} [K_{x,x'},K_{y,y'}]\\
         &=-\i \sum_{y\in \Sigma_2}  \Big[K_{x,x'},\int_{\R} \mathrm{d}s \, W_g(s)\, \e^{\i s\mL_H} \mL_{N_{\Sigma_2^{\mathsf{C}}}} T_y h_0\Big] +  \i \sum_{y'\in \Sigma_2^{\mathsf{C}}} \Big[K_{x,x'},\int_{\R} \mathrm{d}s \, W_g(s)\, \e^{\i s\mL_H} \mL_{N_{\Sigma_2}} T_{y'} h_0 \Big]\,,
    \end{align*}
    where $N_{\Sigma}$ is the number operator restricted to $\Sigma$ for any $\Sigma \subseteq \Z^2$. Due to gauge-invariance we have
    \begin{align*}
        \mL_{N_{\Sigma_2^{\mathsf{C}}}} T_y h_0 = \mL_N T_y h_0 - \mL_{N_{\Sigma_2}} T_y h_0 = - \mL_{N_{\Sigma_2}} T_y h_0\, ,
    \end{align*}
    which yields
    \begin{align*}
        &\sum_{y\in \Sigma_2} \sum_{y'\in \Sigma_2^{\mathsf{C}}} [K_{x,x'},K_{y,y'}]\\
        &= \i \sum_{y\in \Sigma_2}  \Big[K_{x,x'},\int_{\R} \mathrm{d}s \, W_g(s)\, \e^{\i s\mL_H} \mL_{N_{\Sigma_2}} T_y h_0\Big] +  \i \sum_{y'\in \Sigma_2^{\mathsf{C}}} \Big[K_{x,x'},\int_{\R} \mathrm{d}s \, W_g(s)\, \e^{\i s\mL_H} \mL_{N_{\Sigma_2}} T_{y'} h_0\Big]\\
        &= \i \sum_{y \in \Z^2}  \Big[K_{x,x'},\int_{\R} \mathrm{d}s \, W_g(s)\, \e^{\i s\mL_H} \mL_{N_{\Sigma_2}} T_y h_0\Big]\,.
    \end{align*}
    Next we reintroduce the remaining summations and consider the expression
    \begin{align*}
        \i \sum_{x\in \Sigma_1} \sum_{x'\in \Sigma_1} \sum_{y \in \Z^2}  \Big[K_{x,x'},\int_{\R} \mathrm{d}s \, W_g(s)\, \e^{\i s\mL_H} \mL_{N_{\Sigma_2}} T_y h_0\Big]\, ,
    \end{align*}
    where one can again use \cref{cocycles} and \cref{convergence} to show absolute convergence and repeat the same argument from above to arrive at
    \begin{align*}
         \sum_{x\in \Sigma_1} \sum_{x'\in \Sigma_1} \sum_{y\in \Sigma_2} \sum_{y'\in \Sigma_2^{\mathsf{C}}} [K_{x,x'},K_{y,y'}]
        =-\sum_{x \in \Z^2} \sum_{y \in \Z^2} \Big[\int_{\R} \mathrm{d}s \, W_g(s)\, \e^{\i s\mL_H} \mL_{N_{\Sigma_1}} T_x h_0,\int_{\R} \mathrm{d}s \, W_g(s)\, \e^{\i s\mL_H} \mL_{N_{\Sigma_2}} T_y h_0\Big]\,.
    \end{align*}
    We evaluate this expression under the periodic state $\w_0$:
    \begin{align*}
        S \coloneq &\sum_{x\in \Sigma_1} \sum_{x'\in \Sigma_1} \sum_{y\in \Sigma_2} \sum_{y'\in \Sigma_2^{\mathsf{C}}} \i \w_0([K_{x,x'},K_{y,y'}]) \\
        =&\sum_{x \in \Z^2} \sum_{y \in \Z^2} -\i \w_0(\Big[\int_{\R} \mathrm{d}s \, W_g(s)\, \e^{\i s\mL_H} \mL_{N_{\Sigma_1}} T_x h_0,\int_{\R} \mathrm{d}s \, W_g(s)\, \e^{\i s\mL_H} \mL_{N_{\Sigma_2}} T_y h_0\Big])\\
        =&\sum_{x \in \Z^2} \sum_{y \in \Z^2} -\i \w_0(\Big[\int_{\R} \mathrm{d}s \, W_g(s)\, \e^{\i s\mL_H} T_{(x_1,0)}^{-1}\mL_{N_{\Sigma_1}} T_x h_0,\int_{\R} \mathrm{d}s \, W_g(s)\, \e^{\i s\mL_H} \mL_{N_{\Sigma_2}} T_{(x_1,0)}^{-1} T_y h_0\Big]) \, .
    \end{align*}
    Since $H$ is a $T$-periodic $B_\infty$-interaction it holds that $T_{\g+\mu} h_0 = T_\g T_\mu h_0$ for all $\g,\mu \in \Z^2$ (as shown in Proposition \ref{prop: interaction associated to observable}), giving us
    \begin{align*}
        S = \sum_{x \in \Z^2} \sum_{y \in \Z^2} -\i \w_0(\Big[\int_{\R} \mathrm{d}s \, W_g(s)\, \e^{\i s\mL_H} T_{(x_1,0)}^{-1}\mL_{N_{\Sigma_1}} T_{(x_1,0)} T_{(0,x_2)}  h_0,\int_{\R} \mathrm{d}s \, W_g(s)\, \e^{\i s\mL_H} \mL_{N_{\Sigma_2}} T_{(y_1-x_1,y_2)} h_0\Big])\, .
    \end{align*}
    Now an index shift lets us eliminate $x_1$ in the second argument of the commutator, which allows us to sum over $x_1$ in the first argument by using \cref{lem:shifted half-plane} below. This results in
    \begin{align*}
        S &= \sum_{x \in \Z^2} \sum_{y \in \Z^2} -\i \w_0(\Big[\int_{\R} \mathrm{d}s \, W_g(s)\, \e^{\i s\mL_H} T_{(x_1,0)}^{-1}\mL_{N_{\Sigma_1}} T_{(x_1,0)} T_{(0,x_2)}  h_0,\int_{\R} \mathrm{d}s \, W_g(s)\, \e^{\i s\mL_H} \mL_{N_{\Sigma_2}} T_{(y_1,y_2)}  h_0\Big])\\
        &=\sum_{x_2 \in \Z} \sum_{y \in \Z^2} -\i \w_0(\Big[\int_{\R} \mathrm{d}s \, W_g(s)\, \e^{\i s\mL_H} \mL_{X_1} T_{(0,x_2)}  h_0,\int_{\R} \mathrm{d}s \, W_g(s)\, \e^{\i s\mL_H} \mL_{N_{\Sigma_2}} T_{(y_1,y_2)}  h_0\Big])\\
        &=\sum_{x_2 \in \Z} \sum_{y \in \Z^2} -\i \w_0(\Big[\int_{\R} \mathrm{d}s \, W_g(s)\, \e^{\i s\mL_H}   \mL_{X_1} T_{(-y_1,x_2-y_2)}  h_0,\int_{\R} \mathrm{d}s \, W_g(s)\, \e^{\i s\mL_H} T_{(0,y_2)}^{-1}\mL_{N_{\Sigma_2}} T_{(0,y_2)}  h_0\Big])\, ,
    \end{align*}
    where we used gauge-invariance of $h_0$ and periodicity to move the Translation operators in the last step. Another index shift lets us eliminate $y_2$ in the first argument of the commutator and apply \cref{lem:shifted half-plane} to get
    \begin{align*}
        S &= \sum_{x_2 \in \Z} \sum_{y \in \Z^2} -\i \w_0(\Big[\int_{\R} \mathrm{d}s \, W_g(s)\, \e^{\i s\mL_H}   \mL_{X_1} T_{(-y_1,x_2)}  h_0,\int_{\R} \mathrm{d}s \, W_g(s)\, \e^{\i s\mL_H} T_{(0,y_2)}^{-1}\mL_{N_{\Sigma_2}} T_{(0,y_2)}  h_0\Big])\\
        &=\sum_{x_2 \in \Z} \sum_{y_1 \in \Z} -\i \w_0(\Big[\int_{\R} \mathrm{d}s \, W_g(s)\, \e^{\i s\mL_H}   \mL_{X_1} T_{(-y_1,x_2)}  h_0,\int_{\R} \mathrm{d}s \, W_g(s)\, \e^{\i s\mL_H} \mL_{X_2}  h_0\Big])\, .
    \end{align*}
    By gauge-invariance of $h_0$ and periodicity of $H$ we can move the translation in front of the integral
    \begin{align*}
        S &\;= \sum_{x_2 \in \Z} \sum_{y_1 \in \Z} -\i \w_0\left(\Big[T_{(-y_1,x_2)} \int_{\R} \mathrm{d}s \, W_g(s)\, \e^{\i s\mL_H}   \mL_{X_1} h_0,\int_{\R} \mathrm{d}s \, W_g(s)\, \e^{\i s\mL_H} \mL_{X_2}  h_0\Big]\right)\\
        &\;= \sum_{\g \in \Z^2} -\i \w_0\left(\Big[T_\g \int_{\R} \mathrm{d}s \, W_g(s)\, \e^{\i s\mL_H}   \mL_{X_1} h_0,\int_{\R} \mathrm{d}s \, W_g(s)\, \e^{\i s\mL_H} \mL_{X_2}  h_0\Big]\right)\,.
    \end{align*}
    With the definition of the off-diagonal map (Lemma \ref{lem: almost local obs. for OD and inv. liou.}), along with Propositions \ref{liouvillian tools} and \ref{commutator of interactions}, we now find:
    \begin{align*}
        & S = \sum_{\g \in \Z^2} \i \w_0([T_\g (X_1\OD)_0, (X_2\OD)_0])\;= \;\overline{\w_0}(\i[ X_1\OD, X_2\OD])\,. \qedhere
    \end{align*}
\end{proof}

\begin{lemma}\label{lem:shifted half-plane}
    Let $d=2$ and let $T$ be a translation. For $j\in \{1,2\}$ let $N_{\Sigma_j}$ be the number operator restricted to the half plane $\Sigma_j$ and let $A\in D_\infty$. It holds that
    \begin{align*}
       \sum_{y_1 \in\Z} T_{(y_1,0)}^{-1} \mL_{N_{\Sigma_1}}T_{(y_1,0)} A = \mL_{X_1} A \, \qquad\mbox{and}\qquad 
       \sum_{y_2 \in\Z} T_{(0,y_2)}^{-1} \mL_{N_{\Sigma_2}}T_{(0,y_2)} A = \mL_{X_2} A \,.
    \end{align*}
\end{lemma}

\begin{proof}
    We start by considering a local gauge-invariant operator $A\in \mA_0^N$. Let $k\in\N_0$ be such that $A \in \mA_{\La_k}$. We consider
    \begin{align*}
        \mL_{X_1} A &= \sum_{x_2 = -k}^k \sum_{x_1 = -k}^k [x_1n_{(x_1,x_2)},A]
    \end{align*}
    and use the gauge-invariance of $A$ to get
    \begin{align*}
        \mL_{X_1} A \;&= \sum_{x_2 = -k}^k \sum_{x_1 = -k}^k [x_1n_{(x_1,x_2)},A] + (k+1) \mL_N A \;= \sum_{x_2 = -k}^k \sum_{x_1 = -k}^k [(x_1+k+1)n_{(x_1,x_2)},A] \\
        &= \sum_{x_2 = -k}^k \sum_{x_1 = -k}^k \sum_{z=0}^{x_1+k}[n_{(x_1,x_2)},A] \,.
    \end{align*}
    Next we employ the indicator function $\chi$ to change the order of summation 
    \begin{align*}
        \mL_{X_1} A \;
        &= \sum_{x_2 = -k}^k \sum_{x_1 = -k}^k \sum_{z=-k}^{x_1}  [n_{(x_1,x_2)},A] \;= \sum_{x_2 = -k}^k \sum_{x_1 = -k}^k \sum_{z=-k}^{k} \chi_{z \leq x_1} [n_{(x_1,x_2)},A]\\
        &= \sum_{x_2 = -k}^k  \sum_{z=-k}^{k} \sum_{x_1 = z}^{k}  [n_{(x_1,x_2)},A] \,.
    \end{align*}
    Finally, using an index shift and the shape of the support of $A$ we arrive at
    \begin{align*}
        \mL_{X_1} A \;
         =\; \sum_{x_2 = -k}^k  \sum_{z=-k}^{k} \sum_{x_1 = 0}^{k}  [n_{(x_1+z,x_2)},A] \;= \sum_{z=-k}^{k} T_{(-z,0)}^{-1}\mL_{N_{\Sigma_1}} T_{(-z,0)} A \;= \;\sum_{z\in \Z} T_{(z,0)}^{-1}\mL_{N_{\Sigma_1}} T_{(z,0)} A \, .
    \end{align*}
    Now let $A\in D_\infty$. The sum 
    \begin{align*}
        S(A) \coloneq \sum_{y_1 \in\Z} T_{(y_1,0)}^{-1} \mL_{N_{\Sigma_1}}T_{(y_1,0)} A
    \end{align*}
    converges absolutely. This can be seen by splitting it in two parts and using gauge-invariance to get
    \begin{align*}
        S(A) = - \sum_{y_1 = 0 }^\infty \sum_{x\in \Sigma_1^{\mathsf{C}}} [n_{x-(y_1,0)},A] +  \sum_{y_1 = 1 }^\infty \sum_{x\in \Sigma_1} [n_{x+(y_1,0)},A]\, .
    \end{align*}
    The estimates
    \begin{align*}
        &\norm{[n_{x\pm(y_1,0)},A]} \leq 4^{8} \frac{\norm{n_0}_5 \, \norm{A}_5}{(1 + \norm{x\pm(y_1,0)}_\infty)^5}
    \end{align*}
    from \cref{convergence} give us absolute convergence and further tell us that $S(A)$ goes to $0$ if $\norm{A}_5$ goes to $0$. For $k\in\N_0$ we have
    \begin{align*}
        \sum_{y_1 \in\Z} T_{(y_1,0)}^{-1} \mL_{N_{\Sigma_1}}T_{(y_1,0)} A \;&
        = \sum_{y_1 \in\Z} T_{(y_1,0)}^{-1} \mL_{N_{\Sigma_1}}T_{(y_1,0)} \E_{\La_k}A + \sum_{y_1 \in\Z} T_{(y_1,0)}^{-1} \mL_{N_{\Sigma_1}}T_{(y_1,0)} (A -\E_{\La_k}A)\\
        &= \mL_{X_1} A -\mL_{X_1}(A -\E_{\La_k}A) +  S(A -\E_{\La_k}A)\,.
    \end{align*}
    Since $\norm{A -\E_{\La_k}A}_\nu $ goes to $0$ as $k\to \infty$ for all $\nu$ (Lemma \ref{lem:norm properties}), our previous considerations and \cref{lem: sum representation of generator} imply that the last two terms vanish as $k \to \infty$. This shows the claim for $N_{\Sigma_1}$. The claim for $N_{\Sigma_2}$ can be proved analogously.
\end{proof}

\section{Proof of Proposition~\ref{prop:NEASS}}\label{app:NEASS}

The construction of $S_\eps$ was first done in \cite{Teufel2020} for finite systems with bounds uniform in the system size and then in \cite{HenheikTeufel2022,HT20b} for infinitely extended systems in the thermodynamic limit. However, in these works the focus was on the adiabatic theorem and also on the thermodynamic limit itself. Therefore the proofs are not only very technical, but in \cite{HenheikTeufel2022,HT20b} it was also assumed that the ground state $\omega_0$ is unique. Since  uniqueness is very difficult to establish in concrete problems and not needed for our results, we provide here a construction of the NEASS in  infinite volume without assuming that $\omega_0$ is the unique ground state.

The interaction $S_\eps$ is constructed as a suitable resummation of a formal power series $\sum_{\mu=1}^\infty \epsi^\mu K_\mu$ with coefficients $K_\mu\in B_{\infty,T}$ for  $\mu\in\N$. Within this proof we denote by $B_{\infty,T}$ the space of $T$-periodic elements of $B_\infty$.  Roughly speaking, the coefficients $K_\mu$ are determined by the requirement that 
\[
\omega_\eps(\mL_{H_{\eps}}A)
\;= \;\omega_0(\beta_\eps\mL_{ H_\eps}   A) \;= \; \omega_0(\mL_{\beta_\eps H_\eps} \beta_\eps A) 
\;= \; \omega_0(\mL_{(\beta_\eps H_\eps)\OD} \beta_\eps A) \;\stackrel{!}{=} \;\mathcal{O}(\epsi^\infty)\,,
\]
i.e.\ by constructing $\beta_\eps =\e^{\i \mL_{S_{\eps}}} $ such that it block-diagonalizes $H_\eps$ with respect to $\omega_0$ by achieving $(\beta_\eps H_\eps)\OD=\mathcal{O}(\eps^\infty)$. 

More precisely, assume  that $K_\mu\in B_{\infty,T}$ and let
 $\beta_\eps^{(m)}(t) := \e^{\i t \mL_{S_{\eps}^{(m)}}}$ with $S_\eps^{(m)}:= \sum_{\mu=1}^m \epsi^\mu K_\mu$.  Then
\begin{eqnarray*}
\omega_0(\beta_\eps^{(m)}(1)\mL_{H_{\eps}}A) = \omega_0(\beta_\eps^{(m)}(t)\mL_{H_{\eps}} \beta_\eps^{(m)}(-t)\beta_\eps^{(m)}(1)A)|_{t=1} \,.
\end{eqnarray*}
For any $B\in D_\infty$ Taylor expansion at $t=0$ yields
\begin{eqnarray*}\lefteqn{
    \beta_\eps^{(m)}(t)\mL_{H_{\eps}} \beta_\eps^{(m)}(-t) B|_{t=1}\;=}\\&=&
    \sum_{k=0}^m \frac{1}{k!} \frac{\mathrm{d}^k}{\mathrm{d}t^k}\left(\beta_\eps^{(m)}(t)\mL_{H_{\eps}} \beta_\eps^{(m)}(-t) B\right)|_{t=0}\, +\,\frac{1}{(m+1)!} \frac{\mathrm{d}^{m+1}}{\mathrm{d}t^{m+1}}\left(\beta_\eps^{(m)}(t)\mL_{H_{\eps}} \beta_\eps^{(m)}(-t) B\right)|_{t=\tau}\\
    &=&  \sum_{k=0}^m \frac{1}{k!} 
    \mL_{\mathrm{ad}({\i S_\epsi^{(m)}})^k H_\epsi} 
    B\,+\, \frac{1}{(m+1)!}  \beta_\eps^{(m)}(\tau) \mL_{\mathrm{ad}({\i S_\epsi^{(m)}})^{m+1} H_\epsi}  \beta_\eps^{(m)}(-\tau) B
\end{eqnarray*}
for some $\tau\in[0,1]$, where $\mathrm{ad}({\i S_\epsi^{(m)}})^k H_\epsi$ denotes the $k$ times iterated commutator of  $\i S_\epsi^{(m)}$ with $H_\eps$. Now
\begin{eqnarray*}
\sum_{k=0}^m\frac{1}{k!}\,\mathrm{ad}({\i S_\epsi^{(m)}})^k H_\epsi &=& \sum_{k=0}^m\frac{1}{k!}\,
\mathrm{ad}({\i S_\epsi^{(m)}})^k (H + \epsi (X_1+V)) \\
&=:& \sum_{\mu=0}^m \epsi^\mu (H_{\mu} + V_{\mu}) + \epsi^{m+1}R^{(m)}_\eps
\end{eqnarray*}
with
\[
H_{\mu} = \i [K_\mu , H] + L_\mu
\]
and $L_\mu$ being a sum of iterated commutators of $K_j$'s with $H$ for $j<\mu$ and $V_{\mu}$ a sum of iterated commutators of $K_j$'s with   $X_1 + V$ for $j<\mu$. 
In $\epsi^{m+1}R^{(m)}_\eps$ we collect all terms that contain at least $m+1$ factors of $\eps$, so that $ R^{(m)}_\eps$ is a finite sum of iterated commutators of elements of $B_{\infty,T}$ and thus itself an element of $B_{\infty,T}$.

The condition $\w_0(\mL_{H_\mu+V_\mu} \, B)=0$ now becomes
\begin{equation}\label{eq:Kmu}
    \w_0(\mL_{\i[K_\mu , H]}\, B) \,=\, -  \w_0(\mL_{L_\mu + V_{\mu}} \, B) \, ,
\end{equation}
where the right hand side of \eqref{eq:Kmu} depends only on $K_1,\ldots,K_{\mu-1}$ and not on $m$. 
Recalling the definition of $\mathcal{I}$ (Proposition \ref{prop: interaction associated to observable} and Lemma \ref{lem: almost local obs. for OD and inv. liou.}) and applying Lemma \ref{OD-property},
 we can solve \eqref{eq:Kmu} inductively starting with 
\[
K_1 := -\mathcal{I}((X_1+V)) 
\]
and then, given $K_1,\ldots, K_{\mu-1}$ and thus also $L_\mu$ and $V_\mu$, 
\begin{equation}\label{eq:iter}
 K_\mu := -\mathcal{I}( L_\mu + V_{\mu})
\end{equation}
for $\mu=2,\ldots,m$. Note here, that $K_\mu$ is independent of $m\geq\mu$. Hence \eqref{eq:iter} indeed determines a sequence $(K_\mu)_{\mu\in\N}$ in $B_{\infty,T}$. 

We conclude that this choice for $(K_\mu)_{\mu\in\N}$   leads for every $m\in \N$ to 
\begin{align}
 \sum_{k=0}^m \frac{1}{k!}\w_0(\mL_{\mathrm{ad}({\i S_\epsi^{(m)}})^k H_\epsi} \, B) = \epsi^{m+1} \w_0(\mL_{R^{(m)}_\eps} \, B)\, , 
\end{align}
where $R^{(m)}_\eps$ is bounded uniformly in $\eps\in [-1,1]$ in any of the interaction norms $\lVert \cdot \rVert_\nu$. 
Thus, with Lemma~\ref{lem: sum representation of generator},  there exists $c_{m}$ such that for any  $A\in D_\infty$  
\begin{eqnarray}\nonumber
\left|\omega_0\left(\sum_{k=0}^m \frac{1}{k!} 
\mL_{\mathrm{ad}({\i S_\epsi^{(m)}})^k H_\epsi}  A\right)\right| &=& |\eps|^{m+1} \left|\omega_0\left(\mL_{R^{(m)}_\eps} \,  A\right) \right|\leq |\eps|^{m+1} \left\|\mL_{R^{(m)}_\eps}\,A \right\|
\\&\nonumber\stackrel{\mathrm{Lemma}~\ref{lem: sum representation of generator}}{\leq}& 
|\eps|^{m+1} c \|R^{(m)}_\eps\|_{d+1} \,\|A\|_{d+3 }
\\&\leq& 
\lvert \eps \rvert^{m+1} c_{m} \norm{A}_{d+3}\,. \label{A}
\end{eqnarray}
and also  (using Lemma \ref{lem: sum representation of generator} and Lemma \ref{cocycles})
\begin{align}
\left| \omega_0\left( \beta_\eps^{(m)}(\tau) \mL_{\mathrm{ad}({\i S_\epsi^{(m)}})^{m+1} H_\epsi}  \beta_\eps^{(m)}(-\tau) \beta_\eps^{(m)}(1)A)\right) \right| \leq \lvert\eps\rvert^{m+1} c_{m}\|A\|_{d+3}\,.
\end{align}
In summary, we find that for all $A\in D_\infty$
\begin{align}
| \omega_0(\beta_\eps^{(m)}(1)\mL_{H_{\eps}}A) |\leq 2 \,c_m\lvert \eps \rvert^{m+1} \|A\|_{d+3}\,.\label{B}
\end{align}
Finally, we   define 
\begin{equation} \label{eqn:2103}
S_\eps := \sum_{\mu=1}^\infty \chi(\lvert \eps \rvert/\delta_\mu) \,\eps^\mu K_\mu
\end{equation}
with $\chi(x):=\mathbf{1}_{[0,1]}(x)$ the characteristic function of the interval $[0,1]$ and $(\delta_{\mu})_{\mu \in \N}$ the monotonically decreasing sequence defined by setting $\delta_1:=1$ and
\[\delta_\mu := \min \left\{ \frac{1}{2^\mu}, \frac{1}{2\,\max\{1, \|K_1\|_\mu,\ldots,\|K_\mu\|_\mu\}},  \frac{1}{\max\{1, c_{1},\ldots,c_{\mu}\}},\delta_1,\ldots,\delta_{\mu-1} \right\}\,, \quad \mu > 1\, .\]
In~\eqref{eqn:2103}, the $\mu$-th term in the series is included as soon as $\lvert \eps \rvert \leq \delta_\mu$.

Let $\nu \in \N$. For $k\geq \nu$ and $\delta_{k+1} < \lvert \eps \rvert\leq \delta_k$ 
\[
\|S_\eps\|_\nu \leq \sum_{\mu=1}^k \lvert \eps \rvert^\mu \|K_\mu\|_\nu \leq \lvert \eps \rvert \sum_{\mu=1}^k \delta_\nu^{\mu-1} \|K_\mu\|_\nu \leq \lvert \eps \rvert \sum_{\mu=1}^k 2^{1-\mu}\leq 2 \lvert\eps \rvert \, .
\]
For $k < \nu$ and $\delta_{k+1} < \lvert \eps \rvert \leq \delta_k$
\[
\|S_\eps\|_\nu \leq \sum_{\mu=1}^k \lvert \eps \rvert^\mu \|K_\mu\|_\nu \leq \lvert \eps \rvert \sum_{\mu=1}^\nu  \|K_\mu\|_\nu \, .
\]
Therefore,  $\sup_{\eps\in [-1,1]\setminus \{0\}} \frac{1}{\lvert \eps \rvert}\lVert S_\eps \rVert_\nu \leq \max(2, \sum_{\mu=1}^\nu  \|K_\mu\|_\nu) $ for all $\nu\in \N$.

With 
\[
C_n := 2\,\max\left\{\max_{k<n}\left\{ \frac{c_{k}\delta_k^{k+1}}{\delta_{k+1}^n}\right\}, 1  \right\}\,, \quad n \in \N\,,
\]
we find that for $\delta_{k+1} < \lvert\epsi\rvert\leq \delta_k$ and $n\leq k$
\[
c_{k} \lvert\eps\rvert^{k+1} \leq c_{k} \lvert \eps \rvert^{k+1-n} \lvert\eps \rvert^n\leq c_{k} \lvert\eps \rvert  \lvert\eps\rvert^n\leq c_{k} \delta_k  \lvert\eps\rvert^n \leq \lvert\eps\rvert^n\leq \tfrac{1}{2}C_n \lvert\eps\rvert^n\,,
\]
and that for $n>k$
\[
c_{k} \lvert\eps\rvert^{k+1} \leq \frac{c_{k} \delta_k^{k+1}}{\lvert\eps\rvert^n}\lvert\eps\rvert^n\leq \frac{c_{k} \delta_k^{k+1}}{\delta_{k+1}^n}\lvert\eps\rvert^n\leq \tfrac{1}{2}C_n\lvert\eps\rvert^n\,.
\]
Hence, for $\delta_{k+1} < \lvert\eps\rvert\leq \delta_k$ and all $n\in\N$ it follows that
\[
| \omega_0(\beta_\eps\mL_{H_{\eps}}A) | = | \omega_0(\beta_\eps^{(k)}(1)\mL_{H_{\eps}}A) | \leq 2 c_{k} \lvert\eps\rvert^{k+1} \|A\|_{d+3} \leq  C_n\lvert\eps\rvert^n\,\|A\|_{d+3}\,. \qed
\]

\section{Proof of Proposition~\ref{prop:Hofstadter}}\label{sec:proofHofstadter}

We start by recalling that the one-parameter group $\tau_t$ of automorphisms generated by a quadratic interaction of the form $H_0^{b,\mu} = \mathrm{d}\Gamma(\mathfrak{h}_0^{b}-\mu)$ 
on $\mA$ is the group of Bogoliubov transformations $\mA \ni A \mapsto \tau_t(A) = U_t A U_{-t} $ with $ U_t=\Gamma(\mathrm{e}^{\mathrm{i}(\mathfrak{h}_0^{b}-\mu)t})$. For details on the second quantization procedures $\mathrm{d}\Gamma$ and $\Gamma$, see \cite[Chapter 5]{bratteliII}. 
Moreover, according to 
\cite[Example 5.3.20]{bratteliII}, the unique ground state  of the generator of such a Bogoliubov transformation is the quasi-free state $\omega_0$ with two-point function
\[
\omega_0(a_x^* a_y) = P_{(-\infty,\mu)}^{\mathfrak{h}_0^{b}}(x,y)\,,
\]
where $P_{(-\infty,\mu)}^{\mathfrak{h}_0^{b}}(\cdot,\cdot)$ is the integral kernel of the spectral projection $P_{(-\infty,\mu)}^{\mathfrak{h}_0^{b}}$ of $\mathfrak{h}_0^{b}$ onto the set $(-\infty,\mu)$, the so called Fermi projection. In \cite[Example 5.3.20]{bratteliII} it is assumed that  $\mu$ is not an eigenvalue of $\mathfrak{h}_0^{b}$, which holds in our case since $\mu\notin \sigma(\mathfrak{h}_0^{b})$ by assumption.
We will see below that $\omega_0$ is also gapped.

We would like to apply the result of \cite{de2019persistence} showing stability of the spectral gap of free Fermion systems under small perturbations. However, the result of \cite{de2019persistence} holds a priori only for systems on finite lattices $\Z^2/((2k+1)\Z)^2$ for $k\in \N$.
Hence we first need to restrict $\mathfrak{h}_0^{b}$ to $\Lambda_k \cong \Z^2/((2k+1)\Z)^2$ with periodic boundary conditions, which, however, only works if $ b (2k+1) \in 2\pi \Z$. 
So for the moment we assume that $ b=\frac{p}{q}2\pi$ for some odd $q\in\N$ and $p\in\Z$ and consider the sequence $k_n:= \frac{1}{2}(nq-1)$ of system sizes. Let  $H_{\lambda,n}^{b,\mu}:=H_{\lambda}^{b,\mu}|_{\Lambda_{k_n}}$
denote the corresponding finite volume Hamiltonian which is $T^b|_{\Lambda_{k_n}}$-invariant. Here $T^b|_{\Lambda_{k_n}}$ denotes the magnetic translation acting on $\mA_{\La_{k_n}}$ with periodic boundary conditions.

For all   $n\in\N$, the ground state $\omega_{0,n}$ of $H_{0,n}^{b,\mu}$ is the quasi-free state  with corresponding two-point function $ P_{(-\infty,\mu),n}^{\mathfrak{h}^b_0}(x,y)$ which is gapped with gap at least $g := \mathrm{dist}(\mu, \sigma(\mathfrak{h}_0^{b})  ) $ and $T^b|_{\Lambda_{k_n}}$-invariant.\footnote{This is an elementary fact that can be seen as follows: In any finite non-interacting Fermi system with one-body operator $\mathfrak{h}_0$ and Hamiltonian $H_0^\mu = \mathrm{d}\Gamma(\mathfrak{h}_0 - \mu)$ an orthonormal basis of eigenvectors of $H_0^\mu $ is given by Slater determinants of the form $\omega_{j_1,\ldots,j_N} = a^*(f_{j_1})\cdots a^*(f_{j_N}) |0\rangle$, where $|0\rangle$ denotes the Fock vacuum and $f_{j_k}$ are eigenvectors of $\mathfrak{h}_0$. It is straightforward to check that the ground state of $H_0^\mu $ is the unique eigenstate containing exactly the one-body eigenfunctions with eigenvalues smaller than $\mu$ and that all other eigenstates have eigenvalues at least $g$ larger. Uniqueness of the ground state implies that it is also $T^b|_{ \La_{k_n}}$-invariant. The following calculation shows that if $P_0$ is the spectral projection onto the ground state subspace of a self-adjoint operator $H$ with spectral gap of size $g$ defined on a finite dimensional space, then the ground state $\omega_0(\cdot):= \mathrm{tr}(P_0\cdot)$ also satisfies the condition \eqref{eq:gap}: 
\begin{eqnarray*}
\omega_{0}(A^* [H,A]) &=& \mathrm{tr}( P_0 A^* [H ,A])= \mathrm{tr}( P_0 A^* H AP_0) - E_0\mathrm{tr}( P_0 A^*  A) \\
&=& \mathrm{tr}( P_0 A^* P_0HP_0 AP_0)+ \mathrm{tr}( P_0 A^* P_0^\perp HP_0^\perp AP_0)  - E_0\omega_0(A^*A)\\
&=& E_0 (\omega_0(A^*)\omega_0(A) -\omega_0(A^*A)) + {\textstyle\sum_{j>0} }E_j\mathrm{tr}(  P_0 A^* P_j  AP_0 )\\
&\geq&  E_0 (\omega_0(A^*)\omega_0(A) -\omega_0(A^*A)) +   E_1\mathrm{tr}(  P_0 A^* (1-P_0)  AP_0 )\\&=&
 (E_1-E_0) (\omega_0(A^*A)- \omega_0(A^*)\omega_0(A) )\,.
\end{eqnarray*}}
According to \cite{de2019persistence} there exists $\lambda_0>0$, independent of $n$,
such that for $|\lambda|<\lambda_0$ and all $n$ the operator $H_{\lambda,n}^{b,\mu}$ has a unique gapped ground state $\omega_{\lambda,n}$ with  gap at least $g/2$. By uniqueness, also $\omega_{\lambda,n}$ is $T^b|_{\Lambda_{k_n}}$-invariant. By compactness of the set of states on $\mA$ the sequence $\tilde \omega_{\lambda,n} := \omega_{\lambda,n}\circ \E_{\Lambda_{k_n}}$ has weak$^{*}$-limit points and each limit point is a gapped ground state for $H_\lambda^{b,\nu}$ with gap at least $g/2$ by the following lemma.\footnote{We expect that the limit point is unique and that also $H_\lambda^{b,\nu}$ has a unique ground state, but the latter is presumably not easy to prove.}

\medskip

\begin{lemma}\label{lem:gaplimit} Let either $\mA_n=\mA_{\Lambda_{k_n}}$ for some strictly increasing map $k:\N\to\N$, $n\mapsto k_n$ or $\mA_n=\mA$ for all $n\in\N$.
  Let $(\omega_n)_{n\in\N}$ be a sequence of states on $\mA$ and  $(\mL_{H_n})_{n\in\N}$ a sequence of derivations on $\mA_n$ such that for all $n\in\N$ and $A\in \mA_n\cap\mA_0$
  \begin{equation}\label{eq:gap2}
  \w_n( A^* \mL_{H_n} A ) \geq g \left( \w_n(A^* A) - \left| \w_n(A) \right|^2 \right)\,.
    \end{equation}
 If  there exists a state $\omega$ and a derivation $\mL_H$ such that for all $A\in\mA_0$ it holds that $\lim_{n\to \infty}\omega_n(A)= \omega(A)$ and $\lim_{n\to \infty}\|(\mL_{H_n}-\mL_H)A\|=0$, then $\omega$ is a gapped ground state of $\mL_H$ with gap at least $g$.
\end{lemma}

\begin{proof}
    Let $A\in \mA_0$, i.e.\ $A\in \mA_n$ for all $n$ large enough. For the left hand side of  \eqref{eq:gap2} we then find
    \begin{eqnarray*}
|\omega_n(A^*\mL_{H_n}A) - \omega(A^*\mL_{H}A)|&\leq&
|\omega_n(A^*(\mL_{H_n}-\mL_H)A)| + |(\omega_n- \omega)(A^*\mL_{H}A)|\\
&\leq& \|A^*\|\,\|(\mL_{H_n}-\mL_H)A\| + |(\omega_n- \omega)(A^*\mL_{H}A)|\\&\stackrel{n\to\infty}{\rightarrow} &0\,,
     \end{eqnarray*}
     while the right hand side of  \eqref{eq:gap2} converges obviously.
\end{proof}
Let $\omega_\lambda$ be such a limit point. Then 
for any $A\in\mA_0$ and $\g \in \Z^2$ there exists an $n_0 \in \N$ such that for all $n \geq n_0$ both $A$ and $T_\g^b\, A$ lie in $\mA_{\La_{k_n}}$. We thus have 
\[
\omega_\lambda(T_\gamma^bA) = \lim_{n\to\infty}\omega_{\lambda,n}(\E_{\Lambda_{k_n}}T_\gamma^bA) = \lim_{n\to\infty}\omega_{\lambda,n}(T_\gamma^b|_{\Lambda_{k_n}}A)
 = \lim_{n\to\infty}\omega_{\lambda,n}(\E_{\Lambda_{k_n}}A)=\omega_\lambda( A)\,,
\]
where, in order not to overload the notation, we have not made the choice of a convergent subsequence explicit. Thus the limit points are also $T^b$-invariant.

Hence, for each   value of $b$ such that $\mu\in \sigma(\mathfrak{h}_0^b)$ and $b/2\pi = \frac{p}{q}$ with $p\in \Z$ and $q \in \N$ odd,  there exists $\lambda_0>0$ such that for each $|\lambda|<\lambda_0$ the interaction $H_\lambda^{b,\mu}$ possesses at least one gapped $T^b$-invariant ground state $\omega^b_{\lambda}$. Since $\lambda_0$ depends only on the size of the gap $g$ of $H_0^{b,\mu}$  and on certain local norms of $H_0^{b,\mu}$ and $\Phi$ that are insensitive to complex phases, $\lambda_0$ can be chosen uniform for $b$ in sufficiently small intervals where $H_0^{b,\mu}$ has a uniform gap. But this allows to find gapped ground states also for irrational $b/2\pi$ such that $g := \mathrm{dist}(\mu, \sigma(\mathfrak{h}_0^{b})  )>0 $  by looking at a rational sequence $b_m$ with $\lim_{m\to\infty}b_m=b$ and taking weak$^{*}$-limit points $\omega^{b}_{\lambda}$ of the corresponding gapped ground states $\omega^{b_m}_{\lambda}$. Such a limit point is gapped by Lemma~\ref{lem:gaplimit}, and $T^b$-invariant, since $T^{b_m}$ converges strongly to $T^b$. More precisely,   for each $A\in\mA_0$
\begin{eqnarray*}
|\omega^b(T^bA) - \omega^b(A)|&\leq &|(\omega^b- \omega^{b_m})(T^bA)| + |\omega^{b_m}(T^bA - T^{b_m}A)| + |\omega^{b_m}(T^{b_m}A)- \omega^b(A)|\\&\leq & |(\omega^b- \omega^{b_m})(T^bA)| + \|T^bA - T^{b_m}A\| + |\omega^{b_m}(A)- \omega^b(A)|\\&\stackrel{m\to\infty}{\longrightarrow} & 0\,.\hspace{9cm}\qedhere
\end{eqnarray*}

\noindent \textbf{Data Availability.}
Data sharing is not applicable to this article as no datasets were generated or analysed during the current study.

\noindent \textbf{Conflict of interests.}
The authors have no competing interests to declare that are relevant to the content of this article.

\printbibliography

\end{document}